\documentclass[prb,
twocolumn,
nopacs,
nofootinbib,
superscriptaddress]{revtex4-2}

\usepackage{graphicx}
\usepackage{bm}
\usepackage{physics}
\usepackage[colorlinks,
    linkcolor=blue,
    anchorcolor=red,
    citecolor=green
]{hyperref}
\usepackage{wasysym}
\usepackage{MnSymbol}
\usepackage{comment}
\usepackage[dvipsnames]{xcolor}
\usepackage[normalem]{ulem}\usepackage[normalem]{ulem}
\usepackage{pifont}

\usepackage{wasysym}

\renewcommand{\v}[1]{\mathbf{#1}}

\usepackage{amsfonts}

\begin{document}

\title{Topologically enabled superconductivity: possible implications for rhombohedral graphene}
	
\author{Francesca Paoletti}
\affiliation{Institut f\"ur Theoretische Physik und Astrophysik and W\"urzburg-Dresden Cluster of Excellence ct.qmat, Universit\"at W\"urzburg, 97074 W\"urzburg, Germany}

\author{Daniele Guerci}
\affiliation{Department of Physics, Massachusetts Institute of Technology, Cambridge, Massachusetts 02139, USA}

\author{Giorgio Sangiovanni}
\affiliation{Institut f\"ur Theoretische Physik und Astrophysik and W\"urzburg-Dresden Cluster of Excellence ct.qmat, Universit\"at W\"urzburg, 97074 W\"urzburg, Germany}

\author{Urban~F.P.~Seifert}
\affiliation{%
Institut f\"ur Theoretische Physik, Universit\"at zu K\"oln, Z\"ulpicher Str. 77a, 50937 K\"oln, Germany
}%
    
\author{Elio J. K\"onig}
\affiliation{%
Department of Physics, University of Wisconsin-Madison, Madison, Wisconsin 53706, USA
}%
 
\date{\today} 	
	
\begin{abstract}
\textbf{}We present a topological mechanism for superconductivity emerging from Chern-2 insulators. While, naively, time-reversal symmetry breaking is expected to prevent superconductivity, it turns out that the opposite is the case:
An explicit model calculation for a generalized attractive-$U$ Haldane-Hubbard model demonstrates that superconductivity is only stabilized near the quantum anomalous Hall state, but not near a trivial, time-reversal symmetric band insulator.
As standard Bardeen-Cooper-Schrieffer-like mean-field theory fails to capture any superconducting state, we explain this using an effective fractionalized field theory involving fermionic chargeons, bosonic colorons and an emergent U(1) gauge field.
When the chargeons form a gapped topological band structure, the proliferation of single monopoles of this gauge field is forbidden. However, long-ranged monopole-antimonopole correlations emerge, and we argue that those correspond to superconducting order.
Using random phase approximation on top of extensive slave-rotor mean-field calculations we characterize coherence length and stiffness of the superconductor.
Thereby, we deduce the phase diagram in parameter space and furthermore discuss the effect of doping, temperature and an external magnetic field. We complement the fractionalized theory with calculations using an effective spin model and Gutzwiller projected wavefunctions.
While mostly based on a simple toy model, we argue that our findings contribute to a better understanding of superconductivity emerging out of spin- and valley polarized rhombohedral graphene multilayers in a parameter regime with nearby quantum anomalous Hall insulators.
\end{abstract}

\maketitle

\section{Introduction.} 
The discovery of superconductivity in two-dimensional quantum materials~\cite{CaoJarilloHerrero2018,Wang2020}, most notably in graphene devices, challenges theoretical understanding as they transcend the classic Bardeen-Cooper-Schrieffer (BCS) and Eliashberg scenarios~\cite{Marsiglio2020}.
First, the superconductivity emerges at temperature scales comparable to the bandwidth, i.e. it occurs for intermediate coupling strengths and not too distant from the Bose-Einstein condensation physics~\cite{ChenLevin2024}.
Second, the electronic bands in the normal state display non-trivial quantum geometry, or even topology, leading, among others, to topological scenarios for superconductivity~\cite{ChristosScheurer2020,KhalafVishwanath2021,ShavitAlicea2024} as well as rigorous upper bounds~\cite{Hazra2019,Randeria2024,MaoChowdhury2024} and mean-field lower bounds to the superfluid stiffness~\cite{PeottaTorma2015,TormaBernevig2022}. Third, superconductivity occurs close to a ``zoo'' of symmetry-broken states and nearby particle-hole orders have thus to be taken into account in theories of the superconducting state~\cite{ChristosScheurer2023,ChubukovVarma2024,GuerciFu2025}.

In a period where the theoretical fog slowly clears and gives way to few plausible scenarios of superconductivity, it came as a major surprise that superconductivity in rhombohedral tetra- and pentalayer graphene~\cite{HanJu2024} devices was reported to occur in a parameter regime where the system appears to be spin and valley polarized (``quarter'' metal) and thus spontaneously breaks time reversal symmetry in the normal state. Curiously, superconductivity appears at densities where at least pentalayer rhombohedral graphene with aligned substrate forms {quantum} anomalous Hall crystals (QAHC)~\cite{HanJu2024,AronsonAshoori2024,ChoiYoung2024,LuJu2025,WatersYankowitz2025,JC_trithep,JC_ashvin}.
Further, it is particularly striking that in part of the experimentally obtained phase diagrams the superconducting region increases upon the application of a magnetic field $|B|$. For underdoped tetra- and overdoped pentalayer graphene the increase is linear {in the $B$ vs. density plane}, and 
its slope appears consistent with a universal
combination of constants of nature, $\pm h/2 e \approx \pm 20.7 \text{T}/10^{12}$ cm$^{-2}$. This is the same slope as for $C =\pm 2$ insulators in Landau fan diagrams.

\begin{figure*}
    \includegraphics[width=1.\textwidth]{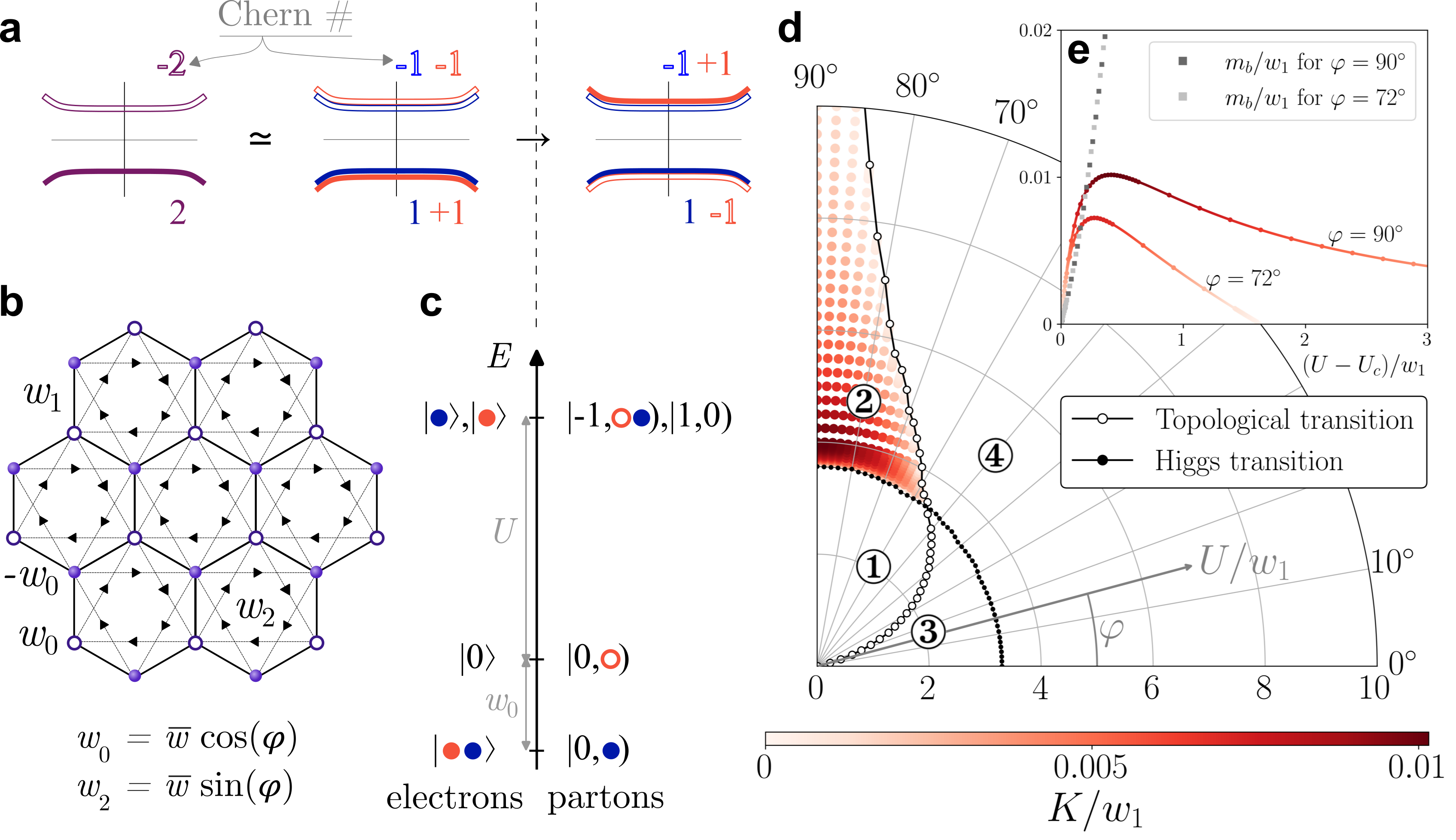}
    \caption{a) Schematic ingredients of the theory: a Chern-2 insulator composed of two colors of Chern-1 bands which are separated in a color neutral and color doublet subspaces (e.g.,~by attractive interactions), panel c). Within slave-rotor parton construction these subspaces are distinguished by bosonic coloron occupation [first entry of the round ket $\vert \dots )$]. The fermionic chargeons (charge indicated by filled/empty circles in the round ket) also carry color degree of freedom allowing them to form a quantum color Hall effect. b) Illustration of the lattice model under consideration, Eq.~\eqref{Eq:free_model}. Note that the energy splitting of empty and doubly occupied states in panel c) is reversed on the opposite sublattice.
    d) Slave-rotor phase diagram of the model illustrated in b), taken at $\bar w = 0.1 w_1$. In phases \ding{172} and \ding{174} the rotor is Higgs-condensed corresponding to a Chern and trivial insulator, respectively. Their counterparts without Higgs condensation are denoted 
    \ding{173}, \ding{175}, respectively. Going beyond mean-field treatment of the slave-rotor theory, \ding{173} is a superconductor with stiffness $K$ (color plot), while phases \ding{174},\ding{175} are adiabatically connected (the Higgs transition at small $\varphi$ becomes a crossover). e) Rotor gap and stiffness as a function of $U$ at two angles corresponding to line cuts in the phase diagram d) illustrating the separation of energy scales at large $U$. }
    \label{fig:SummaryFig}
\end{figure*}

Broadly inspired by these developments, we here present a theory of superconductivity near the Bose-Einstein condensation limit emerging out of Chern-2 insulators, which in the context of rhombohedral graphene could stem from a (failed) QAHC~\cite{DongSenthil2024,DongParker2024,ZhouZhang2023,KwanBernevig2023,ZengCano2024}.
Representing the electronic states by two ``colors'' of Chern-1 bands~\cite{BarkeshliQi2012,Sun2015,Guerci2024}, we employ a fractionalization scheme into bosonic colorons and fermionic chargeons which accounts for strong local color-singlet formation.
So long as the chargeons form topological bands the system displays a strong-coupling superconductor, while the same system with non-topological chargeon bands is an insulator.
After a phenomenological, yet technical, summary we devote the major part of this manuscript to a toy model displaying these features of ``topologically enabled superconductivity''~\cite{RamppSchmalian2022}:
a generalized, attractive-$U$ Haldane-Hubbard model for the quantum anomalous Hall effect with two fermionic colors.
Within this model we employ slave-rotor calculations supplemented with a random phase approximation to determine a rich phase diagram, and in particular to characterize the properties of the superconductor (stiffness, correlation length, transition temperature, response to magnetic field) in terms of microscopic parameters.
We complement these calculations with an effective spin model and inspect correlation functions obtained from Gutzwiller-projected wave functions. We further extract physical characteristics of the superconducting state emerging from the topological mechanism and discuss experimental consequences.

Proposals for topological mechanisms for superconductivity date back to pre-BCS days~\cite{Froehlich1954,Schmalian2010}, and were then discussed in the context of the cuprates~\cite{Wiegmann1992,Wiegmann1994,DungHaiLee1999} as well as anyon~\cite{Wilczek1982,Laughlin1988,BanksLykken1990,LeeFisher1989,Fradkin1990,BarkeshliMcGreevy2014} and skyrmion~\cite{GroverSenthil2008} superconductivity.
Some of those concepts were recently applied to modern two-dimensional materials~\cite{KhalafVishwanath2021,KimWen2024,ShiSenthil2024,DivicVishwanath2024,SahayZalatel2024}.
A common theme of topological mechanisms of superconductivity is that the underlying topological fermionic bandstructure may transmute charge~\cite{GroverSenthil2008,KhalafVishwanath2021} or spin quantum numbers~\cite{RamppSchmalian2022,BollmannKoenig2024} of topological excitations, e.g.,~instantons, magnetic skyrmions, or, as in this paper, monopole excitations of an emergent gauge field.
This is also the foundation of various quantum field theoretical dualities~\cite{SenthilXu2019} of which the charge-vortex duality in 2+1 space-time dimensions is of quintessential relevance for this work~\cite{DumitrescuThorngreen2024,ChesterKomargodski2024}. By means of a partial particle-hole transformation, the present theory of superconductivity is also related to quantum magnetism in topological Mott insulators, specifically Kane-Mele-Hubbard models~\cite{Xu2011,GrisetXu2012,HohenadlerAssaad2013,MaiPhillips2024}, and, on a more technical level, on the emergence of in-plane antiferromagnetism in Dirac quantum spin liquids subject to an external Zeeman field~\cite{RanWen2009}.

\section{Phenomenology.} The first ingredient in this theory is a Chern-2 insulator, which could be composed of two filled Chern-1 bands (with ``color'' index $\alpha = \pm 1 = \text{blue}/\text{red} = \uparrow/\downarrow$) or a single ideal Chern-2 band artificially represented by two colors of Chern-1 bands~\cite{BarkeshliQi2012,Sun2015,Guerci2024}, {cf. Fig.~\ref{fig:SummaryFig} a)}.
The second ingredient is a strong tendency to form local color singlet states, see Fig.~\ref{fig:SummaryFig} c), as, e.g., favored by strong attractive interactions.
We employ a fractionalization scheme to project to this low lying Hilbert space, leading to an effective lattice gauge theory coupled to topological fermion bands.
While we expect the present mechanism for superconductivity to rely only on the band topology of chargeons, we henceforth illustrate it using a concrete continuum field theory of (massive) Dirac fermions interacting with an emergent U(1) gauge field (we set $\hbar = 1$) 

\begin{equation}
    S = \sum_\alpha \int d\tau d^2 x \ \bar \psi_\alpha [D_\tau^{(\alpha)} + h_\alpha(- i D_j^{(\alpha)}) ] \psi_\alpha + \sum_{\mu \nu} \frac{f_{\mu \nu}^2}{4 g_0^2}. \label{eq:BasicQFT}
\end{equation}

\noindent
Here, $\mu = \tau,x,y$ are Euclidean space-time indices, $j = x,y$ are spatial coordinates, and $\psi$ fields are fermionic chargeons. For $\alpha = 1$ ($\alpha = -1$) their Hamiltonian $h(p_i)$ 
describes the dynamics of an extra blue, $\alpha = 1$ particle (red, $\alpha = -1$ hole) on top of a singly occupied reference state containing a single red fermion, cf. Fig.~\ref{fig:SummaryFig} a,c).
The low-energy Hamiltonian $h(p_i)$ leads to Kane-Mele-type mass terms if the lattice model features Chern bands, and trivial mass terms if the starting point was a trivial band insulator.
If no mass terms are present, {Eq.}~\eqref{eq:BasicQFT} corresponds to compact 2+1-dimensional quantum electrodynamics (QED$_3$) with $N_f=4$ flavors.
The minimal coupling to the emergent gauge field $a_\mu$ and the electromagnetic external field $A_\mu$ enter via covariant derivatives $D_\mu^{(\alpha)} = \partial_\mu + i a_\mu + i \alpha A_\mu $.
As usual, the field strength tensor is $f_{\mu \nu} = \partial_\mu a_\nu - \partial_\nu a_\mu$, we here tacitly set the speed of emergent light to unity and denote the emergent gauge coupling by $g_0$. 
Below we include an approximate derivation of this phenomenological field theory and its parameters for a microscopic lattice model. 

While in the present theory the flavor (i.e.~color) degree of freedom is quenched, the charge degree of freedom still fluctuates. This situation is reverse to the case of a Mott insulator, where the charge degree of freedom is quenched, but the flavor (i.e.~spin) degree of freedom still fluctuates.

Upon including fluctuations of the emergent gauge field, the putative fractionalized state with a Chern-insulating band structure of fermionic chargeons becomes instead a \emph{physical} charge-2e superconductor. This is the main object of study of our work.

We first discuss this from the perspective of the fermionic field theory: In the presence of the topological mass terms, the fermionic degrees of freedom are gapped. While one then naively expects monopoles of the compact U(1) emergent gauge field to proliferate (as discussed by Polyakov \cite{Polyakov1987}), this is suppressed as a result of \emph{physical} U(1) charge conservation. 
Technically, this is encoded in fermionic zero-modes at the monopole space-time position imposing linear confinement of \emph{monopole-antimonopole pairs}.
Momentarily considering the (physical charge)-neutral case (i.e.~absent electromagnetic gauge field $A_\mu$), the linearly dispersing photon of the emergent QED$_3$ theory can then be viewed as the linear Goldstone-boson of a superfluid~\cite{RanWen2009}.

As usual, upon coupling to the physical electromagnetic field, this Goldstone-boson is ``eaten up'' by the familiar Anderson-Higgs mechanism in superconductors.
Moreover, note that the off-diagonal long-range order of the superfluid corresponds to long-range order in the monopole-antimonopole correlation function.
Due to the topological nature of chargeon bands with quantized Hall response (Chern number $\alpha = \pm 1$) a single monopole event injects charge $2e$ to the system.
To see this, recall that a monopole in space-time corresponds to the instantaneous creation of an emergent flux quantum, and by means of the Widom-Streda formula~\cite{Streda1982,Widom1982,StredaSmrcka1983} for the quantized Hall response $d n_{\alpha}/d b = \alpha$ directly relates a change in the flux $b = \nabla \times \mathbf{a}$ of the emergent gauge field $a_\mu$ to a change in the density of chargeons $n_{\alpha}$, as they carry a unit of emergent gauge charge. 
In contrast, when fermionic chargeons occupy topologically trivial bands, the monopole fugacity is finite and monopoles proliferate, gapping out the photon and no off-diagonal long-range order emerges.

An alternative description of the same phenomenon appears by explicitly integrating out the fermions in topologically non-trivial bands.
They generate a mixed Chern-Simons theory $S_{CS} = -i \int _{\tau,x}\epsilon_{\mu \nu \rho} a_\mu \partial_\nu A_\rho/\pi$, in addition to renormalizing $g_0 \rightarrow g$ and speed of light in the Maxwell term. 
Purists may stop the proof of superconductivity here and argue that a mixed Chern-Simons theory is nothing but the topological field theory of $\mathbb Z_2$ topological order, a class of order to which superconductivity was argued to belong~\cite{HanssonSondhi2004, footnoteZ2orderSC}.
We prefer to use a duality mapping to a theory of a real compact field $\phi \in [0,2\pi)$, the superfluid phase~\cite{SenthilXu2019,SuppMat}, 
\begin{equation}
    S = \int d\tau d^2 x \ \frac{g^2}{8 \pi} (\partial_\mu \phi - 2 A_\mu)^2 \label{eq:PhaseTheory}
\end{equation}
to identify the superfluid stiffness $K$ in the theory with $g^2/4\pi$. 

\section{Microscopic lattice model.} 
Our mechanism for superconductivity does not explain the origin of attraction but instead highlights the topological stability of off-diagonal long-range order. To illustrate this microscopically, we consider Haldane's honeycomb model\cite{Haldane1988,LiangKou2013,WuKou2015,ZhangZhang2017} with attractive $U>0$ onsite interactions $H = H_{\rm QAH} + H_{\rm int.}$

\begin{subequations}
\begin{align}
    H_{\rm QAH} & =  \sum_\alpha \int_{\v k} c_{\alpha}^\dagger(\v k) h_{[w_0, w_1, w_2]} (\v k) c_{\alpha}(\v k), \label{Eq:free_model}\\
    H_{\rm int.} & = - {U} \sum_{\v{x}} (\hat n_{\v{x}, -1} -1/2)(\hat n_{\v{x}, 1} -1/2) . \label{Eq:interaction}
\end{align}
\label{Eq:Hamiltonian}
Note that fermionic operators $c^{(\dagger)}_{\v x, \alpha}$ and densities $\hat n_{\v x, \alpha}$ contain a color label $\alpha$ and $\v{x}$, the position on the honeycomb lattice. The kinetic term is best presented in momentum space, where the $\v k$ integral runs over the hexagonal Brillouin zone and is normalized $\int_{\v k} = \int \frac{d^2 \v k}{\text{Vol(BZ)}} = 1$ and Fourier transformed fermion fields $c^{(\dagger)}_\alpha(\v k)$ are spinors in sublattice space. We introduced
\begin{equation}
h_{[w_0, w_1, w_2]} (\v k)  = \left (\begin{array}{cc}
w_0 + w_2 m(\v k) & -w_1 s(\v k) \\ 
-w_1 s^*(\v k) & -w_0 - w_2 m(\v k)
\end{array}  \right),
\end{equation}
\label{eq:QAHAttrHubbard}
\end{subequations} 
where $s(\v k) = 1+2 e^{-i{\sqrt{3} k_y}/{2}} \cos \left({k_x}/{2}\right), \quad m(\v k) = 4 \sin \left({k_x}/{2}\right) [\cos \left({k_x}/{2}\right)-\cos \left({\sqrt{3} k_y}/{2}\right)]$.
We mostly use the notation $w_0 = \bar w \cos(\varphi), w_2 = \bar w \sin(\varphi)$, so that for $0< \varphi < { \arctan \left( \frac{1}{\sqrt{27}} \right) }$ 
the non-interacting bandstructure is topologically trivial, while the complementary parameter regime of total Chern number $2$ is the one of greater interest here. Physical electromagnetic gauge potentials were suppressed in Eq.~\eqref{eq:QAHAttrHubbard} for notational convenience but may be readily reintroduced by Peierls' substitution.

In the following, we first solve the problem using conventional mean-field methods at small $U$ and large $U$ leading to the following conundrum: While the large -$U$ calculation does predict superconductivity at a finite range of $0<w_0/w_2 \ll 1$, weak-coupling BCS-like mean-field equations generally do not produce a superconducting solution. 
We then construct and study a fractionalized theory which allows one to resolve this apparent conundrum.

\subsection{Weak-coupling mean-field theory.} 

The most straightforward way to account for interaction effects is through the mean-field approximation, which, for small values of $U$, approximates the self-energy functional using only the Hartree and Fock diagrams.
We focus on the pairing and charge channels. The full calculations are provided in the supplementary material ~\cite{SuppMat}.
We obtain the Bogoliubov-de Gennes Hamiltonian
\begin{align} \label{eq:BdG}
    h_{\rm BdG}(\v k) =  \left (\begin{array}{cc}
        h_{[w_0 {-} \rho, w_1, w_2]} (\v k) & \Delta  \\
        \Delta^*  & -h_{[w_0 {-} \rho, w_1, -w_2]} (\v k)
    \end{array} \right),
\end{align}
where the superconducting gap $\Delta$ and the charge density wave parameter $\rho$ appear, the latter simply modifying the coupling $w_0$.
From this Hamiltonian, we can directly derive the ground-state energy functional.
By calculus of variations, we find the saddle point conditions 
\begin{subequations}
\begin{align}
    \Delta \left( \frac{1}{U} + g(\rho, \Delta)\right) &=0,\\
    \frac{ \rho}{U} +  (\rho - w_0) g(\rho, \Delta) &=0.
\end{align}
\label{eq:MFEq}
\end{subequations}
The function $g(\rho, \Delta)$ (defined in \cite{SuppMat}), entering both Eqs.~\eqref{eq:MFEq} a) and b), is obtained by differentiating the contribution to the energy from filled bands with respect to the two order parameters.
Thus, both conditions cannot be satisfied simultaneously unless $\Delta = 0$ or $w_0 = 0$, i.e.~superconductivity is not recovered at the mean-field level for most of the parameter regime. 

\subsection{Strong-coupling mean-field theory.} 
In the large-$U$ limit, we use a basic superexchange calculation to find an effective Hamiltonian 
\begin{align}
    H &=  I \sum_{\langle \v x, \v x' \rangle} \vec T_{\v x} \cdot \vec T_{\v x'} + \tilde I \sum_{\llangle \v x, \v x' \rrangle} \vec T_{\v x} \Gamma \vec T_{\v x'}- 2 w_0 \sum_{\v x} (-1)^{\v x} \hat e_z \cdot \vec T_{\v x},  
    \label{eq:h-large-U}
\end{align}
where $\vec T_{\v x}$ are pseudospin-1/2 operators acting within the empty/doubly occupied subspace and $\Gamma = \text{diag}(-1,-1,1)$.
The leading order results $I = 16 w_1^2/U$, $\tilde I = 16 w_2^2/U$ illustrate that $w_2$ favors in-plane antiferromagnetic order of pseudospins, which corresponds to superfluidity of the original fermions.
Similarly, the staggered chemical potential $w_0$ stabilizes out-of-plane antiferromagnetism, which signals a charge density wave (CDW) in the attractive-$U$ Hubbard model.
Classically, the transition~\cite{SuppMat} between superfluid and CDW occurs at $w_0 = 6 \tilde I$ or $U = 96 \bar w \sin(\varphi)\tan(\varphi)$.
  
\subsection{Slave-rotor mean-field theory.} In the following, we go beyond a classical analysis of \eqref{eq:h-large-U} and make the direct connection between the fermionic band topology and superconductivity (in-plane ordering of the pseudospin degrees of freedom) explicit.
To this end, we employ a lattice version of the fractionalization scheme underlying \eqref{eq:BasicQFT}, where the electronic degrees are represented in terms of fermionic charge degrees of freedom $\psi_{\v x, \alpha}$ and a bosonic rotor degree of freedom $e^{i \theta_{\v x}}$~\cite{FlorensGeorges1,FlorensGeorges2}. Here, the angular momentum of the rotor $\hat L_{\v x} = -i \partial_{\theta_{\v x}}$ represents the color quantum number, such that a state with $\hat L_{\v x}=0$ implies color-singlet formation (and the opening of a color-gap).
Explicitly, the fermionic operators in Eq.~\eqref{eq:QAHAttrHubbard} can then be expressed as $ c_{\v x, +1} = e^{i \theta_{\v x}} \psi_{\v x, +1} $ and $c_{\v x, -1}  = (-1)^{\v x} e^{-i \theta_{\v x}} \psi^\dagger_{\v x, -1}$.
Note that our parton construction differs from usual slave-rotor constructions: the electromagnetic charge is carried by the fermionic partons, rather than the bosonic degrees of freedom~\cite{FlorensGeorges1,FlorensGeorges2}; and we have performed a partial particle-hole transformation that relates the attractive-$U$ Haldane-Hubbard model to a repulsive-$U$ Kane-Mele-Hubbard model in the presence of spin-orbit coupling (SOC) $w_2$ and a staggered Zeeman field $w_0$.
The parton construction leads to an emergent U(1) gauge structure generated by the on-site constraint $\hat L_{\v x} + \sum_{\alpha} \psi_{\v x, \alpha}^\dagger \psi_{\v x, \alpha} = 1$ which removes unphysical states from the enlarged parton Hilbert space.

A key advantage of the quantum rotor representation is that the original quartic fermion interaction is replaced by \( \frac{U}{2} \hat L_{\v x}^2 \), simplifying the treatment of interactions. However, the kinetic parts of the Hamiltonian, $-w_1 \sum_\alpha \sum_{\langle \v x, \v x' \rangle} [e^{i (\theta_{\v x'} - \theta_{\v x})} \psi^\dagger_{\v x,\alpha}\psi_{\v x',\alpha} + H.c.]$ (and analogously for the $w_2$ term), become a fermion-rotor interaction which we treat using a self-consistent mean-field theory, $H_{\rm MF} = H_\psi + H_\theta$,
\begin{subequations}
\begin{align}
    H_{\psi} & = \sum_\alpha \int_{\v k} \psi_{\alpha}^\dagger(\v k) h_{[w_0, t_1,(-1)^\alpha t_2 ]} (\v k)\psi_{\alpha}(\v k), \label{eq:PsiHam}\\
    H_{\theta} & =  \frac{U}{2} \hat L_{\v x}^2 - \left [J \sum_{\langle \v x, \v x' \rangle} e^{i (\theta_{\v x} - \theta_{\v x'})} + \tilde J \sum_{\llangle \v x, \v x' \rrangle} e^{i (\theta_{\v x} - \theta_{\v x'})} + H.c. \right]. \label{eq:JJArray}
\end{align} \label{eq:SBMeanfield}
\end{subequations}
The self-consistency conditions are $t_{1}/w_1 = \langle e^{i \theta_{\v x} - i \theta_{\v x'}} \rangle$, $J/w_1 = \langle \sum_{\alpha} \psi^\dagger_{\v x,\alpha}\psi_{\v x',\alpha} \rangle$ with $\v x, \v x'$ nearest neighbors (and analogously 
for next nearest neighbor hopping).
It is generally expected that such fractionalized mean-field theories at least qualitatively grasp the main features near Mott transitions of repulsive Hubbard models and, by extension, of the intermediate coupling superconducting transition in the present theory.
We here use a large-$M$ theory~\cite{FlorensGeorges1, FlorensGeorges2, WagnerSangiovanni2024, SuppMat} to solve the effective rotor model, Eq.~\eqref{eq:JJArray}, beyond single-site mean-field theory while in the supplement~\cite{SuppMat} we also present a more lightweight semi-analytical method with qualitatively similar outcome, allowing us to incorporate external electromagnetic fields.

The phase diagram resulting from this approach~\cite{FlorensGeorges1, FlorensGeorges2, WagnerSangiovanni2024, SuppMat} is shown in Fig.~\ref{fig:SummaryFig} d) in radial coordinates.
The angular coordinate \( \varphi \) is given by \( \arctan(w_2/w_0) \), while the radial distance represents the ratio between the Hubbard interaction strength and the hopping amplitude $w_1$.  

For small \( U \), the rotor Higgs-condenses, effectively identifying $c$ and (particle-hole conjugated) $\psi$.
For small values of $\varphi$, the staggered potential $w_0$ (Zeeman field in the language of the repulsive-$U$ model) dominates over Haldane hopping $w_2$ (corresponding to SOC). Reversely, at large $\varphi$, $w_2$ dominates over $w_0$.
Therefore, the system is a trivial band insulator, phase \ding{174} in Fig.~\ref{fig:SummaryFig} d), for small $\varphi$, but at large $\varphi$ the system is topological, \ding{172} in Fig.~\ref{fig:SummaryFig} d), i.e.~a Chern-2 insulator of $c_{\v x, \alpha}$ fermions (or equivalently a ``quantum color Hall'' insulator of $\psi_{\v x, \alpha}$ fermions). 

Increasing $U$ beyond a critical value, the Higgs-field $e^{i \theta_{\v x}}$ in Eq.~\eqref{eq:JJArray} uncondenses and a gap $m_b$ opens in the rotor sector, indicating the suppression of color fluctuations in our attractive-$U$ model (in repulsive-$U$ models the rotor gap signals the freezing of charge fluctuations, i.e.~a Mott transition). {Within large-$M$ theory $m_b$ grows slower than the square root behavior expected from simplest mean-field~\cite{FlorensGeorges2}.} For our choice of \( \bar{w} = 0.1  w_1\), the critical value is \( U_\text{c} \sim 3.4 w_1 \) {at $\varphi=0^\circ$} 
and remains nearly constant upon increasing the angular coordinate, see the solid black circles in Fig.~\ref{fig:SummaryFig} d). For $\varphi \gtrsim 55^\circ$ the Haldane hopping $w_2$ surpasses $w_0$ and induces a topological transition from \ding{175} to \ding{173} (open circles in Fig.~\ref{fig:SummaryFig} d)) in the fermionic sector, so that the $\psi_{\v x, \alpha = +1}$ ($\psi_{\v x, \alpha = -1}$) chargeons occupy a Chern $C = +1$ ($C = -1$) band.
 
We remark that increasing $U$ decreases the segment of $\varphi$ hosting topological fermion bands. This  reflects that correlation effects suppress kinetic terms, including the Haldane hopping $w_2$, relative to the on-site potential $w_0$ so that at $U/w_1 \rightarrow \infty$, the band structure becomes topologically trivial for all $\varphi$.

\subsection{Strongly correlated topological states.}
\label{sec:Zeros}

Thus, at the level of slave-rotor mean-field theory, the phase diagram of this strongly correlated topological insulator~\cite{Rachel2013,WagnerSangiovanni2023,WagnerSangiovanni2024,BollmannKoenig2024b} contains four phases characterized by two binary questions: Is the rotor Higgs-condensed? Is the chargeon bandstructure topological? The arguably most interesting of these four mean-field phases is the topological phase at strong coupling for which two of us recently demonstrated the appearance of Green's function edge zeros~\cite{WagnerSangiovanni2024}. In that study, geared to the repulsive-$U$ Kane-Mele model, the zeros signal a non-local spin-charge separation.
A related color-charge separation is also expected in the present attractive-$U$ Haldane model (now charge edge modes encircle a finite system in the topologically non-trivial phase at large $U$). While we will see shortly that, due to quantum fluctuations, the ground state of the model Eq.~\eqref{Eq:Hamiltonian} in the regime \ding{173} is not a correlated topological insulator, but rather a superfluid with stiffness $K$ (corresponding to an in-plane antiferromagnet in the repulsive case), we anticipate a large temperature window $K \ll k_B T \ll m_b$, cf. Fig.~\ref{fig:SummaryFig} e), in which aspects of the aforementioned features~\cite{WagnerSangiovanni2024} of topological Green's function zeros in paramagnetic topological Mott insulators are expected to persist. 

\subsection{Fractionalized theory beyond mean-field.} We now turn to fluctuations about the saddle-point (i.e.~mean-field) solution.  
These originate from long-range bosonic and fermionic fluctuations, in the simplest treatment within random phase approximation.
To this end, it is sufficient to consider the low-energy sector (continuum field theory) of the lattice model above.

The polarization operators of Higgs boson (coloron) and gapped Dirac fermions (chargeons) are~\cite{SuppMat}
\begin{subequations}
\begin{align}
    \Pi^{\mu \nu,(b)}(q) & = \frac{q^2_b \delta_{\mu \nu}-q_b^\mu q_b^\nu }{24 \pi m_b v_b^2}, \\
    \Pi^{\mu \nu,(f)}_{\alpha, \kappa} (q) & = \frac{q^2_f \delta_{\mu \nu}-q_f^\mu q_f^\nu}{6 \pi \vert m_\kappa \vert v_f^2} - \kappa \alpha \text{sign}(m_\kappa) \frac{\epsilon_{\mu \nu \rho} q_f^\rho}{4\pi v_f^2},
\end{align}
\label{eq:Polops}
\end{subequations}
where $q_{b,f}^\mu = (v_{b,f} q_x, v_{b,f} q_y, \omega)$ and  $v_b = \sqrt{UJ} a$, $v_f = \sqrt{3}t_1 a/2$
($m_b \propto U - U_c$ and $m_\kappa = t_0 +\kappa \sqrt{27} t_2$) are bosonic and fermionic velocities (masses), respectively.
Here, we have assigned ``bare'' microscopic estimates based on the slave-rotor mean-field theory to the continuum field theory parameters (i.e. neglecting higher-order loop corrections). This is expected to be reasonably accurate away from phase transitions in Fig.~\ref{fig:SummaryFig} d).

Note that there are a total of four fermion species (two valleys denoted by subscript $\kappa = \pm$ and two colors $\alpha = \pm$) and that they generate a topological response (mixed CS term) as long as $m_+ m_- <0$.
If we only integrate the bosons out, we obtain Eq.~\eqref{eq:BasicQFT} with the speed of light $c = v_b$ (initially set to unity, but now restored) and $g_0^2 = 12 \pi m_b$.
Integrating out all matter fields including fermions yields a gauge coupling viz.~superfluid stiffness and the speed of the emergent photon~\cite{SuppMat}
\begin{subequations}
\begin{align}
    K &= \frac{g^2}{4\pi} = \frac{3 }{1/m_b + {8}/{\vert m_{+} \vert}  + {8}/{\vert m_{-} \vert}},\\
    c^2& = \frac{{v_b^2}/m_b + {8 v_f^2}/{\vert m_{+} \vert}  + {8 v_f^2}/{\vert m_{+} \vert}}{1/m_b + {8}/{\vert m_{+} \vert}  + {8}/{\vert m_{-} \vert}}  .
\end{align}
\label{eq:StiffnessSpeedOfLight}
\end{subequations}

Crucially, the contributions to the stiffness from fermions and bosons add up as a sum of inverse masses, somewhat reminiscent to the Ioffe-Larkin~\cite{IoffeLarkin1989} addition of inverse conductivities in gauge theories of doped quantum spin liquids.
This ensures that the superfluid stiffness vanishes when either fermionic chargeons or bosonic coloron become massless corresponding to both types of phase boundaries of the superfluid phase.
It is worthwhile to highlight the energy separation of the color gap $\mathcal {O}(U)$ and the stiffness $K$, Fig.~\ref{fig:SummaryFig} e) at large $U$.
This is characteristic of strong-coupling superconductivity~\cite{EmeryKivelson1995,ChubukovSchmalian2005}, which in some cases can be related to Bose-Einstein condensation of Cooper pairs~\cite{Randeria2014}.

This concludes the derivation of the phase diagram of our model, Fig.~\ref{fig:SummaryFig} d).
When $m_+ m_- <0$ and $U<U_c$ the system is a Chern-2 insulator signaled by Higgs condensation in the language of the emergent gauge theory.
For $U>U_c$ the topological theory becomes superconducting by the mechanism prescribed above.
In contrast, when $m_+ m_- > 0$ the system is always adiabatically connected to a trivial insulator, i.e.~phases \ding{174} and \ding{175} are actually the same.
To see this in the gauge theory language, the emergent photon is gapped at low $U$ due to the Anderson-Higgs effect (due to coupling to the rotor degrees of freedom) and at large $U$ due to single monopole proliferation, but, at least in the bulk~\cite{FradkinShenker1979,ThorngrenVerresen2023}, these two phases are indiscernible.

\subsection{``Anti-''Gutzwiller projection.} 
\begin{figure}
    \centering\includegraphics[width=\columnwidth]{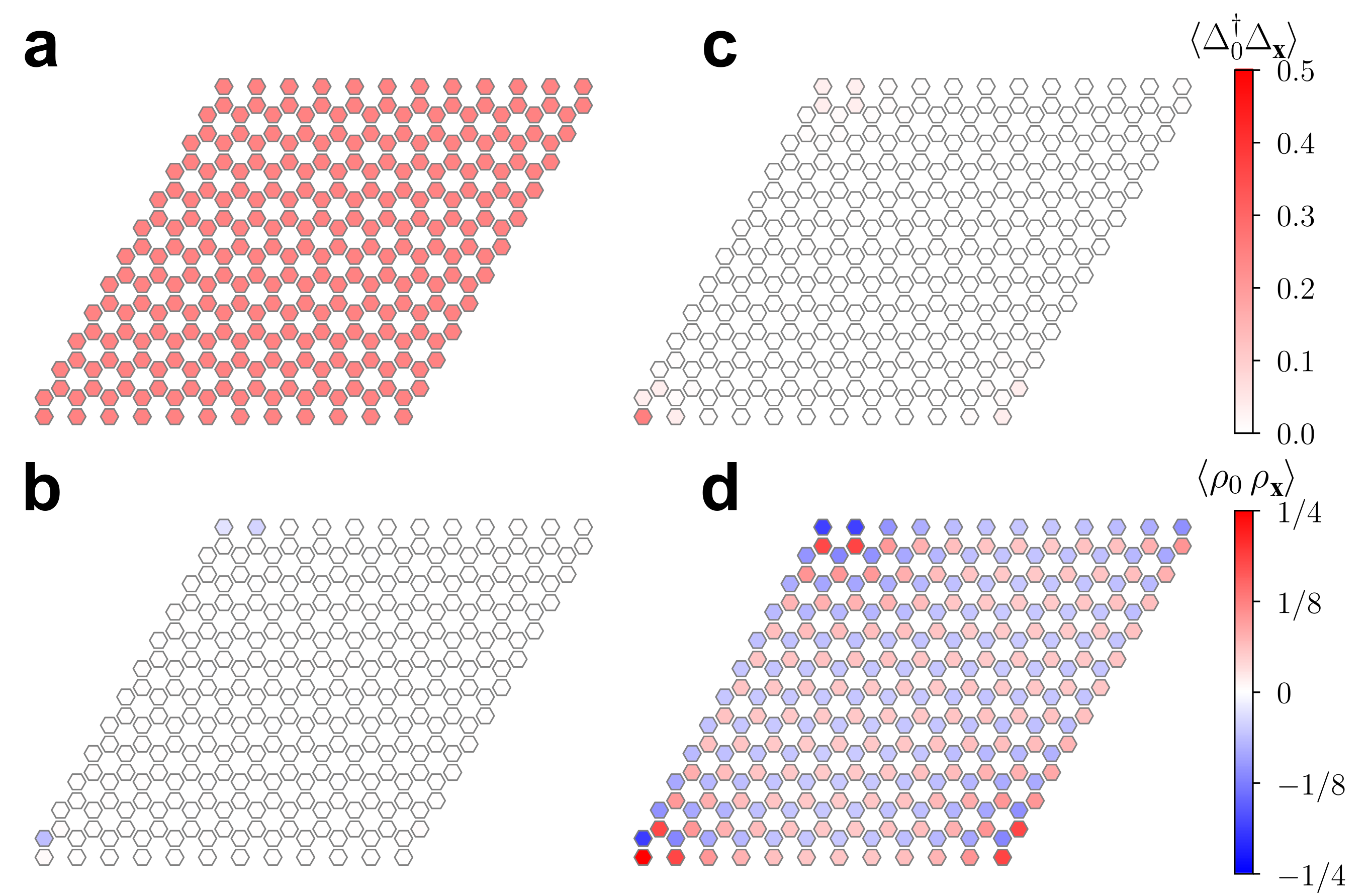}
    \caption{Superconducting [panels a) and c)] and density-density correlators [panels b) and d)] evaluated with anti-Gutzwiller-projected fermionic wave functions. Panels a), b): $w_2/w_1 = 5, w_0/w_1 = 0$. Panels c), d): $w_2/w_1 = 0, w_0/w_1 = 2$. System size: 14$\times$14 unit cells, periodic boundary conditions.}
    \label{fig:Gutzwiller}
\end{figure}

The fractionalization scheme is (formally) exact when projected to the physical Hilbert space.
While an exact treatment of such a projection in an interacting many-body system is not feasible, we may instead consider a free-fermion wavefunction for the chargeon degrees of freedom and perform an ``anti''-Gutzwiller projection to remove (unphysical) contributions with singly-occupied sites. This should be contrasted to the standard Gutzwiller projection in quantum magnetism, wherein one \emph{projects out} empty and doubly-occupied sites.
Observables with respect to such anti-Gutzwiller-projected wavefunctions may then be efficiently evaluated using Monte Carlo methods.
We have evaluated the superconducting correlator $\langle \Delta^\dagger_{\v x} \Delta_{\v x'} \rangle$, the local charge $\langle \rho_{\v x} \rangle$ and charge correlator $\langle \rho_{\v x} \rho_{\v x'} \rangle$ 
with respect to free-fermion ansatz wavefunctions $\ket{\psi_\mathrm{topo}}$ and $\ket{\psi_\mathrm{triv}}$, which correspond to Chern-2 and trivially insulating Slater states, respectively.
As shown in Fig.~\ref{fig:Gutzwiller}, we find that the former (latter) leads to long-ranged superconducting (CDW) correlations while concomitantly displaying vanishing density-density (pair) correlations. 
In order to examine spontaneous symmetry breaking, we probe the finite-size scaling behavior of the superconducting correlator: since the ansatz wavefunctions possess by construction a $U(1)$ symmetry associated to charge conservation (or equivalently to in-plane rotations of the pseudospin), a signature of spontaneous symmetry-breaking is given by a divergent contribution to the pairing correlation structure factor in the thermodynamic limit.
Our finite-size scaling analysis~\cite{SuppMat} 
suggests that for sufficiently large $w_2/w_1$ (and $w_0 \equiv 0$), the projected wavefunction indeed exhibits superconductivity (for more details, and a double scaling analysis for small $w_2/w_1$ see~\cite{SuppMat}).

\section{Physical properties of the topologically enabled superconductor}

Having established the main mechanism for superconductivity in our model starting at half-filling, vanishing temperature and magnetic field, we next discuss physical properties, such as correlation and penetration depth, an estimate of the critical temperature, magnetic field and doping effects.

\subsection{Vortex solutions and topology.}

\label{sec:Vortex}

 Before considering the properties of the charged superconductor, we briefly discuss vortex solutions~\cite{SuppMat}, which are present even for neutral superfluids, i.e.~$A_\mu = 0$.
 As suggested by particle-vortex duality, a vortex, i.e.~a $2\pi$-phase winding of $\phi$ in Eq.~\eqref{eq:PhaseTheory}, corresponds to a local charge accumulation of matter fields, which in turn create a radial emergent electric field.
 Near the Higgs transition separating \ding{172}, \ding{173} of Fig.~\ref{fig:SummaryFig} d), it is energetically preferable to accumulate emergent gauge charge in the bosonic sector.
 In contrast, near the topological transition separating \ding{173}, \ding{175} the ``cheapest'' accumulation is fermionic.
 The spatial extent of the charge accumulation $\xi \sim \text{max}(v_b/m_b, v_f/\vert m_- \vert)$ sets the core size of the vortex and can be interpreted as the coherence length of the superconductor. Crucially, fermionic puddles of emergent gauge charge $\sum_\alpha \bar \psi_\alpha \psi_\alpha$ also imply accumulation of physical fermionic matter, $\sum_\alpha \alpha \bar \psi_\alpha \psi_\alpha$, and a color degeneracy.
 Thus, a proliferation of vortices near the \ding{172} - \ding{175} transition suggests a (semi-)metallic fermionic state. 
It is remarkable that fermionic modes trapped inside the vortex appear:
In contrast, considering a vortex solution of the mean-field Bogoliubov-de Gennes Hamiltonian in Eq.~\eqref{eq:BdG} does not lead to zero modes.
However, in the present strongly interacting context, the mean-field Bogoliubov-de Gennes Hamiltonian Eq.~\eqref{eq:BdG}, describing only Gaussian states, has limited meaning~\cite{TiwariKoenig2024}, but it illustrates the trivial band-topology of Bogoliubov particles.

While a careful study of finite temperature effects is beyond the scope of this paper, finite temperature superconductivity is established via a Berezinskii-Kosterlitz-Thouless (BKT)
transition. In the absence of further energy scales, the temperature can be conjectured to be $k_BT_{\rm BKT} \sim  K$.
As the present superconductor emerges from a band insulator at strong coupling, it is difficult to place it on a standard Uemura plot comparing $T_{\rm BKT}$ and Fermi energy~\cite{Randeria2024}. Instead a comparison of $T_{\rm BKT}$ to the energy scale of the diamagnetic response~\cite{Hazra2019} may be more promising.

\subsection{Effects of magnetic field.}

We next move to the charged superconductor.
Combining Eqs.~\eqref{eq:StiffnessSpeedOfLight},~\eqref{eq:PhaseTheory} allows us to immediately extract the magnetic penetration depth (Pearl length) $\lambda = 2c_0/(\alpha_0 \pi K)$ where $c_0$ is the speed of physical light and $\alpha_0 \simeq 1/137$ is the fine structure constant. 
As we speak about superconductivity in strictly two dimensions, the resulting superconductor is essentially always type-II, and indeed $\lambda/\xi \sim \text{min}(c_0/v_b, c_0/v_f)/\alpha_0 \gg 1$. 
Consistently, we highlight that -- in the rigorously clean limit -- the fractionalized theory has no solutions corresponding to phase \ding{173} (gapped slave rotor and topological fermionic chargeon response) in the presence of a homogeneous $B$ entering the material~\cite{SuppMat}.
However, vortex lattice solutions allow for field penetration in the presence of superconductivity. As usual, the zero temperature critical field can be estimated by equating vortex distance (magnetic length) with vortex core size as $H_{c,2} \sim \Phi_0/\xi^2$, where we restored $\Phi_0$, the superconducting flux quantum.

\subsection{Finite doping, finite magnetic field and finite disorder.} 
\label{sec:Doping}
\begin{figure}
\includegraphics[scale=1]{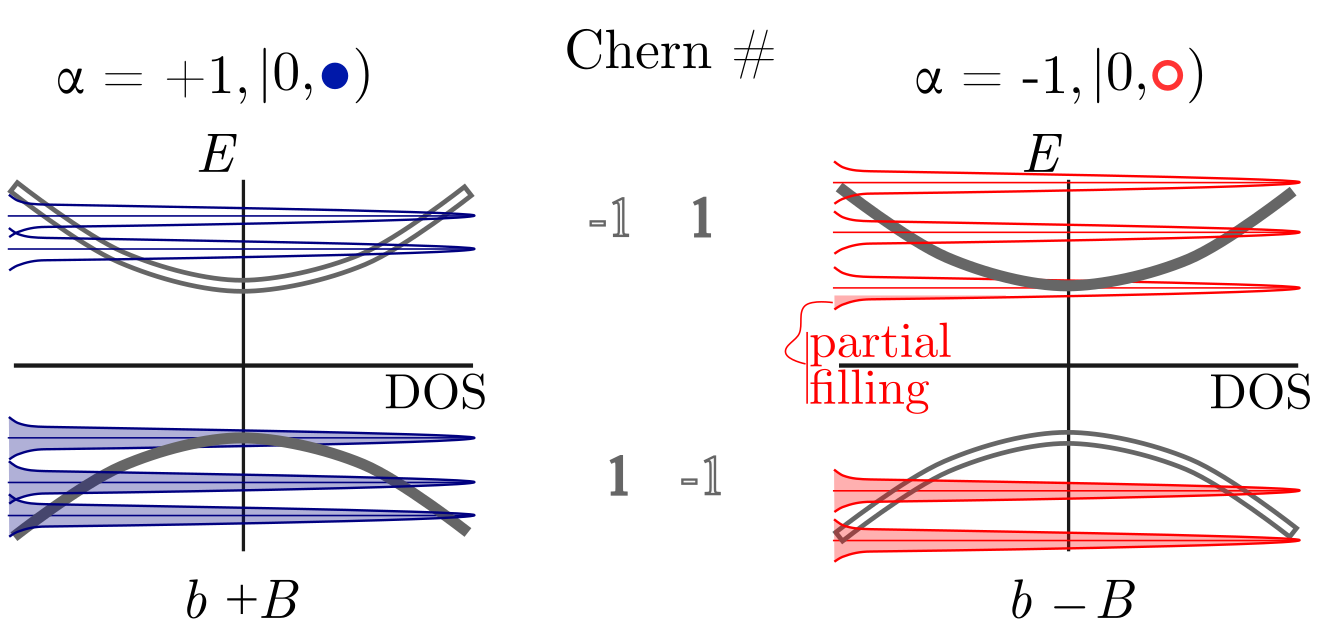}
\caption{Spontaneous formation of Landau levels in the presence of finite doping. In the presence of minimal amount of disorder, the Landau levels broaden allowing for a finite filling fraction of the Landau level closest to zero energy. This figure illustrates the case $n>0$ and $B<0$.}
\label{fig:LLsMaintext}
\end{figure}

In order to accommodate a finite physical charge density $n$ away from charge filling, the energetically preferred mean-field fractionalized state is not, as maybe naively expected, a state with finite chargeon Fermi surface.
Instead, for weak doping, the favored state is obtained by the spontaneous formation of chargeon Landau levels (LL) due to the appearance of a uniform net flux $b\neq 0$ of the emergent gauge field, cf.~Fig.~\ref{fig:LLsMaintext}
~\cite{RanWen2009,SuppMat}, at least on the mean-field level. 

One may additionally include a finite external field $B, \vert B /n \vert \ll 1$, leading to unequal fields $b \pm B$ in the $\alpha = \pm 1$ sectors.
Crucially, to find a superconducting solution with penetrating field, we assumed a small amount of disorder to spread the density of states of Landau levels whilst keeping a mobility gap for all states except a single, delocalized, central Landau level state responsible for the spectral flow. The energetically optimal solution for $n>0$ and in the presence external magnetic field involves a non-zero filling fraction in the Lifshitz tails of the Landau level closest to zero energy. For $n>0$ and $B <0$ ($B>0$) this corresponds to additional particles (holes) of the $\alpha = -1$ ($\alpha = +1$) color. Following the previous discussion, these extra localized charges should be interpreted as pinned (anti-)vortices. Clearly, the occupation of such strongly localized states does not hamper the topological mechanism for superconductivity.
 Technically, here and in the following we concentrate on the limit of large $U$, where $K$ is dominated by $\vert m_{-} \vert = m_+ = m$ (for simplicity, we assume $\varphi = \pi/2$).

Due to the Landau levels, the fermionic gap in the presence of homogeneous magnetic fields $b,B$ is increased as opposed to the case without fields. This affects the gap entering the optical chargeon conductivity, which we use to estimate the stiffness~\cite{SuppMat}.  
We obtain ($\mathcal C$ is a constant of order unity)
\begin{align}
    K(n, B) &\sim \sqrt{m^2 + \mathcal C v_f^2 (\pi \vert n \vert + \vert B \vert)}, \label{eq:StiffnessCorrections}
\end{align}
and  conclude this section with a few comments:
First, perturbatively in doping and magnetic field, the stiffness increases linearly.
Second, we repeat that the superconductor contains randomly pinned vortices, corresponding to a vortex glass~\cite{Fisher1990}.
Third, as both density and magnetic field increase the stiffness, we also expect an increase in $T_{\rm BKT}$.  While finding superconductivity in a model of a metal with attraction is by itself not a surprise, our mechanism leads to a massively larger transition temperature than the BCS expectation $T_c \sim \exp(- w_1^2/w_2 U)$.

\section{Conclusion.} 

\subsection{Summary and Outlook.}
We have discussed a topological mechanism for
superconductivity emerging out of a Chern-2 insulator which -- by means of particle-vortex duality -- is described by emergent QED$_3$ with topological chargeon bands.
Essentially, one may view this superconductor as the projection of the Chern-2-insulator to an empty and doubly-occupied subspace (``anti''-Gutzwiller projection).

We illustrated the mechanism in the context of an attractive-$U$ Haldane model, for which the superconducting phase is entirely absent within a BCS-like mean-field theory. In contrast, slave-rotor theory with the leading quantum fluctuations incorporated on top of the mean-field solution is able to capture the superconducting phase. We have corroborated this effective parton field theory approach 
by complementary calculations using an effective pseudospin model at large $U$ and through the study of anti-Gutzwiller projected wave functions.
Crucially, superconductivity only emerges if the underlying free fermion theory is topological, and we highlight that the present superconducting many-body wave function inherently transcends the Gaussian subspace of the Fock space. While we focused on the problem at half filling (i.e.~a superconductor emerging directly from an insulator upon ramping up attractive interactions), at small doping the superconducting ground state is less surprising, yet it will typically display a much larger transition temperature than what is expected within BCS theory. 

The discussion of finite temperature and, in particular, finite field effects deserves a separate study in a future publication. Another aspect of theoretical interest is the precise nature of the quantum phase transitions.
Here, the \ding{173}-\ding{175} transition is of particular interest, as it corresponds to QED$_3$ with 2 flavors of fermions, which is believed to spontaneously break the SU(2) flavor symmetry by monopole condensation~\cite{DumitrescuThorngreen2024,ChesterKomargodski2024}.

While we have studied concrete microscopic models with a view towards realistic materials, the fundamental mechanism leading to superconductivity in our model lies in the intertwinement of $U(1)$ electric charge and emergent $U(1)$ flux quantum numbers due to the non-trivial topology of the chargeon bands. With recent progress in a non-perturbative understanding of such mixed responses and their field-theoretic description \cite{Senthil15,SenthilXu2019,WangSenthil17}, it appears topical and promising to explore further implications to systems of strongly correlated electrons, e.g., in 3+1-dimensions~\cite{WangSenthil16}, or accounting for crystalline symmetries \cite{ManjunathBarkeshli21,ZhangBarkeshli23}.

\subsection{Experimental implications.}

We return to the experimental inspiration of rhombohedral graphene multilayers~\cite{HanJu2024}
and phenomenologically propose a scenario in which the superconductivity  
may occur due to the vicinity of a $C = \pm 2$ quantum anomalous Hall crystal. While a quantized Hall response is not observed in superconducting devices, we base this conjecture on the enhanced $R_{xy} \sim \text{k} \Omega$ and a slope {of the superconducting phase boundary in magnetic-field vs. density plane which appears consistent with} one superconducting flux quantum, $ dB/dn = \pm h/ 2e \approx \pm 20.7 T/10^{12}cm^{-2}$. 
This is characteristic for $C = \pm 2$ insulators and appears in the phase diagrams of several of the tetralayer and pentalayer samples. Moreover, substrate-aligned rhombohedral graphene has experimentally been found to host Hall crystals in the parameter regime where misaligned samples show superconductivity~\cite{HanJu2024,AronsonAshoori2024,ChoiYoung2024,LuJu2025,WatersYankowitz2025,JC_trithep,JC_ashvin}.

Our scenario is agnostic to the origin of attraction (e.g. phonons, collective modes, ...) 
but explains how superconductivity is enabled by weak doping on top of a $C = \pm 2$ insulator. The interaction energy scale $\sim U$ may be massive as compared to the scale of the topological gap opening $\sim m$ leading to a large separation of scales, Fig.~\ref{fig:SummaryFig} e). In the presence of long-range impurities the stiffness $K \sim m$ for weakest doping and magnetic fields is insufficient to stabilize long-range order. Instead, a strongly correlated anomalous Hall metal with vanishing quasiparticle weight is formed, cf. Sec.~\ref{sec:Zeros}.

\begin{figure}
\includegraphics[scale=.5]{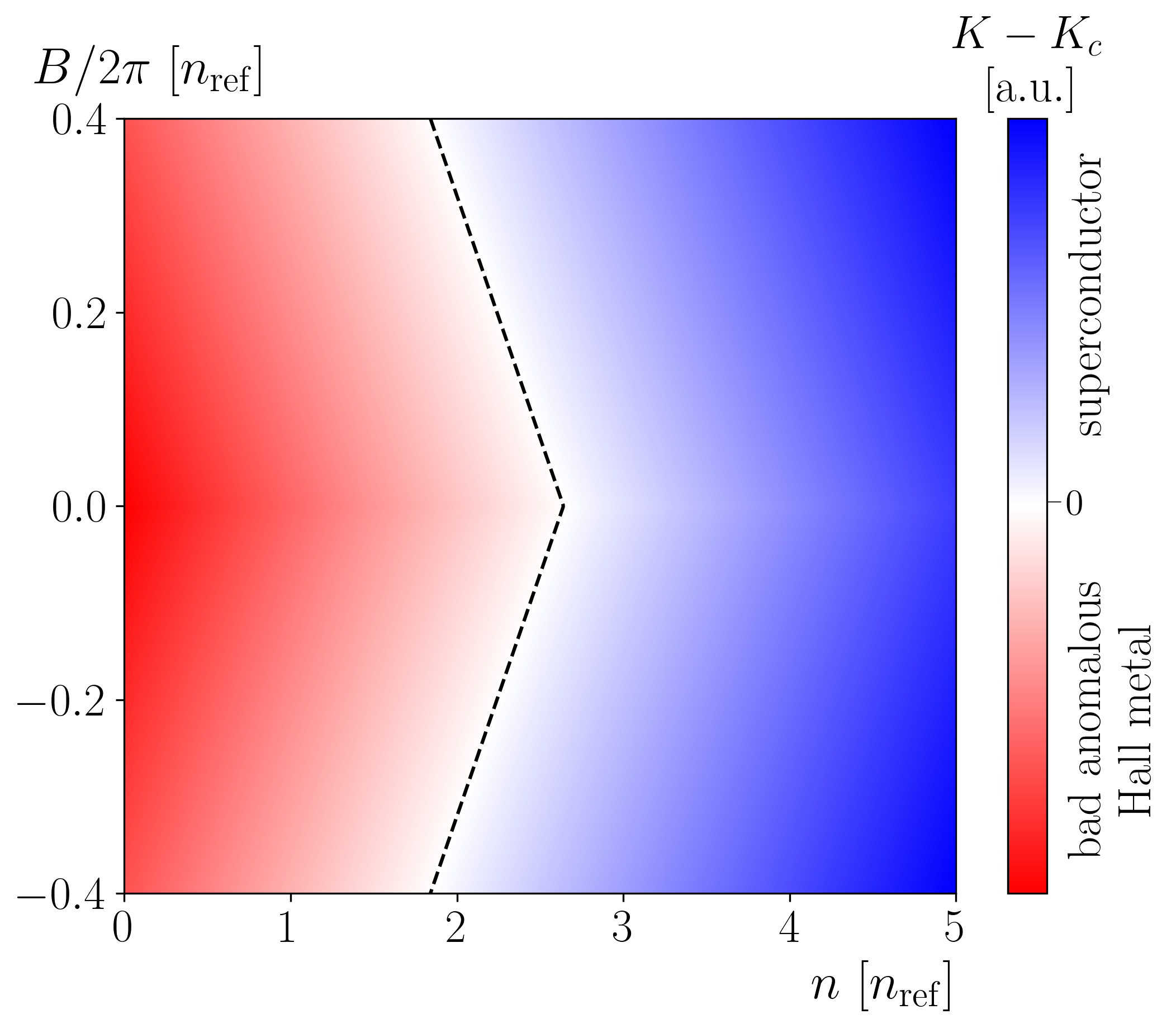}
\caption{{Schematic} plot of the density and magnetic field dependent stiffness relative to a critical stiffness $K(n,B)- K_c$, cf. Eq.~\eqref{eq:StiffnessCorrections}. Here, $n$ and $B$ are expressed in units of $n_{\rm ref} = m^2/\pi v_f^2$ and $K_c  = \mathcal O(m)$.
The dashed line indicates when the density and magnetic field enhanced stiffness surpasses a non-universal critical value, stabilizing the superconductor. Crucially, the slope of the dashed line is universal and in SI units, is $\pm h/2 e = \frac{20.7 T}{10^{12} cm^{-2}}$ (the same as the slope of $C =2$ insulators in Landau fan diagrams)}. 
\label{fig:BqPhasediagram}
\end{figure}

Finally, long-range order is established once $K(n,B)$, which for topological reasons is a function of $ \pi \vert n \vert + \vert B \vert$, exceeds a critical value $K_c$. This is graphically illustrated in Fig.~\ref{fig:BqPhasediagram}, 
note the resemblance to the analogous phase diagram for the arguably best superconductor (SC1 of device T3 in~\cite{HanJu2024}) in rhombohedral graphene. 

We highlight several implications of our theory: First, the superconductor is of strong-pairing type with massively enhanced critical temperature as compared to BCS estimates.
Second, the superconducting state is a color-singlet and transforms trivially under the remaining symmetries of the normal state.
A smoking gun verification would be the experimental discovery of a $C = \pm 2$ (potentially sliding) anomalous Hall crystal in a parameter regime which is close to where superconducting states have previously been found.

\noindent

\textbf{Note added:} During completion of this manuscript we became aware of superconductivity in the vicinity of (quantized) anomalous Hall states in rhombohedral hexalayer graphene ~\cite{MorissetteLi2025} and twisted MoTe$_2$~\cite{XuLi2025}.

\noindent
\section*{Acknowledgments}
It is a pleasure to acknowledge useful discussions with B. M. Andersen, L. Balents, C. Batista, A.Chubukov, M. Crispino, I. Esterlis, Z. Komargodski, M. Scheurer and M. Vavilov. 
FP and GS are indebted to N.~Wagner who has shared his code in the initial stages of the project and acknowledge financial support from the Deutsche Forschungsgemeinschaft (DFG, German Research Foundation)- Project-ID 258499086 - SFB 1170 and through FOR5249-449872909 (Project P5).
UFPS is funded by the DFG under Project~No.~277146847 (SFB 1238, Project C03) and Project No. 544397233 (Emmy Noether program, SE3196/2-1).
The Flatiron Institute is a division of the Simons Foundation. 
Support for this research was provided by the Office of the Vice Chancellor for Research and Graduate Education at the University of Wisconsin–Madison with funding from the Wisconsin Alumni Research Foundation. 
This research was supported in part by grants NSF PHY-1748958 and PHY-2309135 to the Kavli Institute for Theoretical Physics (KITP).
DG, UFPS and EJK acknowledge hospitality by the KITP, where this work was initiated.

\bibliography{TopoEnabledSC}

\begin{thebibliography}{94}%
\makeatletter
\providecommand \@ifxundefined [1]{%
 \@ifx{#1\undefined}
}%
\providecommand \@ifnum [1]{%
 \ifnum #1\expandafter \@firstoftwo
 \else \expandafter \@secondoftwo
 \fi
}%
\providecommand \@ifx [1]{%
 \ifx #1\expandafter \@firstoftwo
 \else \expandafter \@secondoftwo
 \fi
}%
\providecommand \natexlab [1]{#1}%
\providecommand \enquote  [1]{``#1''}%
\providecommand \bibnamefont  [1]{#1}%
\providecommand \bibfnamefont [1]{#1}%
\providecommand \citenamefont [1]{#1}%
\providecommand \href@noop [0]{\@secondoftwo}%
\providecommand \href [0]{\begingroup \@sanitize@url \@href}%
\providecommand \@href[1]{\@@startlink{#1}\@@href}%
\providecommand \@@href[1]{\endgroup#1\@@endlink}%
\providecommand \@sanitize@url [0]{\catcode `\\12\catcode `\$12\catcode `\&12\catcode `\#12\catcode `\^12\catcode `\_12\catcode `\%12\relax}%
\providecommand \@@startlink[1]{}%
\providecommand \@@endlink[0]{}%
\providecommand \url  [0]{\begingroup\@sanitize@url \@url }%
\providecommand \@url [1]{\endgroup\@href {#1}{\urlprefix }}%
\providecommand \urlprefix  [0]{URL }%
\providecommand \Eprint [0]{\href }%
\providecommand \doibase [0]{https://doi.org/}%
\providecommand \selectlanguage [0]{\@gobble}%
\providecommand \bibinfo  [0]{\@secondoftwo}%
\providecommand \bibfield  [0]{\@secondoftwo}%
\providecommand \translation [1]{[#1]}%
\providecommand \BibitemOpen [0]{}%
\providecommand \bibitemStop [0]{}%
\providecommand \bibitemNoStop [0]{.\EOS\space}%
\providecommand \EOS [0]{\spacefactor3000\relax}%
\providecommand \BibitemShut  [1]{\csname bibitem#1\endcsname}%
\let\auto@bib@innerbib\@empty
\bibitem [{\citenamefont {Cao}\ \emph {et~al.}(2018)\citenamefont {Cao}, \citenamefont {Fatemi}, \citenamefont {Fang}, \citenamefont {Watanabe}, \citenamefont {Taniguchi}, \citenamefont {Kaxiras},\ and\ \citenamefont {Jarillo-Herrero}}]{CaoJarilloHerrero2018}%
  \BibitemOpen
  \bibfield  {author} {\bibinfo {author} {\bibfnamefont {Y.}~\bibnamefont {Cao}}, \bibinfo {author} {\bibfnamefont {V.}~\bibnamefont {Fatemi}}, \bibinfo {author} {\bibfnamefont {S.}~\bibnamefont {Fang}}, \bibinfo {author} {\bibfnamefont {K.}~\bibnamefont {Watanabe}}, \bibinfo {author} {\bibfnamefont {T.}~\bibnamefont {Taniguchi}}, \bibinfo {author} {\bibfnamefont {E.}~\bibnamefont {Kaxiras}},\ and\ \bibinfo {author} {\bibfnamefont {P.}~\bibnamefont {Jarillo-Herrero}},\ }\bibfield  {title} {\bibinfo {title} {Unconventional superconductivity in magic-angle graphene superlattices},\ }\href {https://doi.org/10.1038/nature26160} {\bibfield  {journal} {\bibinfo  {journal} {Nature}\ }\textbf {\bibinfo {volume} {556}},\ \bibinfo {pages} {43} (\bibinfo {year} {2018})}\BibitemShut {NoStop}%
\bibitem [{\citenamefont {Wang}\ \emph {et~al.}(2020)\citenamefont {Wang}, \citenamefont {Shih}, \citenamefont {Ghiotto}, \citenamefont {Xian}, \citenamefont {Rhodes}, \citenamefont {Tan}, \citenamefont {Claassen}, \citenamefont {Kennes}, \citenamefont {Bai}, \citenamefont {Kim} \emph {et~al.}}]{Wang2020}%
  \BibitemOpen
  \bibfield  {author} {\bibinfo {author} {\bibfnamefont {L.}~\bibnamefont {Wang}}, \bibinfo {author} {\bibfnamefont {E.-M.}\ \bibnamefont {Shih}}, \bibinfo {author} {\bibfnamefont {A.}~\bibnamefont {Ghiotto}}, \bibinfo {author} {\bibfnamefont {L.}~\bibnamefont {Xian}}, \bibinfo {author} {\bibfnamefont {D.~A.}\ \bibnamefont {Rhodes}}, \bibinfo {author} {\bibfnamefont {C.}~\bibnamefont {Tan}}, \bibinfo {author} {\bibfnamefont {M.}~\bibnamefont {Claassen}}, \bibinfo {author} {\bibfnamefont {D.~M.}\ \bibnamefont {Kennes}}, \bibinfo {author} {\bibfnamefont {Y.}~\bibnamefont {Bai}}, \bibinfo {author} {\bibfnamefont {B.}~\bibnamefont {Kim}}, \emph {et~al.},\ }\bibfield  {title} {\bibinfo {title} {Correlated electronic phases in twisted bilayer transition metal dichalcogenides},\ }\href {https://doi.org/10.1038/s41563-020-0708-6} {\bibfield  {journal} {\bibinfo  {journal} {Nature materials}\ }\textbf {\bibinfo {volume} {19}},\ \bibinfo {pages} {861} (\bibinfo {year} {2020})}\BibitemShut {NoStop}%
\bibitem [{\citenamefont {Marsiglio}(2020)}]{Marsiglio2020}%
  \BibitemOpen
  \bibfield  {author} {\bibinfo {author} {\bibfnamefont {F.}~\bibnamefont {Marsiglio}},\ }\bibfield  {title} {\bibinfo {title} {Eliashberg theory: A short review},\ }\href {https://www.sciencedirect.com/science/article/pii/S000349162030035X} {\bibfield  {journal} {\bibinfo  {journal} {Annals of Physics}\ }\textbf {\bibinfo {volume} {417}},\ \bibinfo {pages} {168102} (\bibinfo {year} {2020})}\BibitemShut {NoStop}%
\bibitem [{\citenamefont {Chen}\ \emph {et~al.}(2024)\citenamefont {Chen}, \citenamefont {Wang}, \citenamefont {Boyack}, \citenamefont {Yang},\ and\ \citenamefont {Levin}}]{ChenLevin2024}%
  \BibitemOpen
  \bibfield  {author} {\bibinfo {author} {\bibfnamefont {Q.}~\bibnamefont {Chen}}, \bibinfo {author} {\bibfnamefont {Z.}~\bibnamefont {Wang}}, \bibinfo {author} {\bibfnamefont {R.}~\bibnamefont {Boyack}}, \bibinfo {author} {\bibfnamefont {S.}~\bibnamefont {Yang}},\ and\ \bibinfo {author} {\bibfnamefont {K.}~\bibnamefont {Levin}},\ }\bibfield  {title} {\bibinfo {title} {When superconductivity crosses over: From {BCS} to {BEC}},\ }\href {https://doi.org/10.1103/RevModPhys.96.025002} {\bibfield  {journal} {\bibinfo  {journal} {Rev. Mod. Phys.}\ }\textbf {\bibinfo {volume} {96}},\ \bibinfo {pages} {025002} (\bibinfo {year} {2024})}\BibitemShut {NoStop}%
\bibitem [{\citenamefont {Christos}\ \emph {et~al.}(2020)\citenamefont {Christos}, \citenamefont {Sachdev},\ and\ \citenamefont {Scheurer}}]{ChristosScheurer2020}%
  \BibitemOpen
  \bibfield  {author} {\bibinfo {author} {\bibfnamefont {M.}~\bibnamefont {Christos}}, \bibinfo {author} {\bibfnamefont {S.}~\bibnamefont {Sachdev}},\ and\ \bibinfo {author} {\bibfnamefont {M.~S.}\ \bibnamefont {Scheurer}},\ }\bibfield  {title} {\bibinfo {title} {Superconductivity, correlated insulators, and {W}ess–{Z}umino–{W}itten terms in twisted bilayer graphene},\ }\href {https://www.pnas.org/doi/abs/10.1073/pnas.2014691117} {\bibfield  {journal} {\bibinfo  {journal} {Proceedings of the National Academy of Sciences}\ }\textbf {\bibinfo {volume} {117}},\ \bibinfo {pages} {29543} (\bibinfo {year} {2020})}\BibitemShut {NoStop}%
\bibitem [{\citenamefont {Khalaf}\ \emph {et~al.}(2021)\citenamefont {Khalaf}, \citenamefont {Chatterjee}, \citenamefont {Bultinck}, \citenamefont {Zaletel},\ and\ \citenamefont {Vishwanath}}]{KhalafVishwanath2021}%
  \BibitemOpen
  \bibfield  {author} {\bibinfo {author} {\bibfnamefont {E.}~\bibnamefont {Khalaf}}, \bibinfo {author} {\bibfnamefont {S.}~\bibnamefont {Chatterjee}}, \bibinfo {author} {\bibfnamefont {N.}~\bibnamefont {Bultinck}}, \bibinfo {author} {\bibfnamefont {M.~P.}\ \bibnamefont {Zaletel}},\ and\ \bibinfo {author} {\bibfnamefont {A.}~\bibnamefont {Vishwanath}},\ }\bibfield  {title} {\bibinfo {title} {Charged skyrmions and topological origin of superconductivity in magic-angle graphene},\ }\href {https://www.science.org/doi/abs/10.1126/sciadv.abf5299} {\bibfield  {journal} {\bibinfo  {journal} {Science advances}\ }\textbf {\bibinfo {volume} {7}} (\bibinfo {year} {2021})}\BibitemShut {NoStop}%
\bibitem [{\citenamefont {Shavit}\ and\ \citenamefont {Alicea}(2024)}]{ShavitAlicea2024}%
  \BibitemOpen
  \bibfield  {author} {\bibinfo {author} {\bibfnamefont {G.}~\bibnamefont {Shavit}}\ and\ \bibinfo {author} {\bibfnamefont {J.}~\bibnamefont {Alicea}},\ }\bibfield  {title} {\bibinfo {title} {Quantum geometric unconventional superconductivity},\ }\href {https://arxiv.org/abs/2411.05071} {\bibfield  {journal} {\bibinfo  {journal} {arXiv:2411.05071}\ } (\bibinfo {year} {2024})}\BibitemShut {NoStop}%
\bibitem [{\citenamefont {Hazra}\ \emph {et~al.}(2019)\citenamefont {Hazra}, \citenamefont {Verma},\ and\ \citenamefont {Randeria}}]{Hazra2019}%
  \BibitemOpen
  \bibfield  {author} {\bibinfo {author} {\bibfnamefont {T.}~\bibnamefont {Hazra}}, \bibinfo {author} {\bibfnamefont {N.}~\bibnamefont {Verma}},\ and\ \bibinfo {author} {\bibfnamefont {M.}~\bibnamefont {Randeria}},\ }\bibfield  {title} {\bibinfo {title} {Bounds on the superconducting transition temperature: Applications to twisted bilayer graphene and cold atoms},\ }\href {https://doi.org/10.1103/PhysRevX.9.031049} {\bibfield  {journal} {\bibinfo  {journal} {Phys. Rev. X}\ }\textbf {\bibinfo {volume} {9}},\ \bibinfo {pages} {031049} (\bibinfo {year} {2019})}\BibitemShut {NoStop}%
\bibitem [{\citenamefont {Randeria}(2024)}]{Randeria2024}%
  \BibitemOpen
  \bibfield  {author} {\bibinfo {author} {\bibfnamefont {M.}~\bibnamefont {Randeria}},\ }\bibfield  {title} {\bibinfo {title} {Bounds on the superconducting transition temperature},\ }\href {https://doi.org/10.1142/S0217984924300047} {\bibfield  {journal} {\bibinfo  {journal} {Modern Physics Letters B}\ }\textbf {\bibinfo {volume} {38}},\ \bibinfo {pages} {2430004} (\bibinfo {year} {2024})}\BibitemShut {NoStop}%
\bibitem [{\citenamefont {Mao}\ and\ \citenamefont {Chowdhury}(2024)}]{MaoChowdhury2024}%
  \BibitemOpen
  \bibfield  {author} {\bibinfo {author} {\bibfnamefont {D.}~\bibnamefont {Mao}}\ and\ \bibinfo {author} {\bibfnamefont {D.}~\bibnamefont {Chowdhury}},\ }\bibfield  {title} {\bibinfo {title} {Upper bounds on superconducting and excitonic phase stiffness for interacting isolated narrow bands},\ }\href {https://doi.org/10.1103/PhysRevB.109.024507} {\bibfield  {journal} {\bibinfo  {journal} {Phys. Rev. B}\ }\textbf {\bibinfo {volume} {109}},\ \bibinfo {pages} {024507} (\bibinfo {year} {2024})}\BibitemShut {NoStop}%
\bibitem [{\citenamefont {Peotta}\ and\ \citenamefont {T{\"o}rm{\"a}}(2015)}]{PeottaTorma2015}%
  \BibitemOpen
  \bibfield  {author} {\bibinfo {author} {\bibfnamefont {S.}~\bibnamefont {Peotta}}\ and\ \bibinfo {author} {\bibfnamefont {P.}~\bibnamefont {T{\"o}rm{\"a}}},\ }\bibfield  {title} {\bibinfo {title} {Superfluidity in topologically nontrivial flat bands},\ }\href {https://doi.org/10.1038/ncomms9944} {\bibfield  {journal} {\bibinfo  {journal} {Nature communications}\ }\textbf {\bibinfo {volume} {6}},\ \bibinfo {pages} {8944} (\bibinfo {year} {2015})}\BibitemShut {NoStop}%
\bibitem [{\citenamefont {T{\"o}rm{\"a}}\ \emph {et~al.}(2022)\citenamefont {T{\"o}rm{\"a}}, \citenamefont {Peotta},\ and\ \citenamefont {Bernevig}}]{TormaBernevig2022}%
  \BibitemOpen
  \bibfield  {author} {\bibinfo {author} {\bibfnamefont {P.}~\bibnamefont {T{\"o}rm{\"a}}}, \bibinfo {author} {\bibfnamefont {S.}~\bibnamefont {Peotta}},\ and\ \bibinfo {author} {\bibfnamefont {B.~A.}\ \bibnamefont {Bernevig}},\ }\bibfield  {title} {\bibinfo {title} {Superconductivity, superfluidity and quantum geometry in twisted multilayer systems},\ }\href {https://doi.org/10.1038/s42254-022-00466-y} {\bibfield  {journal} {\bibinfo  {journal} {Nature Reviews Physics}\ }\textbf {\bibinfo {volume} {4}},\ \bibinfo {pages} {528} (\bibinfo {year} {2022})}\BibitemShut {NoStop}%
\bibitem [{\citenamefont {Christos}\ \emph {et~al.}(2023)\citenamefont {Christos}, \citenamefont {Sachdev},\ and\ \citenamefont {Scheurer}}]{ChristosScheurer2023}%
  \BibitemOpen
  \bibfield  {author} {\bibinfo {author} {\bibfnamefont {M.}~\bibnamefont {Christos}}, \bibinfo {author} {\bibfnamefont {S.}~\bibnamefont {Sachdev}},\ and\ \bibinfo {author} {\bibfnamefont {M.~S.}\ \bibnamefont {Scheurer}},\ }\bibfield  {title} {\bibinfo {title} {Nodal band-off-diagonal superconductivity in twisted graphene superlattices},\ }\href {https://doi.org/10.1038/s41467-023-42471-4} {\bibfield  {journal} {\bibinfo  {journal} {Nature Communications}\ }\textbf {\bibinfo {volume} {14}},\ \bibinfo {pages} {7134} (\bibinfo {year} {2023})}\BibitemShut {NoStop}%
\bibitem [{\citenamefont {Chubukov}\ and\ \citenamefont {Varma}(2025)}]{ChubukovVarma2024}%
  \BibitemOpen
  \bibfield  {author} {\bibinfo {author} {\bibfnamefont {A.~V.}\ \bibnamefont {Chubukov}}\ and\ \bibinfo {author} {\bibfnamefont {C.~M.}\ \bibnamefont {Varma}},\ }\bibfield  {title} {\bibinfo {title} {Quantum criticality and superconductivity in twisted transition metal dichalcogenides},\ }\href {https://doi.org/10.1103/PhysRevB.111.014507} {\bibfield  {journal} {\bibinfo  {journal} {Phys. Rev. B}\ }\textbf {\bibinfo {volume} {111}},\ \bibinfo {pages} {014507} (\bibinfo {year} {2025})}\BibitemShut {NoStop}%
\bibitem [{\citenamefont {Guerci}\ and\ \citenamefont {Fu}(2025)}]{GuerciFu2025}%
  \BibitemOpen
  \bibfield  {author} {\bibinfo {author} {\bibfnamefont {D.}~\bibnamefont {Guerci}}\ and\ \bibinfo {author} {\bibfnamefont {L.}~\bibnamefont {Fu}},\ }\bibfield  {title} {\bibinfo {title} {Spin-polarized superconductivity from excitonic {C}ooper pairs},\ }\href {https://arxiv.org/abs/2503.05863} {\bibfield  {journal} {\bibinfo  {journal} {arXiv:2503.05863}\ } (\bibinfo {year} {2025})}\BibitemShut {NoStop}%
\bibitem [{\citenamefont {Han}\ \emph {et~al.}(2024)\citenamefont {Han}, \citenamefont {Lu}, \citenamefont {Yao}, \citenamefont {Shi}, \citenamefont {Yang}, \citenamefont {Seo}, \citenamefont {Ye}, \citenamefont {Wu}, \citenamefont {Zhou}, \citenamefont {Liu} \emph {et~al.}}]{HanJu2024}%
  \BibitemOpen
  \bibfield  {author} {\bibinfo {author} {\bibfnamefont {T.}~\bibnamefont {Han}}, \bibinfo {author} {\bibfnamefont {Z.}~\bibnamefont {Lu}}, \bibinfo {author} {\bibfnamefont {Y.}~\bibnamefont {Yao}}, \bibinfo {author} {\bibfnamefont {L.}~\bibnamefont {Shi}}, \bibinfo {author} {\bibfnamefont {J.}~\bibnamefont {Yang}}, \bibinfo {author} {\bibfnamefont {J.}~\bibnamefont {Seo}}, \bibinfo {author} {\bibfnamefont {S.}~\bibnamefont {Ye}}, \bibinfo {author} {\bibfnamefont {Z.}~\bibnamefont {Wu}}, \bibinfo {author} {\bibfnamefont {M.}~\bibnamefont {Zhou}}, \bibinfo {author} {\bibfnamefont {H.}~\bibnamefont {Liu}}, \emph {et~al.},\ }\bibfield  {title} {\bibinfo {title} {Signatures of chiral superconductivity in rhombohedral graphene},\ }\href {https://arxiv.org/abs/2408.15233} {\bibfield  {journal} {\bibinfo  {journal} {arXiv:2408.15233}\ } (\bibinfo {year} {2024})}\BibitemShut {NoStop}%
\bibitem [{\citenamefont {Aronson}\ \emph {et~al.}(2024)\citenamefont {Aronson}, \citenamefont {Han}, \citenamefont {Lu}, \citenamefont {Yao}, \citenamefont {Watanabe}, \citenamefont {Taniguchi}, \citenamefont {Ju},\ and\ \citenamefont {Ashoori}}]{AronsonAshoori2024}%
  \BibitemOpen
  \bibfield  {author} {\bibinfo {author} {\bibfnamefont {S.~H.}\ \bibnamefont {Aronson}}, \bibinfo {author} {\bibfnamefont {T.}~\bibnamefont {Han}}, \bibinfo {author} {\bibfnamefont {Z.}~\bibnamefont {Lu}}, \bibinfo {author} {\bibfnamefont {Y.}~\bibnamefont {Yao}}, \bibinfo {author} {\bibfnamefont {K.}~\bibnamefont {Watanabe}}, \bibinfo {author} {\bibfnamefont {T.}~\bibnamefont {Taniguchi}}, \bibinfo {author} {\bibfnamefont {L.}~\bibnamefont {Ju}},\ and\ \bibinfo {author} {\bibfnamefont {R.~C.}\ \bibnamefont {Ashoori}},\ }\bibfield  {title} {\bibinfo {title} {Displacement field-controlled fractional {C}hern insulators and charge density waves in a graphene/h{BN} moir\'e superlattice},\ }\href {https://arxiv.org/abs/2408.11220} {\bibfield  {journal} {\bibinfo  {journal} {arXiv:2408.11220}\ } (\bibinfo {year} {2024})}\BibitemShut {NoStop}%
\bibitem [{\citenamefont {Choi}\ \emph {et~al.}(2024)\citenamefont {Choi}, \citenamefont {Choi}, \citenamefont {Valentini}, \citenamefont {Patterson}, \citenamefont {Holleis}, \citenamefont {Sheekey}, \citenamefont {Stoyanov}, \citenamefont {Cheng}, \citenamefont {Taniguchi}, \citenamefont {Watanabe},\ and\ \citenamefont {Young}}]{ChoiYoung2024}%
  \BibitemOpen
  \bibfield  {author} {\bibinfo {author} {\bibfnamefont {Y.}~\bibnamefont {Choi}}, \bibinfo {author} {\bibfnamefont {Y.}~\bibnamefont {Choi}}, \bibinfo {author} {\bibfnamefont {M.}~\bibnamefont {Valentini}}, \bibinfo {author} {\bibfnamefont {C.~L.}\ \bibnamefont {Patterson}}, \bibinfo {author} {\bibfnamefont {L.~F.~W.}\ \bibnamefont {Holleis}}, \bibinfo {author} {\bibfnamefont {O.~I.}\ \bibnamefont {Sheekey}}, \bibinfo {author} {\bibfnamefont {H.}~\bibnamefont {Stoyanov}}, \bibinfo {author} {\bibfnamefont {X.}~\bibnamefont {Cheng}}, \bibinfo {author} {\bibfnamefont {T.}~\bibnamefont {Taniguchi}}, \bibinfo {author} {\bibfnamefont {K.}~\bibnamefont {Watanabe}},\ and\ \bibinfo {author} {\bibfnamefont {A.~F.}\ \bibnamefont {Young}},\ }\bibfield  {title} {\bibinfo {title} {Electric field control of superconductivity and quantized anomalous {H}all effects in rhombohedral tetralayer graphene},\ }\href {https://arxiv.org/abs/2408.12584} {\bibfield  {journal} {\bibinfo  {journal} {arXiv:2408.12584}\ } (\bibinfo {year}
  {2024})}\BibitemShut {NoStop}%
\bibitem [{\citenamefont {Lu}\ \emph {et~al.}(2025)\citenamefont {Lu}, \citenamefont {Han}, \citenamefont {Yao}, \citenamefont {Hadjri}, \citenamefont {Yang}, \citenamefont {Seo}, \citenamefont {Shi}, \citenamefont {Ye}, \citenamefont {Watanabe}, \citenamefont {Taniguchi} \emph {et~al.}}]{LuJu2025}%
  \BibitemOpen
  \bibfield  {author} {\bibinfo {author} {\bibfnamefont {Z.}~\bibnamefont {Lu}}, \bibinfo {author} {\bibfnamefont {T.}~\bibnamefont {Han}}, \bibinfo {author} {\bibfnamefont {Y.}~\bibnamefont {Yao}}, \bibinfo {author} {\bibfnamefont {Z.}~\bibnamefont {Hadjri}}, \bibinfo {author} {\bibfnamefont {J.}~\bibnamefont {Yang}}, \bibinfo {author} {\bibfnamefont {J.}~\bibnamefont {Seo}}, \bibinfo {author} {\bibfnamefont {L.}~\bibnamefont {Shi}}, \bibinfo {author} {\bibfnamefont {S.}~\bibnamefont {Ye}}, \bibinfo {author} {\bibfnamefont {K.}~\bibnamefont {Watanabe}}, \bibinfo {author} {\bibfnamefont {T.}~\bibnamefont {Taniguchi}}, \emph {et~al.},\ }\bibfield  {title} {\bibinfo {title} {Extended quantum anomalous {H}all states in graphene/h{BN} moir{\'e} superlattices},\ }\href {https://doi.org/10.1038/s41586-024-08470-1} {\bibfield  {journal} {\bibinfo  {journal} {Nature}\ ,\ \bibinfo {pages} {1}} (\bibinfo {year} {2025})}\BibitemShut {NoStop}%
\bibitem [{\citenamefont {Waters}\ \emph {et~al.}(2025)\citenamefont {Waters}, \citenamefont {Okounkova}, \citenamefont {Su}, \citenamefont {Zhou}, \citenamefont {Yao}, \citenamefont {Watanabe}, \citenamefont {Taniguchi}, \citenamefont {Xu}, \citenamefont {Zhang}, \citenamefont {Folk},\ and\ \citenamefont {Yankowitz}}]{WatersYankowitz2025}%
  \BibitemOpen
  \bibfield  {author} {\bibinfo {author} {\bibfnamefont {D.}~\bibnamefont {Waters}}, \bibinfo {author} {\bibfnamefont {A.}~\bibnamefont {Okounkova}}, \bibinfo {author} {\bibfnamefont {R.}~\bibnamefont {Su}}, \bibinfo {author} {\bibfnamefont {B.}~\bibnamefont {Zhou}}, \bibinfo {author} {\bibfnamefont {J.}~\bibnamefont {Yao}}, \bibinfo {author} {\bibfnamefont {K.}~\bibnamefont {Watanabe}}, \bibinfo {author} {\bibfnamefont {T.}~\bibnamefont {Taniguchi}}, \bibinfo {author} {\bibfnamefont {X.}~\bibnamefont {Xu}}, \bibinfo {author} {\bibfnamefont {Y.-H.}\ \bibnamefont {Zhang}}, \bibinfo {author} {\bibfnamefont {J.}~\bibnamefont {Folk}},\ and\ \bibinfo {author} {\bibfnamefont {M.}~\bibnamefont {Yankowitz}},\ }\bibfield  {title} {\bibinfo {title} {Chern insulators at integer and fractional filling in moir\'e pentalayer graphene},\ }\href {https://doi.org/10.1103/PhysRevX.15.011045} {\bibfield  {journal} {\bibinfo  {journal} {Phys. Rev. X}\ }\textbf {\bibinfo {volume} {15}},\ \bibinfo {pages} {011045} (\bibinfo {year}
  {2025})}\BibitemShut {NoStop}%
\bibitem [{\citenamefont {Devakul}(2025)}]{JC_trithep}%
  \BibitemOpen
  \bibfield  {author} {\bibinfo {author} {\bibfnamefont {T.}~\bibnamefont {Devakul}},\ }\href {https://doi.org/10.36471/JCCM_August_2024_01} {\bibinfo {title} {The new quantum anomalous {H}all effects require new concepts}} (\bibinfo {year} {2025})\BibitemShut {NoStop}%
\bibitem [{\citenamefont {Zhang}\ and\ \citenamefont {Vishwanath}(2025)}]{JC_ashvin}%
  \BibitemOpen
  \bibfield  {author} {\bibinfo {author} {\bibfnamefont {Y.-H.}\ \bibnamefont {Zhang}}\ and\ \bibinfo {author} {\bibfnamefont {A.}~\bibnamefont {Vishwanath}},\ }\href {https://doi.org/10.36471/JCCM_February_2025_02} {\bibinfo {title} {Chiral superconductivity in the flat bands of rhombohedral graphene}} (\bibinfo {year} {2025})\BibitemShut {NoStop}%
\bibitem [{\citenamefont {Dong}\ \emph {et~al.}(2024{\natexlab{a}})\citenamefont {Dong}, \citenamefont {Patri},\ and\ \citenamefont {Senthil}}]{DongSenthil2024}%
  \BibitemOpen
  \bibfield  {author} {\bibinfo {author} {\bibfnamefont {Z.}~\bibnamefont {Dong}}, \bibinfo {author} {\bibfnamefont {A.~S.}\ \bibnamefont {Patri}},\ and\ \bibinfo {author} {\bibfnamefont {T.}~\bibnamefont {Senthil}},\ }\bibfield  {title} {\bibinfo {title} {Theory of quantum anomalous {H}all phases in pentalayer rhombohedral graphene moir\'e structures},\ }\href {https://doi.org/10.1103/PhysRevLett.133.206502} {\bibfield  {journal} {\bibinfo  {journal} {Phys. Rev. Lett.}\ }\textbf {\bibinfo {volume} {133}},\ \bibinfo {pages} {206502} (\bibinfo {year} {2024}{\natexlab{a}})}\BibitemShut {NoStop}%
\bibitem [{\citenamefont {Dong}\ \emph {et~al.}(2024{\natexlab{b}})\citenamefont {Dong}, \citenamefont {Wang}, \citenamefont {Wang}, \citenamefont {Soejima}, \citenamefont {Zaletel}, \citenamefont {Vishwanath},\ and\ \citenamefont {Parker}}]{DongParker2024}%
  \BibitemOpen
  \bibfield  {author} {\bibinfo {author} {\bibfnamefont {J.}~\bibnamefont {Dong}}, \bibinfo {author} {\bibfnamefont {T.}~\bibnamefont {Wang}}, \bibinfo {author} {\bibfnamefont {T.}~\bibnamefont {Wang}}, \bibinfo {author} {\bibfnamefont {T.}~\bibnamefont {Soejima}}, \bibinfo {author} {\bibfnamefont {M.~P.}\ \bibnamefont {Zaletel}}, \bibinfo {author} {\bibfnamefont {A.}~\bibnamefont {Vishwanath}},\ and\ \bibinfo {author} {\bibfnamefont {D.~E.}\ \bibnamefont {Parker}},\ }\bibfield  {title} {\bibinfo {title} {Anomalous {H}all crystals in rhombohedral multilayer graphene. {I}. {I}nteraction-driven {C}hern bands and fractional quantum {H}all states at zero magnetic field},\ }\href {http://dx.doi.org/10.1103/PhysRevLett.133.206503} {\bibfield  {journal} {\bibinfo  {journal} {Phys. Rev. Lett.}\ }\textbf {\bibinfo {volume} {133}} (\bibinfo {year} {2024}{\natexlab{b}})}\BibitemShut {NoStop}%
\bibitem [{\citenamefont {Zhou}\ \emph {et~al.}(2024)\citenamefont {Zhou}, \citenamefont {Yang},\ and\ \citenamefont {Zhang}}]{ZhouZhang2023}%
  \BibitemOpen
  \bibfield  {author} {\bibinfo {author} {\bibfnamefont {B.}~\bibnamefont {Zhou}}, \bibinfo {author} {\bibfnamefont {H.}~\bibnamefont {Yang}},\ and\ \bibinfo {author} {\bibfnamefont {Y.-H.}\ \bibnamefont {Zhang}},\ }\bibfield  {title} {\bibinfo {title} {Fractional quantum anomalous {H}all effect in rhombohedral multilayer graphene in the moir\'eless limit},\ }\href {https://doi.org/10.1103/PhysRevLett.133.206504} {\bibfield  {journal} {\bibinfo  {journal} {Phys. Rev. Lett.}\ }\textbf {\bibinfo {volume} {133}},\ \bibinfo {pages} {206504} (\bibinfo {year} {2024})}\BibitemShut {NoStop}%
\bibitem [{\citenamefont {Kwan}\ \emph {et~al.}(2023)\citenamefont {Kwan}, \citenamefont {Yu}, \citenamefont {Herzog-Arbeitman}, \citenamefont {Efetov}, \citenamefont {Regnault},\ and\ \citenamefont {Bernevig}}]{KwanBernevig2023}%
  \BibitemOpen
  \bibfield  {author} {\bibinfo {author} {\bibfnamefont {Y.~H.}\ \bibnamefont {Kwan}}, \bibinfo {author} {\bibfnamefont {J.}~\bibnamefont {Yu}}, \bibinfo {author} {\bibfnamefont {J.}~\bibnamefont {Herzog-Arbeitman}}, \bibinfo {author} {\bibfnamefont {D.~K.}\ \bibnamefont {Efetov}}, \bibinfo {author} {\bibfnamefont {N.}~\bibnamefont {Regnault}},\ and\ \bibinfo {author} {\bibfnamefont {B.~A.}\ \bibnamefont {Bernevig}},\ }\bibfield  {title} {\bibinfo {title} {Moir\'e fractional {C}hern insulators {III}: {H}artree-fock {P}hase diagram, magic angle regime for {C}hern insulator states, the role of the moir\'e potential and {G}oldstone gaps in rhombohedral graphene superlattices},\ }\href {https://arxiv.org/abs/2312.11617} {\bibfield  {journal} {\bibinfo  {journal} {arXiv:2312.11617}\ } (\bibinfo {year} {2023})}\BibitemShut {NoStop}%
\bibitem [{\citenamefont {Zeng}\ \emph {et~al.}(2024)\citenamefont {Zeng}, \citenamefont {Guerci}, \citenamefont {Cr\'epel}, \citenamefont {Millis},\ and\ \citenamefont {Cano}}]{ZengCano2024}%
  \BibitemOpen
  \bibfield  {author} {\bibinfo {author} {\bibfnamefont {Y.}~\bibnamefont {Zeng}}, \bibinfo {author} {\bibfnamefont {D.}~\bibnamefont {Guerci}}, \bibinfo {author} {\bibfnamefont {V.}~\bibnamefont {Cr\'epel}}, \bibinfo {author} {\bibfnamefont {A.~J.}\ \bibnamefont {Millis}},\ and\ \bibinfo {author} {\bibfnamefont {J.}~\bibnamefont {Cano}},\ }\bibfield  {title} {\bibinfo {title} {Sublattice structure and topology in spontaneously crystallized electronic states},\ }\href {https://doi.org/10.1103/PhysRevLett.132.236601} {\bibfield  {journal} {\bibinfo  {journal} {Phys. Rev. Lett.}\ }\textbf {\bibinfo {volume} {132}},\ \bibinfo {pages} {236601} (\bibinfo {year} {2024})}\BibitemShut {NoStop}%
\bibitem [{\citenamefont {Barkeshli}\ and\ \citenamefont {Qi}(2012)}]{BarkeshliQi2012}%
  \BibitemOpen
  \bibfield  {author} {\bibinfo {author} {\bibfnamefont {M.}~\bibnamefont {Barkeshli}}\ and\ \bibinfo {author} {\bibfnamefont {X.-L.}\ \bibnamefont {Qi}},\ }\bibfield  {title} {\bibinfo {title} {Topological nematic states and non-abelian lattice dislocations},\ }\href {https://doi.org/10.1103/PhysRevX.2.031013} {\bibfield  {journal} {\bibinfo  {journal} {Phys. Rev. X}\ }\textbf {\bibinfo {volume} {2}},\ \bibinfo {pages} {031013} (\bibinfo {year} {2012})}\BibitemShut {NoStop}%
\bibitem [{\citenamefont {Wu}\ \emph {et~al.}(2015{\natexlab{a}})\citenamefont {Wu}, \citenamefont {Jain},\ and\ \citenamefont {Sun}}]{Sun2015}%
  \BibitemOpen
  \bibfield  {author} {\bibinfo {author} {\bibfnamefont {Y.-H.}\ \bibnamefont {Wu}}, \bibinfo {author} {\bibfnamefont {J.~K.}\ \bibnamefont {Jain}},\ and\ \bibinfo {author} {\bibfnamefont {K.}~\bibnamefont {Sun}},\ }\bibfield  {title} {\bibinfo {title} {Fractional topological phases in generalized {H}ofstadter bands with arbitrary {C}hern numbers},\ }\href {https://doi.org/10.1103/PhysRevB.91.041119} {\bibfield  {journal} {\bibinfo  {journal} {Phys. Rev. B}\ }\textbf {\bibinfo {volume} {91}},\ \bibinfo {pages} {041119} (\bibinfo {year} {2015}{\natexlab{a}})}\BibitemShut {NoStop}%
\bibitem [{\citenamefont {Guerci}\ \emph {et~al.}(2024)\citenamefont {Guerci}, \citenamefont {Wang},\ and\ \citenamefont {Mora}}]{Guerci2024}%
  \BibitemOpen
  \bibfield  {author} {\bibinfo {author} {\bibfnamefont {D.}~\bibnamefont {Guerci}}, \bibinfo {author} {\bibfnamefont {J.}~\bibnamefont {Wang}},\ and\ \bibinfo {author} {\bibfnamefont {C.}~\bibnamefont {Mora}},\ }\bibfield  {title} {\bibinfo {title} {Layer skyrmions for ideal chern bands and twisted bilayer graphene},\ }\href {https://arxiv.org/abs/2408.12652} {\bibfield  {journal} {\bibinfo  {journal} {arXiv:2408.12652}\ } (\bibinfo {year} {2024})}\BibitemShut {NoStop}%
\bibitem [{\citenamefont {Rampp}\ \emph {et~al.}(2022)\citenamefont {Rampp}, \citenamefont {K\"onig},\ and\ \citenamefont {Schmalian}}]{RamppSchmalian2022}%
  \BibitemOpen
  \bibfield  {author} {\bibinfo {author} {\bibfnamefont {M.~A.}\ \bibnamefont {Rampp}}, \bibinfo {author} {\bibfnamefont {E.~J.}\ \bibnamefont {K\"onig}},\ and\ \bibinfo {author} {\bibfnamefont {J.}~\bibnamefont {Schmalian}},\ }\bibfield  {title} {\bibinfo {title} {Topologically enabled superconductivity},\ }\href {https://doi.org/10.1103/PhysRevLett.129.077001} {\bibfield  {journal} {\bibinfo  {journal} {Phys. Rev. Lett.}\ }\textbf {\bibinfo {volume} {129}},\ \bibinfo {pages} {077001} (\bibinfo {year} {2022})}\BibitemShut {NoStop}%
\bibitem [{\citenamefont {Fr{\"o}hlich}(1954)}]{Froehlich1954}%
  \BibitemOpen
  \bibfield  {author} {\bibinfo {author} {\bibfnamefont {H.}~\bibnamefont {Fr{\"o}hlich}},\ }\bibfield  {title} {\bibinfo {title} {On the theory of superconductivity: the one-dimensional case},\ }\href {https://doi.org/10.1098/rspa.1954.0116} {\bibfield  {journal} {\bibinfo  {journal} {Proceedings of the Royal Society of London. Series A. Mathematical and Physical Sciences}\ }\textbf {\bibinfo {volume} {223}},\ \bibinfo {pages} {296} (\bibinfo {year} {1954})}\BibitemShut {NoStop}%
\bibitem [{\citenamefont {Schmalian}(2010)}]{Schmalian2010}%
  \BibitemOpen
  \bibfield  {author} {\bibinfo {author} {\bibfnamefont {J.}~\bibnamefont {Schmalian}},\ }\bibfield  {title} {\bibinfo {title} {Failed theories of superconductivity},\ }\href {https://doi.org/10.1142/S0217984910025280} {\bibfield  {journal} {\bibinfo  {journal} {Modern Physics Letters B}\ }\textbf {\bibinfo {volume} {24}},\ \bibinfo {pages} {2679} (\bibinfo {year} {2010})}\BibitemShut {NoStop}%
\bibitem [{\citenamefont {Wiegmann}(1992)}]{Wiegmann1992}%
  \BibitemOpen
  \bibfield  {author} {\bibinfo {author} {\bibfnamefont {P.}~\bibnamefont {Wiegmann}},\ }\bibfield  {title} {\bibinfo {title} {Topological superconductivity},\ }\href {https://academic.oup.com/ptps/article/doi/10.1143/PTPS.107.243/1870547} {\bibfield  {journal} {\bibinfo  {journal} {Progress of Theoretical Physics Supplement}\ }\textbf {\bibinfo {volume} {107}},\ \bibinfo {pages} {243} (\bibinfo {year} {1992})}\BibitemShut {NoStop}%
\bibitem [{\citenamefont {Wiegmann}(2007)}]{Wiegmann1994}%
  \BibitemOpen
  \bibfield  {author} {\bibinfo {author} {\bibfnamefont {P.}~\bibnamefont {Wiegmann}},\ }\bibfield  {title} {\bibinfo {title} {Topological mechanism of superconductivity},\ }in\ \href {https://arxiv.org/pdf/cond-mat/9504094} {\emph {\bibinfo {booktitle} {Field Theory, Topology and Condensed Matter Physics: Proceedings of the Ninth Chris Engelbrecht Summer School in Theoretical Physics Held at Storms River Mouth, Tsitsikamma National Park, South Africa, 17--28 January 1994}}}\ (\bibinfo  {publisher} {Springer},\ \bibinfo {year} {2007})\ pp.\ \bibinfo {pages} {177--206}\BibitemShut {NoStop}%
\bibitem [{\citenamefont {Lee}(1999)}]{DungHaiLee1999}%
  \BibitemOpen
  \bibfield  {author} {\bibinfo {author} {\bibfnamefont {D.-H.}\ \bibnamefont {Lee}},\ }\bibfield  {title} {\bibinfo {title} {Pairing via index theorem},\ }\href {https://doi.org/10.1103/PhysRevB.60.12429} {\bibfield  {journal} {\bibinfo  {journal} {Phys. Rev. B}\ }\textbf {\bibinfo {volume} {60}},\ \bibinfo {pages} {12429} (\bibinfo {year} {1999})}\BibitemShut {NoStop}%
\bibitem [{\citenamefont {Wilczek}(1982)}]{Wilczek1982}%
  \BibitemOpen
  \bibfield  {author} {\bibinfo {author} {\bibfnamefont {F.}~\bibnamefont {Wilczek}},\ }\bibfield  {title} {\bibinfo {title} {Quantum mechanics of fractional-spin particles},\ }\href {https://doi.org/10.1103/PhysRevLett.49.957} {\bibfield  {journal} {\bibinfo  {journal} {Phys. Rev. Lett.}\ }\textbf {\bibinfo {volume} {49}},\ \bibinfo {pages} {957} (\bibinfo {year} {1982})}\BibitemShut {NoStop}%
\bibitem [{\citenamefont {Laughlin}(1988)}]{Laughlin1988}%
  \BibitemOpen
  \bibfield  {author} {\bibinfo {author} {\bibfnamefont {R.~B.}\ \bibnamefont {Laughlin}},\ }\bibfield  {title} {\bibinfo {title} {Superconducting ground state of noninteracting particles obeying fractional statistics},\ }\href {https://doi.org/10.1103/PhysRevLett.60.2677} {\bibfield  {journal} {\bibinfo  {journal} {Phys. Rev. Lett.}\ }\textbf {\bibinfo {volume} {60}},\ \bibinfo {pages} {2677} (\bibinfo {year} {1988})}\BibitemShut {NoStop}%
\bibitem [{\citenamefont {Banks}\ and\ \citenamefont {Lykken}(1990)}]{BanksLykken1990}%
  \BibitemOpen
  \bibfield  {author} {\bibinfo {author} {\bibfnamefont {T.}~\bibnamefont {Banks}}\ and\ \bibinfo {author} {\bibfnamefont {J.~D.}\ \bibnamefont {Lykken}},\ }\bibfield  {title} {\bibinfo {title} {Landau-{G}inzburg description of anyonic superconductors},\ }\href {https://www.sciencedirect.com/science/article/pii/055032139090439K} {\bibfield  {journal} {\bibinfo  {journal} {Nuclear Physics B}\ }\textbf {\bibinfo {volume} {336}},\ \bibinfo {pages} {500} (\bibinfo {year} {1990})}\BibitemShut {NoStop}%
\bibitem [{\citenamefont {Lee}\ and\ \citenamefont {Fisher}(1989)}]{LeeFisher1989}%
  \BibitemOpen
  \bibfield  {author} {\bibinfo {author} {\bibfnamefont {D.-H.}\ \bibnamefont {Lee}}\ and\ \bibinfo {author} {\bibfnamefont {M.~P.~A.}\ \bibnamefont {Fisher}},\ }\bibfield  {title} {\bibinfo {title} {Anyon superconductivity and the fractional quantum {H}all effect},\ }\href {https://doi.org/10.1103/PhysRevLett.63.903} {\bibfield  {journal} {\bibinfo  {journal} {Phys. Rev. Lett.}\ }\textbf {\bibinfo {volume} {63}},\ \bibinfo {pages} {903} (\bibinfo {year} {1989})}\BibitemShut {NoStop}%
\bibitem [{\citenamefont {Fradkin}(1990)}]{Fradkin1990}%
  \BibitemOpen
  \bibfield  {author} {\bibinfo {author} {\bibfnamefont {E.}~\bibnamefont {Fradkin}},\ }\bibfield  {title} {\bibinfo {title} {Superfluidity of the lattice anyon gas and topological invariance},\ }\href {https://doi.org/10.1103/PhysRevB.42.570} {\bibfield  {journal} {\bibinfo  {journal} {Phys. Rev. B}\ }\textbf {\bibinfo {volume} {42}},\ \bibinfo {pages} {570} (\bibinfo {year} {1990})}\BibitemShut {NoStop}%
\bibitem [{\citenamefont {Barkeshli}\ and\ \citenamefont {McGreevy}(2014)}]{BarkeshliMcGreevy2014}%
  \BibitemOpen
  \bibfield  {author} {\bibinfo {author} {\bibfnamefont {M.}~\bibnamefont {Barkeshli}}\ and\ \bibinfo {author} {\bibfnamefont {J.}~\bibnamefont {McGreevy}},\ }\bibfield  {title} {\bibinfo {title} {Continuous transition between fractional quantum {H}all and superfluid states},\ }\href {https://doi.org/10.1103/PhysRevB.89.235116} {\bibfield  {journal} {\bibinfo  {journal} {Phys. Rev. B}\ }\textbf {\bibinfo {volume} {89}},\ \bibinfo {pages} {235116} (\bibinfo {year} {2014})}\BibitemShut {NoStop}%
\bibitem [{\citenamefont {Grover}\ and\ \citenamefont {Senthil}(2008)}]{GroverSenthil2008}%
  \BibitemOpen
  \bibfield  {author} {\bibinfo {author} {\bibfnamefont {T.}~\bibnamefont {Grover}}\ and\ \bibinfo {author} {\bibfnamefont {T.}~\bibnamefont {Senthil}},\ }\bibfield  {title} {\bibinfo {title} {Topological spin {H}all states, charged skyrmions, and superconductivity in two dimensions},\ }\href {https://doi.org/10.1103/PhysRevLett.100.156804} {\bibfield  {journal} {\bibinfo  {journal} {Phys. Rev. Lett.}\ }\textbf {\bibinfo {volume} {100}},\ \bibinfo {pages} {156804} (\bibinfo {year} {2008})}\BibitemShut {NoStop}%
\bibitem [{\citenamefont {Kim}\ \emph {et~al.}(2025)\citenamefont {Kim}, \citenamefont {Timmel}, \citenamefont {Ju},\ and\ \citenamefont {Wen}}]{KimWen2024}%
  \BibitemOpen
  \bibfield  {author} {\bibinfo {author} {\bibfnamefont {M.}~\bibnamefont {Kim}}, \bibinfo {author} {\bibfnamefont {A.}~\bibnamefont {Timmel}}, \bibinfo {author} {\bibfnamefont {L.}~\bibnamefont {Ju}},\ and\ \bibinfo {author} {\bibfnamefont {X.-G.}\ \bibnamefont {Wen}},\ }\bibfield  {title} {\bibinfo {title} {Topological chiral superconductivity beyond pairing in a {F}ermi liquid},\ }\href {https://doi.org/10.1103/PhysRevB.111.014508} {\bibfield  {journal} {\bibinfo  {journal} {Phys. Rev. B}\ }\textbf {\bibinfo {volume} {111}},\ \bibinfo {pages} {014508} (\bibinfo {year} {2025})}\BibitemShut {NoStop}%
\bibitem [{\citenamefont {Shi}\ and\ \citenamefont {Senthil}(2024)}]{ShiSenthil2024}%
  \BibitemOpen
  \bibfield  {author} {\bibinfo {author} {\bibfnamefont {Z.~D.}\ \bibnamefont {Shi}}\ and\ \bibinfo {author} {\bibfnamefont {T.}~\bibnamefont {Senthil}},\ }\bibfield  {title} {\bibinfo {title} {Doping a fractional quantum anomalous {H}all insulator},\ }\href {https://doi.org/10.48550/arXiv.2409.20567} {\bibfield  {journal} {\bibinfo  {journal} {arXiv:2409.20567}\ } (\bibinfo {year} {2024})}\BibitemShut {NoStop}%
\bibitem [{\citenamefont {Divic}\ \emph {et~al.}(2024)\citenamefont {Divic}, \citenamefont {Crépel}, \citenamefont {Soejima}, \citenamefont {Song}, \citenamefont {Millis}, \citenamefont {Zaletel},\ and\ \citenamefont {Vishwanath}}]{DivicVishwanath2024}%
  \BibitemOpen
  \bibfield  {author} {\bibinfo {author} {\bibfnamefont {S.}~\bibnamefont {Divic}}, \bibinfo {author} {\bibfnamefont {V.}~\bibnamefont {Crépel}}, \bibinfo {author} {\bibfnamefont {T.}~\bibnamefont {Soejima}}, \bibinfo {author} {\bibfnamefont {X.-Y.}\ \bibnamefont {Song}}, \bibinfo {author} {\bibfnamefont {A.}~\bibnamefont {Millis}}, \bibinfo {author} {\bibfnamefont {M.~P.}\ \bibnamefont {Zaletel}},\ and\ \bibinfo {author} {\bibfnamefont {A.}~\bibnamefont {Vishwanath}},\ }\bibfield  {title} {\bibinfo {title} {Anyon superconductivity from topological criticality in a {H}ofstadter-{H}ubbard model},\ }\href {https://arxiv.org/abs/2410.18175} {\bibfield  {journal} {\bibinfo  {journal} {arXiv:2410.18175}\ } (\bibinfo {year} {2024})}\BibitemShut {NoStop}%
\bibitem [{\citenamefont {Sahay}\ \emph {et~al.}(2024)\citenamefont {Sahay}, \citenamefont {Divic}, \citenamefont {Parker}, \citenamefont {Soejima}, \citenamefont {Anand}, \citenamefont {Hauschild}, \citenamefont {Aidelsburger}, \citenamefont {Vishwanath}, \citenamefont {Chatterjee}, \citenamefont {Yao},\ and\ \citenamefont {Zaletel}}]{SahayZalatel2024}%
  \BibitemOpen
  \bibfield  {author} {\bibinfo {author} {\bibfnamefont {R.}~\bibnamefont {Sahay}}, \bibinfo {author} {\bibfnamefont {S.}~\bibnamefont {Divic}}, \bibinfo {author} {\bibfnamefont {D.~E.}\ \bibnamefont {Parker}}, \bibinfo {author} {\bibfnamefont {T.}~\bibnamefont {Soejima}}, \bibinfo {author} {\bibfnamefont {S.}~\bibnamefont {Anand}}, \bibinfo {author} {\bibfnamefont {J.}~\bibnamefont {Hauschild}}, \bibinfo {author} {\bibfnamefont {M.}~\bibnamefont {Aidelsburger}}, \bibinfo {author} {\bibfnamefont {A.}~\bibnamefont {Vishwanath}}, \bibinfo {author} {\bibfnamefont {S.}~\bibnamefont {Chatterjee}}, \bibinfo {author} {\bibfnamefont {N.~Y.}\ \bibnamefont {Yao}},\ and\ \bibinfo {author} {\bibfnamefont {M.~P.}\ \bibnamefont {Zaletel}},\ }\bibfield  {title} {\bibinfo {title} {Superconductivity in a topological lattice model with strong repulsion},\ }\href {https://doi.org/10.1103/PhysRevB.110.195126} {\bibfield  {journal} {\bibinfo  {journal} {Phys. Rev. B}\ }\textbf {\bibinfo {volume} {110}},\ \bibinfo {pages} {195126}
  (\bibinfo {year} {2024})}\BibitemShut {NoStop}%
\bibitem [{\citenamefont {Bollmann}\ \emph {et~al.}(2024{\natexlab{a}})\citenamefont {Bollmann}, \citenamefont {V\"ayrynen},\ and\ \citenamefont {K\"onig}}]{BollmannKoenig2024}%
  \BibitemOpen
  \bibfield  {author} {\bibinfo {author} {\bibfnamefont {S.}~\bibnamefont {Bollmann}}, \bibinfo {author} {\bibfnamefont {J.~I.}\ \bibnamefont {V\"ayrynen}},\ and\ \bibinfo {author} {\bibfnamefont {E.~J.}\ \bibnamefont {K\"onig}},\ }\bibfield  {title} {\bibinfo {title} {Topological {K}ondo effect with spinful {M}ajorana fermions},\ }\href {https://doi.org/10.1103/PhysRevB.110.035136} {\bibfield  {journal} {\bibinfo  {journal} {Phys. Rev. B}\ }\textbf {\bibinfo {volume} {110}},\ \bibinfo {pages} {035136} (\bibinfo {year} {2024}{\natexlab{a}})}\BibitemShut {NoStop}%
\bibitem [{\citenamefont {Senthil}\ \emph {et~al.}(2019)\citenamefont {Senthil}, \citenamefont {Son}, \citenamefont {Wang},\ and\ \citenamefont {Xu}}]{SenthilXu2019}%
  \BibitemOpen
  \bibfield  {author} {\bibinfo {author} {\bibfnamefont {T.}~\bibnamefont {Senthil}}, \bibinfo {author} {\bibfnamefont {D.~T.}\ \bibnamefont {Son}}, \bibinfo {author} {\bibfnamefont {C.}~\bibnamefont {Wang}},\ and\ \bibinfo {author} {\bibfnamefont {C.}~\bibnamefont {Xu}},\ }\bibfield  {title} {\bibinfo {title} {Duality between (2+1) d quantum critical points},\ }\href {https://doi.org/10.1016/j.physrep.2019.09.001} {\bibfield  {journal} {\bibinfo  {journal} {Physics Reports}\ }\textbf {\bibinfo {volume} {827}},\ \bibinfo {pages} {1} (\bibinfo {year} {2019})}\BibitemShut {NoStop}%
\bibitem [{\citenamefont {Dumitrescu}\ \emph {et~al.}(2024)\citenamefont {Dumitrescu}, \citenamefont {Niro},\ and\ \citenamefont {Thorngren}}]{DumitrescuThorngreen2024}%
  \BibitemOpen
  \bibfield  {author} {\bibinfo {author} {\bibfnamefont {T.~T.}\ \bibnamefont {Dumitrescu}}, \bibinfo {author} {\bibfnamefont {P.}~\bibnamefont {Niro}},\ and\ \bibinfo {author} {\bibfnamefont {R.}~\bibnamefont {Thorngren}},\ }\bibfield  {title} {\bibinfo {title} {{Symmetry Breaking from monopole condensation in {QED}$_3$}},\ }\href {https://doi.org/10.48550/arXiv.2410.05366} {\bibfield  {journal} {\bibinfo  {journal} {arXiv:2410.05366}\ } (\bibinfo {year} {2024})}\BibitemShut {NoStop}%
\bibitem [{\citenamefont {Chester}\ and\ \citenamefont {Komargodski}(2024)}]{ChesterKomargodski2024}%
  \BibitemOpen
  \bibfield  {author} {\bibinfo {author} {\bibfnamefont {S.~M.}\ \bibnamefont {Chester}}\ and\ \bibinfo {author} {\bibfnamefont {Z.}~\bibnamefont {Komargodski}},\ }\bibfield  {title} {\bibinfo {title} {{Symmetry Enhancement, {SPT} Absorption, and Duality in {QED}$_3$}},\ }\href {https://doi.org/10.48550/arXiv.2409.17913} {\bibfield  {journal} {\bibinfo  {journal} {arXiv:2409.17913}\ } (\bibinfo {year} {2024})}\BibitemShut {NoStop}%
\bibitem [{\citenamefont {Xu}(2011)}]{Xu2011}%
  \BibitemOpen
  \bibfield  {author} {\bibinfo {author} {\bibfnamefont {C.}~\bibnamefont {Xu}},\ }\bibfield  {title} {\bibinfo {title} {Quantum spin {H}all, triplet superconductor, and topological liquids on the honeycomb lattice},\ }\href {https://doi.org/10.1103/PhysRevB.83.024408} {\bibfield  {journal} {\bibinfo  {journal} {Phys. Rev. B}\ }\textbf {\bibinfo {volume} {83}},\ \bibinfo {pages} {024408} (\bibinfo {year} {2011})}\BibitemShut {NoStop}%
\bibitem [{\citenamefont {Griset}\ and\ \citenamefont {Xu}(2012)}]{GrisetXu2012}%
  \BibitemOpen
  \bibfield  {author} {\bibinfo {author} {\bibfnamefont {C.}~\bibnamefont {Griset}}\ and\ \bibinfo {author} {\bibfnamefont {C.}~\bibnamefont {Xu}},\ }\bibfield  {title} {\bibinfo {title} {Phase diagram of the {K}ane-{M}ele-{H}ubbard model},\ }\href {https://doi.org/10.1103/PhysRevB.85.045123} {\bibfield  {journal} {\bibinfo  {journal} {Phys. Rev. B}\ }\textbf {\bibinfo {volume} {85}},\ \bibinfo {pages} {045123} (\bibinfo {year} {2012})}\BibitemShut {NoStop}%
\bibitem [{\citenamefont {Hohenadler}\ and\ \citenamefont {Assaad}(2013)}]{HohenadlerAssaad2013}%
  \BibitemOpen
  \bibfield  {author} {\bibinfo {author} {\bibfnamefont {M.}~\bibnamefont {Hohenadler}}\ and\ \bibinfo {author} {\bibfnamefont {F.~F.}\ \bibnamefont {Assaad}},\ }\bibfield  {title} {\bibinfo {title} {Correlation effects in two-dimensional topological insulators},\ }\href {https://doi.org/10.1088/0953-8984/25/14/143201} {\bibfield  {journal} {\bibinfo  {journal} {Journal of Physics: Condensed Matter}\ }\textbf {\bibinfo {volume} {25}},\ \bibinfo {pages} {143201} (\bibinfo {year} {2013})}\BibitemShut {NoStop}%
\bibitem [{\citenamefont {Mai}\ \emph {et~al.}(2024)\citenamefont {Mai}, \citenamefont {Zhao},\ and\ \citenamefont {Phillips}}]{MaiPhillips2024}%
  \BibitemOpen
  \bibfield  {author} {\bibinfo {author} {\bibfnamefont {P.}~\bibnamefont {Mai}}, \bibinfo {author} {\bibfnamefont {J.}~\bibnamefont {Zhao}},\ and\ \bibinfo {author} {\bibfnamefont {P.~W.}\ \bibnamefont {Phillips}},\ }\bibfield  {title} {\bibinfo {title} {Incipient quantum spin {H}all insulator under strong correlations},\ }\href {https://doi.org/10.48550/arXiv.2409.07557} {\bibfield  {journal} {\bibinfo  {journal} {arXiv:2409.07557}\ } (\bibinfo {year} {2024})}\BibitemShut {NoStop}%
\bibitem [{\citenamefont {Ran}\ \emph {et~al.}(2009)\citenamefont {Ran}, \citenamefont {Ko}, \citenamefont {Lee},\ and\ \citenamefont {Wen}}]{RanWen2009}%
  \BibitemOpen
  \bibfield  {author} {\bibinfo {author} {\bibfnamefont {Y.}~\bibnamefont {Ran}}, \bibinfo {author} {\bibfnamefont {W.-H.}\ \bibnamefont {Ko}}, \bibinfo {author} {\bibfnamefont {P.~A.}\ \bibnamefont {Lee}},\ and\ \bibinfo {author} {\bibfnamefont {X.-G.}\ \bibnamefont {Wen}},\ }\bibfield  {title} {\bibinfo {title} {Spontaneous spin ordering of a {D}irac spin liquid in a magnetic field},\ }\href {https://doi.org/10.1103/PhysRevLett.102.047205} {\bibfield  {journal} {\bibinfo  {journal} {Phys. Rev. Lett.}\ }\textbf {\bibinfo {volume} {102}},\ \bibinfo {pages} {047205} (\bibinfo {year} {2009})}\BibitemShut {NoStop}%
\bibitem [{\citenamefont {Polyakov}(1987)}]{Polyakov1987}%
  \BibitemOpen
  \bibfield  {author} {\bibinfo {author} {\bibfnamefont {A.~M.}\ \bibnamefont {Polyakov}},\ }\href@noop {} {\emph {\bibinfo {title} {{G}auge fields and strings}}}\ (\bibinfo  {publisher} {Harwood Academic Publishers, Switzerland},\ \bibinfo {year} {1987})\BibitemShut {NoStop}%
\bibitem [{\citenamefont {Streda}(1982)}]{Streda1982}%
  \BibitemOpen
  \bibfield  {author} {\bibinfo {author} {\bibfnamefont {P.}~\bibnamefont {Streda}},\ }\bibfield  {title} {\bibinfo {title} {Theory of quantised {H}all conductivity in two dimensions},\ }\href {https://dx.doi.org/10.1088/0022-3719/15/22/005} {\bibfield  {journal} {\bibinfo  {journal} {Journal of Physics C: Solid State Physics}\ }\textbf {\bibinfo {volume} {15}},\ \bibinfo {pages} {L717} (\bibinfo {year} {1982})}\BibitemShut {NoStop}%
\bibitem [{\citenamefont {Widom}(1982)}]{Widom1982}%
  \BibitemOpen
  \bibfield  {author} {\bibinfo {author} {\bibfnamefont {A.}~\bibnamefont {Widom}},\ }\bibfield  {title} {\bibinfo {title} {Thermodynamic derivation of the {H}all effect current},\ }\href {https://doi.org/https://doi.org/10.1016/0375-9601(82)90401-7} {\bibfield  {journal} {\bibinfo  {journal} {Physics Letters A}\ }\textbf {\bibinfo {volume} {90}},\ \bibinfo {pages} {474} (\bibinfo {year} {1982})}\BibitemShut {NoStop}%
\bibitem [{\citenamefont {Streda}\ and\ \citenamefont {Smrcka}(1983)}]{StredaSmrcka1983}%
  \BibitemOpen
  \bibfield  {author} {\bibinfo {author} {\bibfnamefont {P.}~\bibnamefont {Streda}}\ and\ \bibinfo {author} {\bibfnamefont {L.}~\bibnamefont {Smrcka}},\ }\bibfield  {title} {\bibinfo {title} {Thermodynamic derivation of the hall current and the thermopower in quantising magnetic field},\ }\href {https://doi.org/10.1088/0022-3719/16/24/005} {\bibfield  {journal} {\bibinfo  {journal} {Journal of Physics C: Solid State Physics}\ }\textbf {\bibinfo {volume} {16}},\ \bibinfo {pages} {L895} (\bibinfo {year} {1983})}\BibitemShut {NoStop}%
\bibitem [{\citenamefont {Hansson}\ \emph {et~al.}(2004)\citenamefont {Hansson}, \citenamefont {Oganesyan},\ and\ \citenamefont {Sondhi}}]{HanssonSondhi2004}%
  \BibitemOpen
  \bibfield  {author} {\bibinfo {author} {\bibfnamefont {T.}~\bibnamefont {Hansson}}, \bibinfo {author} {\bibfnamefont {V.}~\bibnamefont {Oganesyan}},\ and\ \bibinfo {author} {\bibfnamefont {S.~L.}\ \bibnamefont {Sondhi}},\ }\bibfield  {title} {\bibinfo {title} {Superconductors are topologically ordered},\ }\href@noop {} {\bibfield  {journal} {\bibinfo  {journal} {Annals of Physics}\ }\textbf {\bibinfo {volume} {313}},\ \bibinfo {pages} {497} (\bibinfo {year} {2004})}\BibitemShut {NoStop}%
\bibitem [{foo()}]{footnoteZ2orderSC}%
  \BibitemOpen
  \href@noop {} {}\bibinfo {note} {Clearly, physical, two-dimensional superconductors and ${\mathbf Z}_2$ topologically ordered states occuring in 2+1 dimensional gauge theories with mixed Chern-Simons terms are not exactly the same. First, physical superconducting samples are coupled to electromagnetic fields which are free to explore three spatial dimensions. Second, superconductors sustain vortices of any winding, while ${\mathbf Z}_2$ topologically ordered states only knows about a single vison. The second difference is related to the non-compactness of physical QED.}\BibitemShut {Stop}%
\bibitem [{Sup()}]{SuppMat}%
  \BibitemOpen
  \href@noop {} {}\bibinfo {note} {See supplemental Materials.}\BibitemShut {Stop}%
\bibitem [{\citenamefont {Haldane}(1988)}]{Haldane1988}%
  \BibitemOpen
  \bibfield  {author} {\bibinfo {author} {\bibfnamefont {F.~D.~M.}\ \bibnamefont {Haldane}},\ }\bibfield  {title} {\bibinfo {title} {Model for a quantum {H}all effect without {L}andau levels: Condensed-matter realization of the ``parity anomaly"},\ }\href {https://doi.org/10.1103/PhysRevLett.61.2015} {\bibfield  {journal} {\bibinfo  {journal} {Phys. Rev. Lett.}\ }\textbf {\bibinfo {volume} {61}},\ \bibinfo {pages} {2015} (\bibinfo {year} {1988})}\BibitemShut {NoStop}%
\bibitem [{\citenamefont {Liang}\ \emph {et~al.}(2013)\citenamefont {Liang}, \citenamefont {He}, \citenamefont {Wu}, \citenamefont {Zhu},\ and\ \citenamefont {Kou}}]{LiangKou2013}%
  \BibitemOpen
  \bibfield  {author} {\bibinfo {author} {\bibfnamefont {Y.}~\bibnamefont {Liang}}, \bibinfo {author} {\bibfnamefont {J.}~\bibnamefont {He}}, \bibinfo {author} {\bibfnamefont {Y.-J.}\ \bibnamefont {Wu}}, \bibinfo {author} {\bibfnamefont {Y.-X.}\ \bibnamefont {Zhu}},\ and\ \bibinfo {author} {\bibfnamefont {S.-P.}\ \bibnamefont {Kou}},\ }\bibfield  {title} {\bibinfo {title} {Topological superconductors in correlated topological insulators on the honeycomb lattice},\ }\href {https://doi.org/10.1140/epjb/e2013-40521-5} {\bibfield  {journal} {\bibinfo  {journal} {The European Physical Journal B}\ }\textbf {\bibinfo {volume} {86}},\ \bibinfo {pages} {1} (\bibinfo {year} {2013})}\BibitemShut {NoStop}%
\bibitem [{\citenamefont {Wu}\ \emph {et~al.}(2015{\natexlab{b}})\citenamefont {Wu}, \citenamefont {Li},\ and\ \citenamefont {Kou}}]{WuKou2015}%
  \BibitemOpen
  \bibfield  {author} {\bibinfo {author} {\bibfnamefont {Y.-J.}\ \bibnamefont {Wu}}, \bibinfo {author} {\bibfnamefont {N.}~\bibnamefont {Li}},\ and\ \bibinfo {author} {\bibfnamefont {S.-P.}\ \bibnamefont {Kou}},\ }\bibfield  {title} {\bibinfo {title} {Chiral topological superfluids in the attractive {H}aldane-{H}ubbard model with opposite {Z}eeman energy at two sublattice sites},\ }\href {https://doi.org/10.1140/epjb/e2015-60412-y} {\bibfield  {journal} {\bibinfo  {journal} {The European Physical Journal B}\ }\textbf {\bibinfo {volume} {88}},\ \bibinfo {pages} {1} (\bibinfo {year} {2015}{\natexlab{b}})}\BibitemShut {NoStop}%
\bibitem [{\citenamefont {Zhang}\ \emph {et~al.}(2017)\citenamefont {Zhang}, \citenamefont {Xu},\ and\ \citenamefont {Zhang}}]{ZhangZhang2017}%
  \BibitemOpen
  \bibfield  {author} {\bibinfo {author} {\bibfnamefont {Y.-C.}\ \bibnamefont {Zhang}}, \bibinfo {author} {\bibfnamefont {Z.}~\bibnamefont {Xu}},\ and\ \bibinfo {author} {\bibfnamefont {S.}~\bibnamefont {Zhang}},\ }\bibfield  {title} {\bibinfo {title} {Topological superfluids and the {BEC}-{BCS} crossover in the attractive {H}aldane-{H}ubbard model},\ }\href {https://doi.org/10.1103/PhysRevA.95.043640} {\bibfield  {journal} {\bibinfo  {journal} {Phys. Rev. A}\ }\textbf {\bibinfo {volume} {95}},\ \bibinfo {pages} {043640} (\bibinfo {year} {2017})}\BibitemShut {NoStop}%
\bibitem [{\citenamefont {Florens}\ and\ \citenamefont {Georges}(2002)}]{FlorensGeorges1}%
  \BibitemOpen
  \bibfield  {author} {\bibinfo {author} {\bibfnamefont {S.}~\bibnamefont {Florens}}\ and\ \bibinfo {author} {\bibfnamefont {A.}~\bibnamefont {Georges}},\ }\bibfield  {title} {\bibinfo {title} {Quantum impurity solvers using a slave rotor representation},\ }\href {https://doi.org/10.1103/PhysRevB.66.165111} {\bibfield  {journal} {\bibinfo  {journal} {Phys. Rev. B}\ }\textbf {\bibinfo {volume} {66}},\ \bibinfo {pages} {165111} (\bibinfo {year} {2002})}\BibitemShut {NoStop}%
\bibitem [{\citenamefont {Florens}\ and\ \citenamefont {Georges}(2004)}]{FlorensGeorges2}%
  \BibitemOpen
  \bibfield  {author} {\bibinfo {author} {\bibfnamefont {S.}~\bibnamefont {Florens}}\ and\ \bibinfo {author} {\bibfnamefont {A.}~\bibnamefont {Georges}},\ }\bibfield  {title} {\bibinfo {title} {Slave-rotor mean-field theories of strongly correlated systems and the {M}ott transition in finite dimensions},\ }\href {https://doi.org/10.1103/PhysRevB.70.035114} {\bibfield  {journal} {\bibinfo  {journal} {Phys. Rev. B}\ }\textbf {\bibinfo {volume} {70}},\ \bibinfo {pages} {035114} (\bibinfo {year} {2004})}\BibitemShut {NoStop}%
\bibitem [{\citenamefont {Wagner}\ \emph {et~al.}(2024)\citenamefont {Wagner}, \citenamefont {Guerci}, \citenamefont {Millis},\ and\ \citenamefont {Sangiovanni}}]{WagnerSangiovanni2024}%
  \BibitemOpen
  \bibfield  {author} {\bibinfo {author} {\bibfnamefont {N.}~\bibnamefont {Wagner}}, \bibinfo {author} {\bibfnamefont {D.}~\bibnamefont {Guerci}}, \bibinfo {author} {\bibfnamefont {A.~J.}\ \bibnamefont {Millis}},\ and\ \bibinfo {author} {\bibfnamefont {G.}~\bibnamefont {Sangiovanni}},\ }\bibfield  {title} {\bibinfo {title} {Edge zeros and boundary spinons in topological {M}ott insulators},\ }\href {https://doi.org/10.1103/PhysRevLett.133.126504} {\bibfield  {journal} {\bibinfo  {journal} {Phys. Rev. Lett.}\ }\textbf {\bibinfo {volume} {133}},\ \bibinfo {pages} {126504} (\bibinfo {year} {2024})}\BibitemShut {NoStop}%
\bibitem [{\citenamefont {Rachel}(2013)}]{Rachel2013}%
  \BibitemOpen
  \bibfield  {author} {\bibinfo {author} {\bibfnamefont {S.}~\bibnamefont {Rachel}},\ }\bibfield  {title} {\bibinfo {title} {Quantum phase transitions of topological insulators without gap closing},\ }\href {https://api.semanticscholar.org/CorpusID:3766691} {\bibfield  {journal} {\bibinfo  {journal} {Journal of Physics: Condensed Matter}\ }\textbf {\bibinfo {volume} {28}} (\bibinfo {year} {2013})}\BibitemShut {NoStop}%
\bibitem [{\citenamefont {Wagner}\ \emph {et~al.}(2023)\citenamefont {Wagner}, \citenamefont {Crippa}, \citenamefont {Amaricci}, \citenamefont {Hansmann}, \citenamefont {Klett}, \citenamefont {K{\"o}nig}, \citenamefont {Sch{\"a}fer}, \citenamefont {Sante}, \citenamefont {Cano}, \citenamefont {Millis} \emph {et~al.}}]{WagnerSangiovanni2023}%
  \BibitemOpen
  \bibfield  {author} {\bibinfo {author} {\bibfnamefont {N.}~\bibnamefont {Wagner}}, \bibinfo {author} {\bibfnamefont {L.}~\bibnamefont {Crippa}}, \bibinfo {author} {\bibfnamefont {A.}~\bibnamefont {Amaricci}}, \bibinfo {author} {\bibfnamefont {P.}~\bibnamefont {Hansmann}}, \bibinfo {author} {\bibfnamefont {M.}~\bibnamefont {Klett}}, \bibinfo {author} {\bibfnamefont {E.}~\bibnamefont {K{\"o}nig}}, \bibinfo {author} {\bibfnamefont {T.}~\bibnamefont {Sch{\"a}fer}}, \bibinfo {author} {\bibfnamefont {D.~D.}\ \bibnamefont {Sante}}, \bibinfo {author} {\bibfnamefont {J.}~\bibnamefont {Cano}}, \bibinfo {author} {\bibfnamefont {A.}~\bibnamefont {Millis}}, \emph {et~al.},\ }\bibfield  {title} {\bibinfo {title} {Mott insulators with boundary zeros},\ }\href {https://doi.org/10.1038/s41467-023-42773-7} {\bibfield  {journal} {\bibinfo  {journal} {Nature Communications}\ }\textbf {\bibinfo {volume} {14}},\ \bibinfo {pages} {7531} (\bibinfo {year} {2023})}\BibitemShut {NoStop}%
\bibitem [{\citenamefont {Bollmann}\ \emph {et~al.}(2024{\natexlab{b}})\citenamefont {Bollmann}, \citenamefont {Setty}, \citenamefont {Seifert},\ and\ \citenamefont {K\"onig}}]{BollmannKoenig2024b}%
  \BibitemOpen
  \bibfield  {author} {\bibinfo {author} {\bibfnamefont {S.}~\bibnamefont {Bollmann}}, \bibinfo {author} {\bibfnamefont {C.}~\bibnamefont {Setty}}, \bibinfo {author} {\bibfnamefont {U.~F.~P.}\ \bibnamefont {Seifert}},\ and\ \bibinfo {author} {\bibfnamefont {E.~J.}\ \bibnamefont {K\"onig}},\ }\bibfield  {title} {\bibinfo {title} {Topological {G}reen's function zeros in an exactly solved model and beyond},\ }\href {https://doi.org/10.1103/PhysRevLett.133.136504} {\bibfield  {journal} {\bibinfo  {journal} {Phys. Rev. Lett.}\ }\textbf {\bibinfo {volume} {133}},\ \bibinfo {pages} {136504} (\bibinfo {year} {2024}{\natexlab{b}})}\BibitemShut {NoStop}%
\bibitem [{\citenamefont {Ioffe}\ and\ \citenamefont {Larkin}(1989)}]{IoffeLarkin1989}%
  \BibitemOpen
  \bibfield  {author} {\bibinfo {author} {\bibfnamefont {L.~B.}\ \bibnamefont {Ioffe}}\ and\ \bibinfo {author} {\bibfnamefont {A.~I.}\ \bibnamefont {Larkin}},\ }\bibfield  {title} {\bibinfo {title} {Gapless fermions and gauge fields in dielectrics},\ }\href {https://doi.org/10.1103/PhysRevB.39.8988} {\bibfield  {journal} {\bibinfo  {journal} {Phys. Rev. B}\ }\textbf {\bibinfo {volume} {39}},\ \bibinfo {pages} {8988} (\bibinfo {year} {1989})}\BibitemShut {NoStop}%
\bibitem [{\citenamefont {Emery}\ and\ \citenamefont {Kivelson}(1995)}]{EmeryKivelson1995}%
  \BibitemOpen
  \bibfield  {author} {\bibinfo {author} {\bibfnamefont {V.}~\bibnamefont {Emery}}\ and\ \bibinfo {author} {\bibfnamefont {S.}~\bibnamefont {Kivelson}},\ }\bibfield  {title} {\bibinfo {title} {Importance of phase fluctuations in superconductors with small superfluid density},\ }\href@noop {} {\bibfield  {journal} {\bibinfo  {journal} {Nature}\ }\textbf {\bibinfo {volume} {374}},\ \bibinfo {pages} {434} (\bibinfo {year} {1995})}\BibitemShut {NoStop}%
\bibitem [{\citenamefont {Chubukov}\ and\ \citenamefont {Schmalian}(2005)}]{ChubukovSchmalian2005}%
  \BibitemOpen
  \bibfield  {author} {\bibinfo {author} {\bibfnamefont {A.~V.}\ \bibnamefont {Chubukov}}\ and\ \bibinfo {author} {\bibfnamefont {J.}~\bibnamefont {Schmalian}},\ }\bibfield  {title} {\bibinfo {title} {Superconductivity due to massless boson exchange in the strong-coupling limit},\ }\href {https://doi.org/10.1103/PhysRevB.72.174520} {\bibfield  {journal} {\bibinfo  {journal} {Phys. Rev. B}\ }\textbf {\bibinfo {volume} {72}},\ \bibinfo {pages} {174520} (\bibinfo {year} {2005})}\BibitemShut {NoStop}%
\bibitem [{\citenamefont {Randeria}\ and\ \citenamefont {Taylor}(2014)}]{Randeria2014}%
  \BibitemOpen
  \bibfield  {author} {\bibinfo {author} {\bibfnamefont {M.}~\bibnamefont {Randeria}}\ and\ \bibinfo {author} {\bibfnamefont {E.}~\bibnamefont {Taylor}},\ }\bibfield  {title} {\bibinfo {title} {Crossover from bardeen-cooper-schrieffer to bose-einstein condensation and the unitary fermi gas},\ }\href@noop {} {\bibfield  {journal} {\bibinfo  {journal} {Annu. Rev. Condens. Matter Phys.}\ }\textbf {\bibinfo {volume} {5}},\ \bibinfo {pages} {209} (\bibinfo {year} {2014})}\BibitemShut {NoStop}%
\bibitem [{\citenamefont {Fradkin}\ and\ \citenamefont {Shenker}(1979)}]{FradkinShenker1979}%
  \BibitemOpen
  \bibfield  {author} {\bibinfo {author} {\bibfnamefont {E.}~\bibnamefont {Fradkin}}\ and\ \bibinfo {author} {\bibfnamefont {S.~H.}\ \bibnamefont {Shenker}},\ }\bibfield  {title} {\bibinfo {title} {Phase diagrams of lattice gauge theories with {H}iggs fields},\ }\href {https://doi.org/10.1103/PhysRevD.19.3682} {\bibfield  {journal} {\bibinfo  {journal} {Phys. Rev. D}\ }\textbf {\bibinfo {volume} {19}},\ \bibinfo {pages} {3682} (\bibinfo {year} {1979})}\BibitemShut {NoStop}%
\bibitem [{\citenamefont {Thorngren}\ \emph {et~al.}(2023)\citenamefont {Thorngren}, \citenamefont {Rakovszky}, \citenamefont {Verresen},\ and\ \citenamefont {Vishwanath}}]{ThorngrenVerresen2023}%
  \BibitemOpen
  \bibfield  {author} {\bibinfo {author} {\bibfnamefont {R.}~\bibnamefont {Thorngren}}, \bibinfo {author} {\bibfnamefont {T.}~\bibnamefont {Rakovszky}}, \bibinfo {author} {\bibfnamefont {R.}~\bibnamefont {Verresen}},\ and\ \bibinfo {author} {\bibfnamefont {A.}~\bibnamefont {Vishwanath}},\ }\bibfield  {title} {\bibinfo {title} {Higgs condensates are symmetry-protected topological phases: {II}. $ {U} (1) $ gauge theory and superconductors},\ }\href {https://doi.org/10.48550/arXiv.2303.08136} {\bibfield  {journal} {\bibinfo  {journal} {arXiv:2303.08136}\ } (\bibinfo {year} {2023})}\BibitemShut {NoStop}%
\bibitem [{\citenamefont {Tiwari}\ \emph {et~al.}(2024)\citenamefont {Tiwari}, \citenamefont {Bollmann}, \citenamefont {Paeckel},\ and\ \citenamefont {K{\"o}nig}}]{TiwariKoenig2024}%
  \BibitemOpen
  \bibfield  {author} {\bibinfo {author} {\bibfnamefont {D.}~\bibnamefont {Tiwari}}, \bibinfo {author} {\bibfnamefont {S.}~\bibnamefont {Bollmann}}, \bibinfo {author} {\bibfnamefont {S.}~\bibnamefont {Paeckel}},\ and\ \bibinfo {author} {\bibfnamefont {E.~J.}\ \bibnamefont {K{\"o}nig}},\ }\bibfield  {title} {\bibinfo {title} {Quantum restored symmetry protected topological phases},\ }\href {https://doi.org/10.48550/arXiv.2410.02689} {\bibfield  {journal} {\bibinfo  {journal} {arXiv:2410.02689}\ } (\bibinfo {year} {2024})}\BibitemShut {NoStop}%
\bibitem [{\citenamefont {Fisher}(1990)}]{Fisher1990}%
  \BibitemOpen
  \bibfield  {author} {\bibinfo {author} {\bibfnamefont {M.~P.~A.}\ \bibnamefont {Fisher}},\ }\bibfield  {title} {\bibinfo {title} {Quantum phase transitions in disordered two-dimensional superconductors},\ }\href {https://doi.org/10.1103/PhysRevLett.65.923} {\bibfield  {journal} {\bibinfo  {journal} {Phys. Rev. Lett.}\ }\textbf {\bibinfo {volume} {65}},\ \bibinfo {pages} {923} (\bibinfo {year} {1990})}\BibitemShut {NoStop}%
\bibitem [{\citenamefont {Senthil}(2015)}]{Senthil15}%
  \BibitemOpen
  \bibfield  {author} {\bibinfo {author} {\bibfnamefont {T.}~\bibnamefont {Senthil}},\ }\bibfield  {title} {\bibinfo {title} {Symmetry-protected topological phases of quantum matter},\ }\href {https://doi.org/https://doi.org/10.1146/annurev-conmatphys-031214-014740} {\bibfield  {journal} {\bibinfo  {journal} {Annual Review of Condensed Matter Physics}\ }\textbf {\bibinfo {volume} {6}},\ \bibinfo {pages} {299} (\bibinfo {year} {2015})}\BibitemShut {NoStop}%
\bibitem [{\citenamefont {Wang}\ \emph {et~al.}(2017)\citenamefont {Wang}, \citenamefont {Nahum}, \citenamefont {Metlitski}, \citenamefont {Xu},\ and\ \citenamefont {Senthil}}]{WangSenthil17}%
  \BibitemOpen
  \bibfield  {author} {\bibinfo {author} {\bibfnamefont {C.}~\bibnamefont {Wang}}, \bibinfo {author} {\bibfnamefont {A.}~\bibnamefont {Nahum}}, \bibinfo {author} {\bibfnamefont {M.~A.}\ \bibnamefont {Metlitski}}, \bibinfo {author} {\bibfnamefont {C.}~\bibnamefont {Xu}},\ and\ \bibinfo {author} {\bibfnamefont {T.}~\bibnamefont {Senthil}},\ }\bibfield  {title} {\bibinfo {title} {Deconfined quantum critical points: Symmetries and dualities},\ }\href {https://doi.org/10.1103/PhysRevX.7.031051} {\bibfield  {journal} {\bibinfo  {journal} {Phys. Rev. X}\ }\textbf {\bibinfo {volume} {7}},\ \bibinfo {pages} {031051} (\bibinfo {year} {2017})}\BibitemShut {NoStop}%
\bibitem [{\citenamefont {Wang}\ and\ \citenamefont {Senthil}(2016)}]{WangSenthil16}%
  \BibitemOpen
  \bibfield  {author} {\bibinfo {author} {\bibfnamefont {C.}~\bibnamefont {Wang}}\ and\ \bibinfo {author} {\bibfnamefont {T.}~\bibnamefont {Senthil}},\ }\bibfield  {title} {\bibinfo {title} {Time-reversal symmetric $u(1)$ quantum spin liquids},\ }\href {https://doi.org/10.1103/PhysRevX.6.011034} {\bibfield  {journal} {\bibinfo  {journal} {Phys. Rev. X}\ }\textbf {\bibinfo {volume} {6}},\ \bibinfo {pages} {011034} (\bibinfo {year} {2016})}\BibitemShut {NoStop}%
\bibitem [{\citenamefont {Manjunath}\ and\ \citenamefont {Barkeshli}(2021)}]{ManjunathBarkeshli21}%
  \BibitemOpen
  \bibfield  {author} {\bibinfo {author} {\bibfnamefont {N.}~\bibnamefont {Manjunath}}\ and\ \bibinfo {author} {\bibfnamefont {M.}~\bibnamefont {Barkeshli}},\ }\bibfield  {title} {\bibinfo {title} {Crystalline gauge fields and quantized discrete geometric response for abelian topological phases with lattice symmetry},\ }\href {https://doi.org/10.1103/PhysRevResearch.3.013040} {\bibfield  {journal} {\bibinfo  {journal} {Phys. Rev. Res.}\ }\textbf {\bibinfo {volume} {3}},\ \bibinfo {pages} {013040} (\bibinfo {year} {2021})}\BibitemShut {NoStop}%
\bibitem [{\citenamefont {Zhang}\ \emph {et~al.}(2023)\citenamefont {Zhang}, \citenamefont {Manjunath}, \citenamefont {Nambiar},\ and\ \citenamefont {Barkeshli}}]{ZhangBarkeshli23}%
  \BibitemOpen
  \bibfield  {author} {\bibinfo {author} {\bibfnamefont {Y.}~\bibnamefont {Zhang}}, \bibinfo {author} {\bibfnamefont {N.}~\bibnamefont {Manjunath}}, \bibinfo {author} {\bibfnamefont {G.}~\bibnamefont {Nambiar}},\ and\ \bibinfo {author} {\bibfnamefont {M.}~\bibnamefont {Barkeshli}},\ }\bibfield  {title} {\bibinfo {title} {Quantized charge polarization as a many-body invariant in $(2+1)\mathrm{D}$ crystalline topological states and hofstadter butterflies},\ }\href {https://doi.org/10.1103/PhysRevX.13.031005} {\bibfield  {journal} {\bibinfo  {journal} {Phys. Rev. X}\ }\textbf {\bibinfo {volume} {13}},\ \bibinfo {pages} {031005} (\bibinfo {year} {2023})}\BibitemShut {NoStop}%
\bibitem [{\citenamefont {Morissette}\ \emph {et~al.}(2025)\citenamefont {Morissette}, \citenamefont {Qin}, \citenamefont {Wu}, \citenamefont {Watanabe}, \citenamefont {Taniguchi},\ and\ \citenamefont {Li}}]{MorissetteLi2025}%
  \BibitemOpen
  \bibfield  {author} {\bibinfo {author} {\bibfnamefont {E.}~\bibnamefont {Morissette}}, \bibinfo {author} {\bibfnamefont {P.}~\bibnamefont {Qin}}, \bibinfo {author} {\bibfnamefont {H.}~\bibnamefont {Wu}}, \bibinfo {author} {\bibfnamefont {K.}~\bibnamefont {Watanabe}}, \bibinfo {author} {\bibfnamefont {T.}~\bibnamefont {Taniguchi}},\ and\ \bibinfo {author} {\bibfnamefont {J.}~\bibnamefont {Li}},\ }\bibfield  {title} {\bibinfo {title} {Superconductivity, anomalous hall effect, and stripe order in rhombohedral hexalayer graphene},\ }\href {https://arxiv.org/abs/2504.05129} {\bibfield  {journal} {\bibinfo  {journal} {arXiv:2504.05129}\ } (\bibinfo {year} {2025})}\BibitemShut {NoStop}%
\bibitem [{\citenamefont {Xu}\ \emph {et~al.}(2025)\citenamefont {Xu}, \citenamefont {Sun}, \citenamefont {Li}, \citenamefont {Zheng}, \citenamefont {Xu}, \citenamefont {Gao}, \citenamefont {Jia}, \citenamefont {Watanabe}, \citenamefont {Taniguchi}, \citenamefont {Tong}, \citenamefont {Lu}, \citenamefont {Jia}, \citenamefont {Shi}, \citenamefont {Jiang}, \citenamefont {Zhang}, \citenamefont {Zhang}, \citenamefont {Lei}, \citenamefont {Liu},\ and\ \citenamefont {Li}}]{XuLi2025}%
  \BibitemOpen
  \bibfield  {author} {\bibinfo {author} {\bibfnamefont {F.}~\bibnamefont {Xu}}, \bibinfo {author} {\bibfnamefont {Z.}~\bibnamefont {Sun}}, \bibinfo {author} {\bibfnamefont {J.}~\bibnamefont {Li}}, \bibinfo {author} {\bibfnamefont {C.}~\bibnamefont {Zheng}}, \bibinfo {author} {\bibfnamefont {C.}~\bibnamefont {Xu}}, \bibinfo {author} {\bibfnamefont {J.}~\bibnamefont {Gao}}, \bibinfo {author} {\bibfnamefont {T.}~\bibnamefont {Jia}}, \bibinfo {author} {\bibfnamefont {K.}~\bibnamefont {Watanabe}}, \bibinfo {author} {\bibfnamefont {T.}~\bibnamefont {Taniguchi}}, \bibinfo {author} {\bibfnamefont {B.}~\bibnamefont {Tong}}, \bibinfo {author} {\bibfnamefont {L.}~\bibnamefont {Lu}}, \bibinfo {author} {\bibfnamefont {J.}~\bibnamefont {Jia}}, \bibinfo {author} {\bibfnamefont {Z.}~\bibnamefont {Shi}}, \bibinfo {author} {\bibfnamefont {S.}~\bibnamefont {Jiang}}, \bibinfo {author} {\bibfnamefont {Y.}~\bibnamefont {Zhang}}, \bibinfo {author} {\bibfnamefont {Y.}~\bibnamefont {Zhang}}, \bibinfo {author} {\bibfnamefont
  {S.}~\bibnamefont {Lei}}, \bibinfo {author} {\bibfnamefont {X.}~\bibnamefont {Liu}},\ and\ \bibinfo {author} {\bibfnamefont {T.}~\bibnamefont {Li}},\ }\bibfield  {title} {\bibinfo {title} {Signatures of unconventional superconductivity near reentrant and fractional quantum anomalous hall insulators},\ }\href {https://arxiv.org/abs/2504.06972} {\bibfield  {journal} {\bibinfo  {journal} {2504.06972}\ } (\bibinfo {year} {2025})}\BibitemShut {NoStop}%
\bibitem [{\citenamefont {Sachdev}(1999)}]{SachdevBook}%
  \BibitemOpen
  \bibfield  {author} {\bibinfo {author} {\bibfnamefont {S.}~\bibnamefont {Sachdev}},\ }\bibfield  {title} {\bibinfo {title} {Quantum phase transitions},\ }\href@noop {} {\bibfield  {journal} {\bibinfo  {journal} {Physics world}\ }\textbf {\bibinfo {volume} {12}},\ \bibinfo {pages} {33} (\bibinfo {year} {1999})}\BibitemShut {NoStop}%
\bibitem [{\citenamefont {Efetov}(1999)}]{EfetovBook}%
  \BibitemOpen
  \bibfield  {author} {\bibinfo {author} {\bibfnamefont {K.}~\bibnamefont {Efetov}},\ }\href@noop {} {\emph {\bibinfo {title} {Supersymmetry in disorder and chaos}}}\ (\bibinfo  {publisher} {Cambridge university press},\ \bibinfo {year} {1999})\BibitemShut {NoStop}%
\bibitem [{\citenamefont {Tong}(2018)}]{Tong2018}%
  \BibitemOpen
  \bibfield  {author} {\bibinfo {author} {\bibfnamefont {D.}~\bibnamefont {Tong}},\ }\href {http://www.damtp.cam.ac.uk/user/tong/gaugetheory.html} {\bibinfo {title} {Lectures on gauge theory}} (\bibinfo {year} {2018})\BibitemShut {NoStop}%
\bibitem [{\citenamefont {Kaneko}\ \emph {et~al.}(2016)\citenamefont {Kaneko}, \citenamefont {Tocchio}, \citenamefont {Valent\'{\i}}, \citenamefont {Becca},\ and\ \citenamefont {Gros}}]{KanekoGros16}%
  \BibitemOpen
  \bibfield  {author} {\bibinfo {author} {\bibfnamefont {R.}~\bibnamefont {Kaneko}}, \bibinfo {author} {\bibfnamefont {L.~F.}\ \bibnamefont {Tocchio}}, \bibinfo {author} {\bibfnamefont {R.}~\bibnamefont {Valent\'{\i}}}, \bibinfo {author} {\bibfnamefont {F.}~\bibnamefont {Becca}},\ and\ \bibinfo {author} {\bibfnamefont {C.}~\bibnamefont {Gros}},\ }\bibfield  {title} {\bibinfo {title} {Spontaneous symmetry breaking in correlated wave functions},\ }\href {https://doi.org/10.1103/PhysRevB.93.125127} {\bibfield  {journal} {\bibinfo  {journal} {Phys. Rev. B}\ }\textbf {\bibinfo {volume} {93}},\ \bibinfo {pages} {125127} (\bibinfo {year} {2016})}\BibitemShut {NoStop}%
\bibitem [{\citenamefont {Wu}\ \emph {et~al.}(2019)\citenamefont {Wu}, \citenamefont {Gong},\ and\ \citenamefont {Sheng}}]{WuShen19}%
  \BibitemOpen
  \bibfield  {author} {\bibinfo {author} {\bibfnamefont {H.-Q.}\ \bibnamefont {Wu}}, \bibinfo {author} {\bibfnamefont {S.-S.}\ \bibnamefont {Gong}},\ and\ \bibinfo {author} {\bibfnamefont {D.~N.}\ \bibnamefont {Sheng}},\ }\bibfield  {title} {\bibinfo {title} {Randomness-induced spin-liquid-like phase in the spin-$\frac{1}{2}$ ${J}_{1}\ensuremath{-}{J}_{2}$ triangular heisenberg model},\ }\href {https://doi.org/10.1103/PhysRevB.99.085141} {\bibfield  {journal} {\bibinfo  {journal} {Phys. Rev. B}\ }\textbf {\bibinfo {volume} {99}},\ \bibinfo {pages} {085141} (\bibinfo {year} {2019})}\BibitemShut {NoStop}%
\bibitem [{\citenamefont {Neuberger}\ and\ \citenamefont {Ziman}(1989)}]{NeubergerZiman89}%
  \BibitemOpen
  \bibfield  {author} {\bibinfo {author} {\bibfnamefont {H.}~\bibnamefont {Neuberger}}\ and\ \bibinfo {author} {\bibfnamefont {T.}~\bibnamefont {Ziman}},\ }\bibfield  {title} {\bibinfo {title} {Finite-size effects in heisenberg antiferromagnets},\ }\href {https://doi.org/10.1103/PhysRevB.39.2608} {\bibfield  {journal} {\bibinfo  {journal} {Phys. Rev. B}\ }\textbf {\bibinfo {volume} {39}},\ \bibinfo {pages} {2608} (\bibinfo {year} {1989})}\BibitemShut {NoStop}%
\end{thebibliography}%

\clearpage

\setcounter{equation}{0}
\setcounter{figure}{0}
\setcounter{section}{0}
\setcounter{table}{0}
\setcounter{page}{1}
\makeatletter
\renewcommand{\theequation}{S\arabic{equation}}
\renewcommand{\thesection}{S\arabic{section}}
\renewcommand{\thefigure}{S\arabic{figure}}
\renewcommand{\thepage}{S\arabic{page}}

\begin{widetext}
\begin{center}
Supplementary materials on \\
\textbf{"Topologically enabled superconductivity: possible implications for rhombohedral graphene"}\\
{Francesca Paoletti$^1$,} 
{Daniele Guerci$^2$,} 
{Giorgio Sangiovanni$^1$}, 
{Urban~F.P.~Seifert$^3$},
{Elio J.~K\"onig$^4$}\\
{%
$^1${Institut f\"ur Theoretische Physik und Astrophysik and W\"urzburg-Dresden Cluster of Excellence ct.qmat, Universit\"at W\"urzburg, 97074 W\"urzburg, Germany}\\%
$^2${Department of Physics, Massachusetts Institute of Technology, Cambridge, Massachusetts 02139, USA}\\%
$^3${Institut f\"ur Theoretische Physik, Universit\"at zu K\"oln, Z\"ulpicher Str. 77a, 50937 K\"oln, Germany}\\%
$^4$ Department of Physics, University of Wisconsin-Madison, Madison, Wisconsin 53706, USA
}\\%
\end{center}
\end{widetext}

This supplement contains details on the weak-coupling Hartree-Fock solution (Sec.~\ref{sec:WeakCoupling}), the classical solution of the spin model of the main text (Sec.~\ref{sec:ClassicalSpin}), details about the slave-rotor treatment, (Sec.~\ref{sec:SB}), about the effective field theory (Sec.~\ref{sec:QFT}), and about the anti-Gutzwiller projection (Sec.~\ref{sec:gutzwiller-appendix}).

\section{Weak coupling Mean-Field theory}

\label{sec:WeakCoupling}

{In this part of the supplement, we outline the Hartree-Fock solution of Eq. \eqref{Eq:Hamiltonian} from the main text. The inability of this approach to detect the superconducting phase serves as motivation to employ more advanced solution techniques.} 

We apply the Hubbard-Stratonovich transformation to decouple the quartic fermionic interaction in Eq. \eqref{Eq:interaction}. This replaces it with a quadratic interaction involving auxiliary bosonic fields, which capture the fluctuations of an emergent order parameter. In our case the charge and the pairing channels are considered.

We begin by decoupling the charge terms
\begin{equation}
\begin{split}
    -U \sum_\v{x} n_{\v{x},1} n_{\v{x},-1} \rightarrow -U \sum_{\v{x}} & [ \langle n_{\v{x},1} \rangle n_{\v{x},-1} + n_{\v{x},1}  \langle  n_{\v{x},-1} \rangle +
    \\& -\langle n_{\v{x},1} \rangle \langle n_{\v{x},-1}\rangle].
\end{split}
\end{equation}

Since our lattice is bipartite and we are at half-filling, we consider the charge modulation $\rho = (-1)^{\v x} U \langle n_{\v x, \alpha} \rangle$ with opposite sign on $A$ and $B$ sublattice. 
After Fourier transforming in k-space, we end up with the charge density wave contribution
\begin{equation}
    H_{\text{CDW}} = - \sum_{\mathbf{k}} \rho (n_{\mathbf{k}, A,1} - n_{\mathbf{k}, B,1} +n_{\mathbf{k}, A,-1} - n_{\mathbf{k}, B, -1}) + N \rho^2/U,
\end{equation}
which simply modifies the diagonal elements of the free model in Eq. \eqref{Eq:free_model}.
Here $N$ is the total number of sites and we introduced the index $A,B$ to distinguish different sublattices.

To study superconductivity, we focus on the Cooper pair formation, which involves electron pairs of opposite {color}: 

\begin{equation}
\begin{split}
    -U \sum_\v{x} n_{\v{x},1} & n_{\v{x},-1} \rightarrow -U 
    \sum_\v{x} [ \langle c^\dagger_{\v{x},1} c^\dagger_{\v{x},-1} \rangle c_{\v{x},-1} c_{\v{x},1} + \\ & + c^\dagger_{\v{x},1} c^\dagger_{\v{x},-1}\langle c_{\v{x},-1} c_{\v{x},1} \rangle  - \langle c^\dagger_{\v{x},1} c^\dagger_{\v{x},-1} \rangle \langle c_{\v{x},-1} c_{\v{x},1} \rangle ].
\end{split}
\end{equation}

After introducing the parameters $\Delta = U\langle c_{\v{x},-1} c_{\v{x},1} \rangle $ and $\bar{\Delta} = U\langle c^\dagger_{\v{x},1} c^\dagger_{\v{x},-1} \rangle$, and performing the Fourier transformation, 
we obtain
\begin{equation}
\begin{split}
    H_{\text{SC}} =& - \sum_{\mathbf{k}} [\Delta (c^\dagger_{\mathbf{k}, A, 1} c^\dagger_{-\mathbf{k}, A, -1} + c^\dagger_{\mathbf{k}, B, 1} c^\dagger_{-\mathbf{k}, B, -1}) + \\ & \qquad\bar{\Delta} (c_{-\mathbf{k}, A, -1} c_{\mathbf{k}, A, 1} + c_{-\mathbf{k}, B, -1} c_{\mathbf{k}, B, 1})] + N  \vert \Delta \vert^2/U.
\end{split}
\end{equation}

The Bogoliubov-deGennes Hamiltonian in the Nambu spinor basis
$\psi_\mathbf{k} = \{ c_{\mathbf{k}, A, 1} \  c_{\mathbf{k}, B, 1} \ c^\dagger_{-\mathbf{k}, A, -1} \  c^\dagger_{-\mathbf{k}, B, -1} \}$ finally is presented in Eq.~\eqref{eq:BdG} of the main text and we henceforth use 
$\tilde{w}_0 = w_0 -\rho$. The lower diagonal block is related to the upper one by particle-hole transformation $-h^*(-\v k)$, and we have $m(-\v{k}) = -m(\v{k})$ and $s(-\v{k})= s^*(\v{k})$.

At the lowest order, i.e.~in the mean-field approximation, the bosonic fields $\rho$ and $\Delta$ are treated as time-independent and spatially uniform quantities, and we deal with a simple single-particle problem. {For simplicity we also consider the superconductive gap $\Delta$ as a real number}. At zero temperature the free energy density is:
\begin{equation}
    F[\rho,\Delta] = \frac{\rho^2}{U} + \frac{\Delta^2}{U}  + \frac{1}{N} \sum_{\mathbf{k},l=\pm} \lambda_l (\mathbf{k}; \rho,\Delta)
\end{equation} 
with 
\begin{equation}
\begin{split}
\lambda_\pm = - & \biggl [ w_2^2 m^2(\mathbf{k}) + w_1^2 |s(\mathbf{k})|^2 + \tilde{w}_0^2 + \Delta^2 \\ & \pm 2 \sqrt{w_2^2 m^2(\mathbf{k}) (\tilde{w}_0^2 + \Delta^2)} \biggr ] ^{1/2} 
\end{split}
\end{equation} 
$ \left. \frac{\partial F}{\partial \rho} \right|_{\rho,\Delta} = 0 \ \text{and} \ \left. \frac{\partial F}{\partial \Delta}\right|_{\rho,\Delta} = 0 $ 
, we obtain the set of conditions
\begin{subequations}
\begin{align}
    \Delta \left( \frac{1}{U} + g(\rho, \Delta)\right) &=0. \label{eq:mf_delta}\\
    \frac{ \rho}{U} +  (\rho - w_0) g(\rho, \Delta) &=0, \label{eq:mf_rho}
\end{align}
\end{subequations}
with 
\begin{equation} 
    g(\rho, \Delta) = \frac{1}{N} \sum_{\mathbf{k}, l=\pm} \frac{1 +l \frac{w_2^2 m^2(\mathbf{k})}{\sqrt{w_2^2 m^2(\mathbf{k}) (\tilde{w}_0^2 + |\Delta|^2)}}}{2 \lambda_l (\mathbf{k}; \rho,\Delta)}.
\end{equation}

If we seek a finite solution for $\Delta$, then from Eq. \eqref{eq:mf_delta}, we obtain $g(\rho, \Delta) = -U^{-1}$. Substituting this into Eq. \eqref{eq:mf_rho} leads to $\frac{\rho}{U} + \frac{(w_0 - \rho)}{U} = 0$. Thus, we conclude that a superconducting solution is not possible at the mean-field level for small $U$. {We highlight that this contradicts the claim in Ref. \cite{LiangKou2013}, where the charge density wave channel is overlooked.}

\section{Large-S theory for the large-$U$ limit}
\label{sec:ClassicalSpin}

In this section we present details about the classical solution for the effective spin model, Eq.~\eqref{eq:h-large-U} of the main text.

To get a feeling about the strong coupling phases, we project to the large $U$ subspace corresponding to empty/doubly occupied states of the attractive QAH Hubbard model.
Simple superexchange leads to Eq.~\eqref{eq:h-large-U} of the main text where, $\vec T = \vec \sigma/2$ are spin-1/2 operators. 

Technically taking large $S$ limits (by first replacing $w_0 \rightarrow w_0 S$) we estimate entirely classically the energies of this spin-model ($\varepsilon$ are energies normalized by number of unit cells and spin length $S^2$)

\begin{figure}
\centering\includegraphics[width = .45\textwidth]{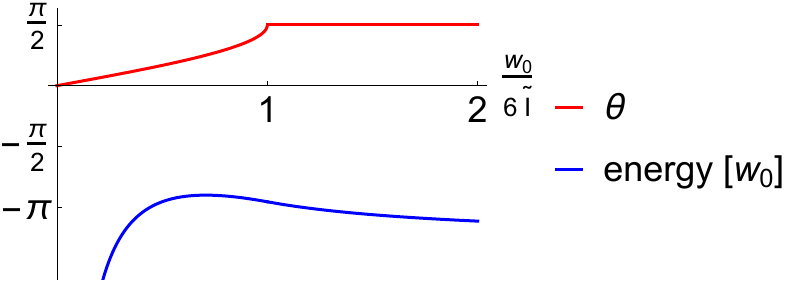}
    \caption{Classical (large-$S$) solution for the spin model, Eq.~\eqref{eq:h-large-U} using the parametrization outlined in Sec.~\ref{sec:ClassicalSpin}. For $w_0 < 6 \tilde I_0$ the model displays in-plane antiferromagnetic order corresponding to $0<\theta< \pi/2$.}
    \label{fig:ClassicSpin}
\end{figure}

\begin{itemize}
    \item A slanted antiferromagnet (AFM) $\vec S_{\v x} = (-1)^{\v x} S (\cos(\theta), 0, \sin (\theta))^T$: 
    \begin{equation}
        \varepsilon_{\rm AFM} (\theta)  = -4w_0 \sin(\theta) - 3 I - 6 \tilde I \cos (2\theta).
    \end{equation}
    \item A slanted ferromagnet (FM) $\vec S_{\v x} = S (\cos(\theta), 0, \sin (\theta))^T$
     \begin{equation}
        \varepsilon_{\rm FM} (\theta)  = 3 I - 6 \tilde I \cos (2\theta).
    \end{equation}
\end{itemize}
We optimize the corresponding energies leading to the condition
\begin{itemize}
    \item slanted AFM: 
    \begin{align}
        \frac{\sin(2\theta)}{2\cos(\theta)} & = \frac{w_0}{6\tilde I}  
    \end{align}
    so we find 
    \begin{align}
        \theta = \left . \arctan \left ( \frac{- \sqrt{1-{y}^2}}{{y}} \right) \right \vert_{{y} = \frac{w_0}{6\tilde I}} + 
        \frac{\pi}{2}
    \end{align}
    for ${w_0}<{6\tilde I}$ and $\theta = \pi/2$ otherwise. So the energy is (using some trigonometric identities)
    \begin{align}
        \varepsilon_{\rm AFM} & = - \frac{2 w_0^2}{3 \tilde I} - 3 I - 6 \tilde I \left (1 - 2\frac{w_0^2}{36 \tilde I^2} \right ) \\
        &= -3I - 6 \tilde I  - \frac{ w_0^2}{3 \tilde I} \text{ for } w_0< 6 \tilde I\\
        \varepsilon_{\rm AFM} & = - 4 w_0 - 3I + 6 \tilde I\text{ for } w_0> 6 \tilde I
    \end{align}
    \item slanted FM: The calculations are similar. The associated energies are
    \begin{align}
        \varepsilon_{\rm FM} & =  + 3 I - 6 \tilde I 
    \end{align}
    The FM state is never favorable. 
\end{itemize}
In summary, we find the boundary for spontaneous XY breaking at $w_0 = 6 \tilde I$, see Fig.~\ref{fig:ClassicSpin}, where we plot the classical ground state energy (shifted by $3I$) along with the angle $\theta$.

\begin{figure}
\centering\includegraphics[width = .45\textwidth]{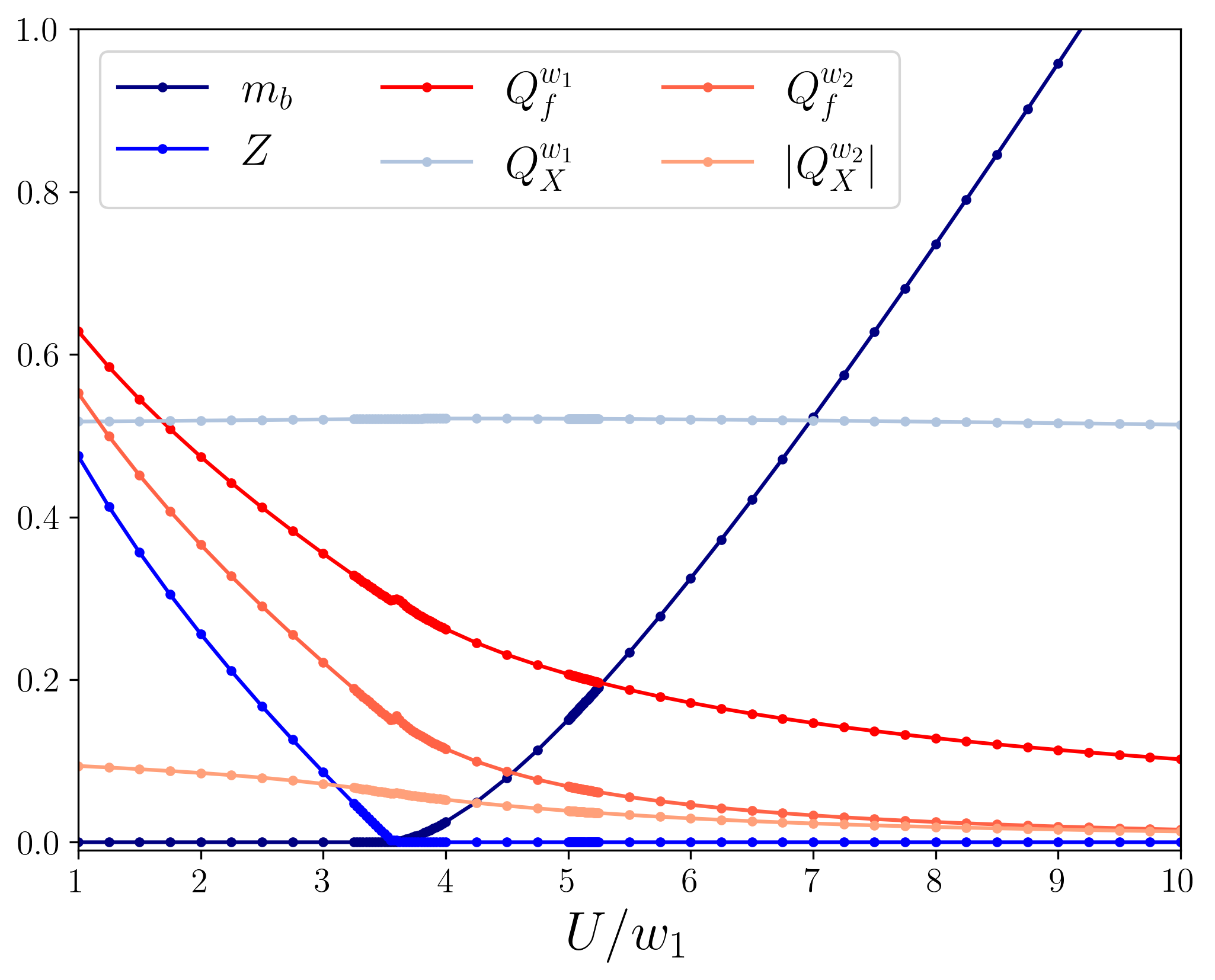}
    \caption{Slave-rotor mean field result for the renormalization of various quantities.   Here, $w_0 = 0.03090 w_1, w_2 = 0.09511 w_1$.}
    \label{fig:PlotSimilNik}
\end{figure}

\section{Slave rotor mean-field theory}
\label{sec:SB}

In this section we present details about the evaluation of the slave-rotor mean-field Hamiltonian, Eq.~\eqref{eq:SBMeanfield} of the main text, supplemented with the Lagrange multiplier
\begin{equation}
    \delta H = -\sum_{\v x} \lambda_{\v x} [\hat L_{\v x} + \sum_{\alpha} \psi^\dagger_{\v x, \alpha} \psi_{\v x, \alpha} -1].
\end{equation}

\subsection{Numerical details}

Most of our numerical data is obtained using a generalization of the code implemented in Ref.~\cite{WagnerSangiovanni2024} and we refer to the corresponding Supplemental material for additional details. 

{To map out the phase diagram in Fig. \ref{fig:SummaryFig}, we numerically solve the model Hamiltonian in Eq. \eqref{Eq:Hamiltonian} using the slave rotors technique while varying the model parameters. Specifically, we construct a uniform grid of $\varphi$ values ranging from 0 to $\pi/2$, obtaining the corresponding \((w_0, w_2)\) pairs for the free model. For each of these sets, we run the simulation by varying the Hubbard interaction \(U\) from 0 to 10, initially sampling uniformly and then refining the sampling near the two transition lines.  

An example of our self-consistent calculation for \(w_0 = 0.03090 w_1\) and \(w_2 = 0.09511 w_1\) is presented in Fig. \ref{fig:PlotSimilNik}, illustrating the evolution of the converged parameters of the effective rotor and spinon Hamiltonians, Eq.~\eqref{eq:SBMeanfield},
as \(U\) increases. 
Here, we use the notation $Q_X^{w_1} = J/w_1, Q_f^{w_i} = t_1/w_1$ and analogously for next nearest hopping (denoted by superscript $w_2$.
We also show the quasiparticle weight \(Z\), which decreases to zero at the critical interaction \(U_c\), signaling a Mott-like transition. Correspondingly, the rotor gap, or bosonic mass \(m_b\), vanishes, indicating the suppression of charge fluctuations.  

Notably, the rotor gap increases linearly with $(U - U_c)$ after the ``Mott'' transition. Within large-$M$ theory, \(m_b\) grows more slowly than the square-root behavior predicted by a naive mean-field approach. In our formulation, \(m_b = \rho + \min(\epsilon_X)\), where \(\epsilon_X\) represents the rotor dispersion and $\rho$ is the Lagrange multiplier enforcing the constraint \(|X_\mathbf{x}(\tau)|^2 = 1\), with \(X_\mathbf{x} = e^{i\theta_\mathbf{x}}\). Meanwhile, the Lagrange parameter $\lambda$, associated with the angular momentum constraint, is found to be zero, a condition that is first verified and then enforced in all calculations to ensure faster convergence.
}

{The topological transition occurs at one of the $K$ points when $\sqrt{27} t_2$ drops below $w_0$~\cite{Rachel2013}.}

\subsection{Approximate solution.}

\begin{figure}
\centering\includegraphics[width = .45\textwidth]{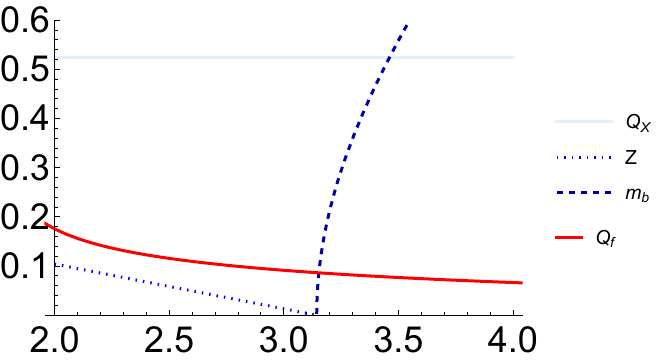}
    \caption{Results of approximate slave rotor mean-field theory using single-site-U(1) mean-field theory. The position of mean-field transition and $Q_X$, and in part $Q_f$ are similar to the one obtained by the more sophisticated large $M$ solution, Fig.~\ref{fig:PlotSimilNik}, yet $Z$ and $m_b$ behave only qualitatively similarly.}
    \label{fig:ApproxMF}
\end{figure}

\subsubsection{Summary of mean-field equations.}

As a lightweight method which allows to incorporate magnetic fields (see Sec.~\ref{sec:BField} below) we here present an alternative approach. First, note that the slave rotor part of the Hamiltonian, Eq.~\eqref{eq:JJArray}, effectively analogous to a  Josephson-junction array, can be solved by mean-field decoupling of the kinetic term~\cite{SachdevBook} using a field $\Psi$ leading to
\begin{equation}
    H_{\theta, \text{MF}}  = \sum_{\v x} \left [\frac{U}{2} \hat L_{\v x^2} - \lambda \hat L_{\v x} - (\Psi e^{- i \theta_{\v x}} + H.c.) \right]. \label{eq:MFHam}
\end{equation}

It is worthwhile to emphasize that this is the main distinction as compared to the numerical large-$M$ solution employed above and in Ref.~\cite{WagnerSangiovanni2024}. 
We use the notation $Q_f \equiv  \frac{t_1}{w_1}, Q_X \equiv  \frac{J}{w_1}$. Then, the mean field equations become

\begin{subequations}
\begin{align}
   Q_f  & = \frac{1}{6 \mathcal N} \sum_{\langle \v x, \v x' \rangle} \langle e^{i \theta_{\v x} - i \theta_{\v x'}} + H.c. \rangle_{Q_f, Q_X, \Psi, \lambda}, \\
  Q_X  & = \frac{1}{6 \mathcal N} \sum_{\langle \v x, \v x' \rangle, \alpha} \langle \psi_{\v x, \alpha}^\dagger \psi_{\v x', \alpha}+ H.c. \rangle_{Q_f, Q_X, \Psi, \lambda}, \label{eq:QXEq} \\
    \Psi & = 3 J \langle e^{i \theta_{\v x}} \rangle_{Q_f, Q_X, \Psi, \lambda} , \label{eq:PsiEq}\\
    1 & = \langle \hat L_{\v x} \rangle_{Q_f, Q_X, \Psi, \lambda} + \sum_{\alpha} \langle \psi^\dagger_{\v x,\alpha} \psi_{\v x, \alpha} \rangle _{Q_f, Q_X, \Psi, \lambda}. \label{eq:lambdaEq}
\end{align} 
\label{eq:MFSpelledOut}
\end{subequations}
Here, $\mathcal N$ is the number of unit cells in the system. To be precise, there are two more equations for next-nearest neighbor terms, but we henceforth assume $w_2 \ll w_1$ in which case their effect can be neglected.  

The notation $\langle \dots \rangle_{Q_f, Q_X, \Psi, \lambda}$ not only indicates that the quantum mechanical average ought to be taken self-consistently with a mean-field wave function which depends on the four parameters ${Q_f, Q_X, \Psi, \lambda}$. It also indicates that the right hand side of Eqs.~\eqref{eq:SBMeanfield} is in general a complicated function of these four parameters which requires a numerical approach. 

\subsubsection{Main simplification}

In the presence of particle-hole symmetry, as given in the problem at hand, Eq.~\eqref{Eq:Hamiltonian} of the main text, it is apriori clear that $\lambda = 0$ is consistent with Eq.~\eqref{eq:lambdaEq}. We will check this at the very end of the calculation, but for now simply state that the number of undetermined mean-field parameters is reduced to three. 

Next, the major simplification occurs: It turns out that (within our level of approximation keeping only zeroth order in $w_2/w_1, w_0/w_1$) the right-hand side of Eq.~\eqref{eq:QXEq} is a number which is itself $Q_X,Q_f,\Psi$ independent. Indeed

\begin{align}
    Q_X &= - \frac{1}{6 t_1} \sum_{\alpha} E_{\rm Fermi}^{(\alpha)}  = \frac{1}{3} \int_{\v k} \vert s(\v k) \vert  \approx 0.524. \label{eq:QX}
\end{align}
Here, $E_{\rm Fermi}^{(\alpha)}$ is the energy per unit cell of the filled state of $\psi_\alpha$ state for the free fermion Hamiltonian Eq.~\eqref{eq:PsiHam}. Recall that $\int_{\v k}$ is the integral over the Brillouin zone normalized such that $\int_{\v k} = \int \frac{d^2 \v k}{\text{Vol(BZ)}} = 1$.

This reduces the number of undetermined mean-field parameters to two.

\subsubsection{Superfluid-Mott transition in the effective Josephson junction array}

Next, we solve Eq.~\eqref{eq:PsiEq}. Note that the mean-field wave function is given by a direct product $\ket{\{Q_f,\lambda \}}_{\psi} \otimes \ket{\{Q_X, \Psi,\lambda \}}_{\theta}$ of fermion and boson sectors. The expectation value in Eq.~\eqref{eq:PsiEq} is taken with respect to the bosonic sector, and (given that $Q_X = 0.524$ and $\lambda = 0$ independently of $Q_f$) the right-hand side is completely $Q_f$ independent. 

At this point, we only need to solve the slave-rotor-array on the honeycomb lattice with repulsion $U/2$ and coupling $J = 0.524 w_1$.
We use perturbation theory in $\Psi$~\cite{SachdevBook} to derive the effective energy per unit cell associated with Eq.~\eqref{eq:MFHam}
\begin{equation}
    E_{\rm Bose}  = 8 \left [ \vert \Psi \vert^2 \left ( \frac{1}{6 J} - \frac{U}{U^2 - 4 \lambda^2}\right) + \frac{7}{U^3} f(\lambda/U) \vert \Psi \vert^4 \right].
\end{equation}
For later reference $\vert \lambda \vert \ll U$ was restored in this expression and 
\begin{equation}
    f(\lambda/U) = 1 + \frac{111}{7} (\lambda/U)^2 + \mathcal O(\lambda^4/U^4) .
\end{equation}
Similarly, the wave function on a given site is
\begin{align}
    \ket{GS} &= \left (1 - \frac{2U \vert \Psi \vert^2 }{U^2 - 4\lambda^2}\right) \ket{0} \notag \\
    &+ \frac{\Psi^* }{U/2 + \lambda} \ket {1} + \frac{\Psi }{U/2 - \lambda} \ket {-1} + \sum_{\pm}\mathcal C_\pm \vert \Psi \vert^2 \ket{\pm 2},
\end{align}
where the number in the ket is the $\hat L$ eigenvalue and $\mathcal C_
\pm$ are constants of no importance. Using $\ket{GS}$, the occupation of bosons to leading non-trivial order in $\Psi$ is
\begin{equation}
    \langle \hat L_{\v x} \rangle = 32 \frac{\lambda \vert \Psi \vert^2}{U^3} [1 + \mathcal O(\lambda^2/U^2)]. \label{eq:Lexp}
\end{equation}
This justifies that $\lambda = 0$ solves the constraint Eq.~\eqref{eq:lambdaEq}.

Returning to the case $\lambda = 0$ the mean field transition is $U = 6J = 6 w_1 Q_X \simeq 3.144 w_1$ and the vacuum expectation value in the condensate is
\begin{align}
    \vert \Psi \vert^2 &= \frac{U^3}{14} \left ( \frac{1}{U}-\frac{1}{6J}\right), \label{eq:PsiMF}\\
    \Rightarrow Z =  \vert \langle e^{i \theta_{\v x}} \rangle \vert^2 & \simeq \frac{2}{7}  \left ( 1-\frac{U}{6J}\right), \\
    m_b & = \frac{U}{2} \sqrt{1- \frac{6J}{U}},
\end{align}
where $Z$ is the quasiparticle weight. We remark that the position of the transition is captured reasonably well within this approach, yet the quasiparticle weight substantially underestimated, cf.~Fig.~\ref{fig:ApproxMF}.

\subsection{Homogeneous Magnetic fields.}
\label{sec:BField}

We will see that the impact of a finite magnetic field effect stems from nearby the Fermi energy and we can use an effective low-energy Hamiltonian density takes the form

\begin{equation}
    \mathcal H = \sum_{\alpha = \pm 1} \sum_\kappa \psi_{\alpha, \kappa}^\dagger[v_f (- i D_i^{(\alpha)} \sigma^i \pm \alpha m_\kappa) - \lambda] \psi_{\alpha, \kappa},
\end{equation}
where, as before, $D_i^{(\alpha)} = \partial_i + i a_i +i \alpha A_i$.
Seeking the most general homogeneous mean field solution, we also allowed for an emergent internal field $b = \epsilon_{ij} \partial_i a_j$. 

\subsubsection{Solution for a single cone}

We consider a Hamiltonian 
\begin{equation}
h(\v p)  = \left (\begin{array}{cc}
m & v p \\ v \bar p & -m
\end{array} \right),
\end{equation}
which has eigenenergies $\epsilon(\v p) = \pm \sqrt{(v p)^2 + m^2}$ in the absence of a magnetic field. When there is a magnetic field $\mathcal B$ the free particle states are $E_l = \pm \sqrt{2 v^2 \mathcal B l + m^2}$ where the $l = 0$ state is of one sign only (``zeroth" Landau level). The degeneracy of each Landau level (LL) is $\mathcal B/2\pi$.

The particle number  and energy density in the LL state for the situation when the zeroth LL is below $0$ is
\begin{align}
n & = \frac{\Lambda}{4\pi v^2} + \frac{\mathcal B}{4\pi}, \\
\epsilon & = - \frac{\Lambda ( \Lambda^2 + m^2)}{8 \sqrt{\pi} v^2} - \frac{\mathcal B}{2\pi} \sqrt{2 v^2 \mathcal B} \zeta(-1/2, \tilde m^2). \label{eq:EpsLandau}
\end{align}
Here, $\tilde m^2 = m^2/(2 v^2 \mathcal B)$ and $\zeta(-1/2,x)$ is the Hurwitz zeta function. The  term involving the UV cut-off $\Lambda$ reflects the contribution of the filled Fermi sea at half-filling and without fields. This reference value will be dropped unless explicitly mentioned.

Analogously for the case when the zeroth LL is above zero

\begin{align}
n & = \frac{\Lambda}{4\pi v^2} - \frac{\mathcal B}{4\pi}, \\
\epsilon & = - \frac{\Lambda ( \Lambda^2 + m^2)}{8 \sqrt{\pi} v^2} - \frac{\mathcal B}{2\pi} \sqrt{2 v^2 \mathcal B} \zeta(-1/2,  1+\tilde m^2).
\end{align}

\subsubsection{Topological case $m_+ m_- <0$ and constraints and effective chemical potential}

\begin{figure*}
    \centering
    \includegraphics{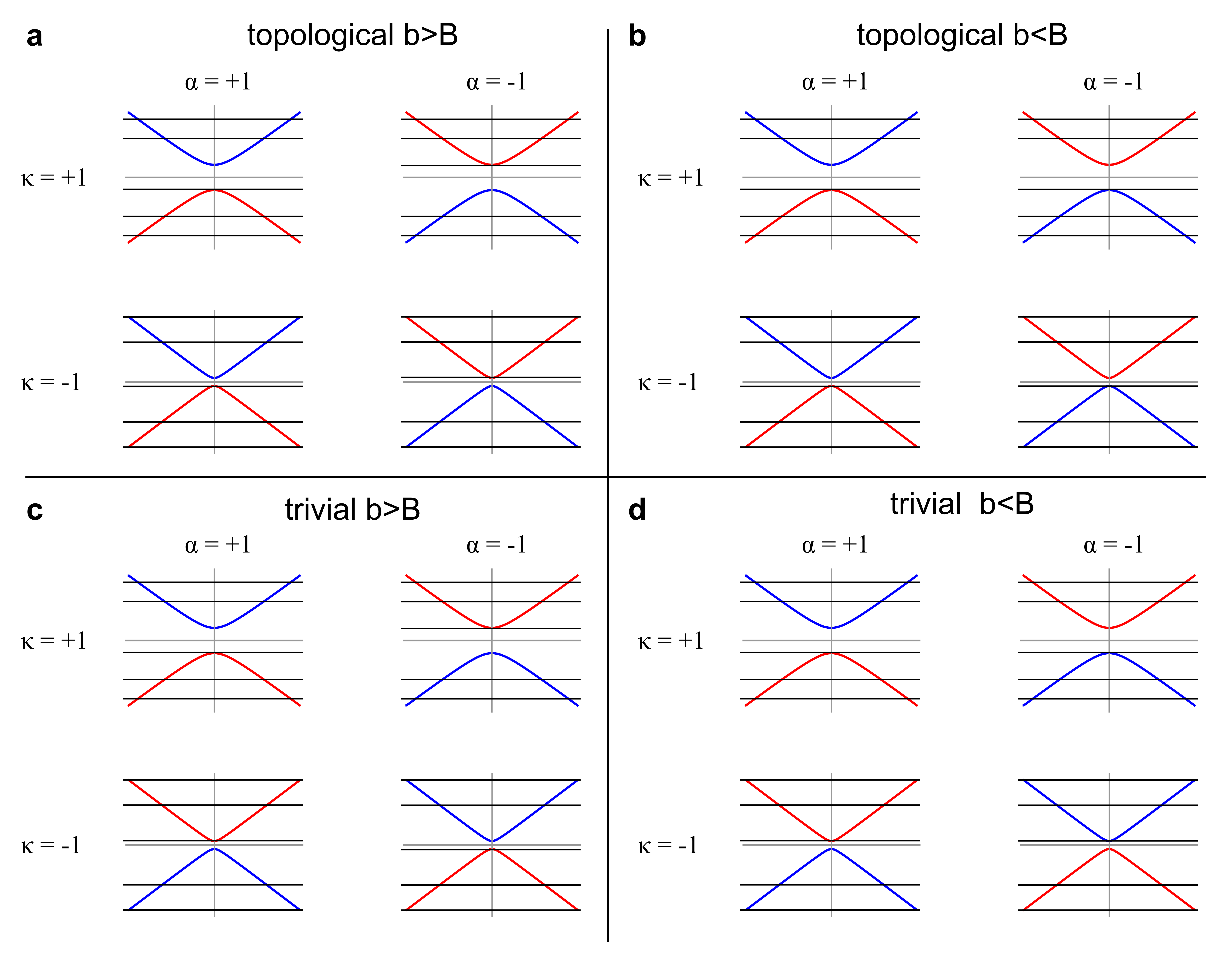}
    \caption{Schematic Landau level spectrum for applied emergent $b$ and external gauge fields $B$. Different Chern numbers are represented by different colors. (Not to scale. Degeneracies other than those of the zeroth Landau level are accidental.)}
    \label{fig:LandauLevelSpectra}
\end{figure*}

We next switch to the situation of a set of four gapped Dirac Hamiltonians with masses and magnetic fields, cf.~Fig.\ref{fig:LandauLevelSpectra}.

\begin{center}
\begin{tabular}{c|c|c|c|c}
    $(\kappa,\alpha) $ & mass & field & $E_{l = 0}$ for $b>B$ & $E_{l = 0}$ for $b<B$\\
    \hline \hline
    (+,+) & $m_+>0$ & $\mathcal B_{+}  = b + B$ & $-m_+<0$ &$-m_+<0$ , \\ \hline
    (+,-) & $- m_+<0$& $\mathcal B_{-} = b - B$ & $m_+ >0$& $-m_+ <0$, \\ \hline
    (-,+) & $-m_->0$& $\mathcal B_{+} = b + B$& $m_-<0$ &$m_-<0$ , \\ \hline
    (-,-) & $m_-<0$& $\mathcal B_{-} = b - B$& $-m_->0$ &$m_-<0$ . 
\end{tabular}
\end{center}

We here assumed positive fields $b>0, B>0$. 
The LL spectrum spectrum shows additional flavor SU(2) symmetries when $B = 0$ or $b = 0$, but clearly the zeroth LL always has emergent SU(2) symmetry when $b<B$.

The constraints to be satisfied are
\begin{align}
\langle \hat L_{\v x} \rangle &= -A_{\rm u.c.}\sum_\alpha  n_\alpha, \\
0 & = 2\sum_\alpha \alpha n_\alpha,
\end{align}
where the second line corresponds to the physical density relative to half-filling and $A_{\rm u.c.}$ is the area of the unit cell. Since the two colors experience different fields $b \pm B$, we conclude that only $b=0$ is generally consistent with zero charge density.

\subsubsection{Case $\vert \lambda\vert < m_- $ applicable to the vicinity of the $U$-driven superconducting transition.}

Here, we assume small $\vert \lambda \vert \ll \vert m_\pm \vert$. This implies that the fermionic occupation $\langle \sum_\alpha \psi^\dagger_{\v x, \alpha} \psi_{\v x,\alpha} \rangle - 1  \equiv \sum_\alpha A_{\rm u.c.} n_\alpha = \Phi_B/\pi$ (where $\Phi_B = B A_{\rm u.c.}$ is the flux through the unit cell), see Fig.~\ref{fig:LandauLevelSpectra} b). To compensate this, we expect a small, negative rotor occupation, or equivalently a small negative $\lambda$.
Note that generally in Bose-Hubbard or rotor models, there is no Mott-insulating state unless we allow for integer rotor occupation. Instead the system is always in the Higgs phase (i.e.~a slave-rotor superfluid). 

As we identify the Mott phase of the rotor model with a superconductor, this is equivalent to saying that any internal homogeneous field $\Phi_B$ in the systems is inconsistent with the existence of a superconductor (=Mei{\ss}ner effect). 

Using \eqref{eq:Lexp} the gauge constraint becomes
\begin{align}
    \frac{\vert \Psi \vert^2}{U^2} = - \frac{U}{32 \pi \lambda} \Phi_B,
\end{align}
and we find the bosonic energy per unit cell
\begin{align}
    E_{\rm Bose}  = 8 U \left [ \frac{U}{32 \pi \vert\lambda \vert} \Phi_B \left ( \frac{U}{6 J} - 1 - \frac{4\lambda^2}{U^2} \right) + \frac{7  }{(32\pi)^2} \frac{U^2}{\lambda^2} \Phi_B^2
    \right ].
\end{align}

We are interested in the regime $U\sim U_c$, so the bosonic energy is minimized for the maximal $\vert \lambda \vert$ under consideration, i.e.~$\lambda = -m_- + 0^+$, so that 
\begin{align}
    E_{\rm Bose}  = 8 U \left [ \frac{U}{32 \pi m_-} \Phi_B \left ( \frac{U}{6 J} - 1 - \frac{4m_-^2}{U^2} \right) + \frac{7  }{(32\pi)^2} \frac{U^2}{m_-^2} \Phi_B^2
    \right ].
\end{align}

The field dependent fermionic contribution to the free energy per unity cell is
\begin{align}
   \delta E_{\rm Fermi} & = \frac{2  \sqrt{3}}{\pi} t_1 \sum_{\pm } \frac{m_\pm^3}{t_1^3}\left [ \frac{\zeta(-1/2, \tilde m_\pm^2)}{\tilde m_\pm^3} - \left .\frac{\zeta(-1/2, \tilde m_\pm^2)}{\tilde m_\pm^3} \right  \vert_{B =0 }\right] \notag \\
    & \simeq \frac{2\sqrt{3}}{\pi} \sum_\pm m_\pm \frac{v^2 B}{t_1^2} \left (1 + \frac{v^2 B    }{6 m_\pm^2}\right)\\
    &\simeq \Phi_B\sum_{\pm} {m_\pm} \left (1 + \Phi_B^2\frac{t_1^2}{12 \sqrt{3} m_
    \pm^2}\right) + \mathcal O(\Phi_B^6).
\end{align}
As expected, the energy of the Fermi sea increases with $\Phi_B$ (diamagnetism) and we used the approximation
\begin{equation}
    \frac{\zeta(-1/2, \tilde m_\pm^2)}{\tilde m_\pm^3} \simeq -\frac{2}{3} + \frac{1}{2 \tilde m_\pm^2} + \frac{1}{24 \tilde m_\pm^4}.
\end{equation}
Note that the increase in $\delta E_{\rm Fermi}$ leads to a decrease in boson hopping.
\begin{equation}
    Q_X = Q_X\vert_{B = 0} - \frac{E_{\rm Fermi}}{6 t_1}.
\end{equation}

We summarize that for $-m_- < \lambda <0$ we find to leading order in $\Phi_B$ and $U\sim U_c = 6J$
\begin{align}
    E_{\rm Bose}  &=  \Phi_B \frac{3J}{2 \pi m_-}  {(U- U_c)},\\
    E_{\rm Fermi} & = \Phi_B \sum_{\pm} m_{\pm}.
\end{align}

\subsubsection{Case $-\sqrt{m_-^2 + 2v^2 B}<\lambda<-m_-$ applicable to the vicinity of the $\varphi$-driven superconducting transition.}

Assuming that $\lambda$ sits between $l = -1$ and $l = 0$ Landau levels of the $\kappa = -1$ valley, we find that $\sum_\alpha  n_\alpha =0 $ so that we can search for $\langle \hat L_{\v x} \rangle = 0$ solutions (i.e.~$\vert \Psi \vert = 0$ by means of the constraint and Eq.~\eqref{eq:Lexp}). Note that this also implies a vanishing $C_{+1} - C_{-1} = 0$, so that the topological mechanism of superconductivity is inefficient.

In this sector and independently on the sign of $U-U_c$, $E_{\rm Bose} =0 $, but
\begin{equation}
    \delta E_{\rm Fermi} = \Phi_B (m_+ + 3 m_-).
\end{equation}

Effective doping beyond $\lambda = -\sqrt{m_-^2 + 2v^2 B}$ appears unfavorable, as it implies energy cost stemming from both fermion sector and boson sector.

Combining the result of this and the previous section, we thus estimate the energy in the presence of a magnetic field as
\begin{align}
    \delta E (\Phi_B) &= \Phi_B \sum_{\pm} m_{\pm} +\Phi_B \text{min} \left [ \frac{3 J(U- U_c) }{2\pi m_-}, 2m_-\right ],
\end{align}
where $U>U_c$ and $J/m_- \sim w_1/w_0 \gg 1$.

In summary the system is never a superconductor if there is a homogeneous $B$-field, consistent with standard expectations.

\subsection{Finite doping}
\label{sec:FiniteDoping}

In analogy to \cite{RanWen2009}, doping into the system may be achieved by spontaneous creation of a small, homogeneous $b$ field. Concentrating on the topological regime and following Fig.~\ref{fig:LandauLevelSpectra} a), we see that small $b$ (at $B = 0$) leads to finite doping, yet no emergent gauge charge.

\subsubsection{Spontaneous Landau levels formation in the absence of external magnetic field}

In this section we argue on the mean-field level that the spontaneous formation of Landau levels is favored of creating pockets. 

First we consider the state with Fermi surfaces (i.e. 4 pocket state without internal magnetic field $b = 0$). Note that this state breaks the topology behind the present mechanism for superconductivity.
We find for density and energy density
\begin{align}
n &=p_F^2 /\pi \\
\epsilon_{\rm FS} & = 2 \frac{(m^2 + {v^2 n \pi})^{3/2}}{3\pi v^2} \\
& = \frac{2 \sqrt{\pi} v n^{3/2}}{3} \left ( 1 + 2 \tilde m^2 \right )^{3/2}
\end{align}
where $\tilde m = m^2/(2 v^2 n \pi)$. For simplicity we here took $\vert m_+ \vert = \vert m_- \vert = m$, in the case when the masses in different color sector are different, we still have the same Fermi momenta, but we have to symmetrize the result over up and down sectors.

Next we consider the state with internal magnetic field $b$. In this case we have two sets of negative Landau levels filled, one set has the zeroth LL below, the other above $0$ leading to vanishing emergent gauge charge but
\begin{align}
n & = b/\pi \\
\epsilon_{LL} & = -2 \frac{ b}{2\pi} \sqrt{2 v^2 g b} [\zeta(-1/2, \tilde m^2) + \zeta(-1/2, \tilde m^2 + 1)]\\
& = - n m - v n^{3/2} \sqrt{8 \pi} \zeta(-1/2, 1 + \tilde m^2) \\
& = \frac{v n^{3/2}}{\sqrt{2\pi}} \left (- 2 \pi \tilde m - 4\pi \zeta(-1/2, 1 + \tilde m^2) \right ) \\
& \stackrel{m \rightarrow 0}{\longrightarrow} \frac{v n^{3/2} \zeta(3/2)}{\sqrt{2\pi}},
\end{align}
and again  $\tilde m = m^2/(2 v^2 n \pi)$, but with $b$ dependent $n$.
This coincides with the result by \cite{RanWen2009}. Again, this situation was discussed for the case of equal masses. When they are unequal, the expressions for physical and gauge charge are unchanged, while energy expression need to be adapted following Eq.~\eqref{eq:EpsLandau}.

\begin{figure}
\includegraphics[scale=.55]{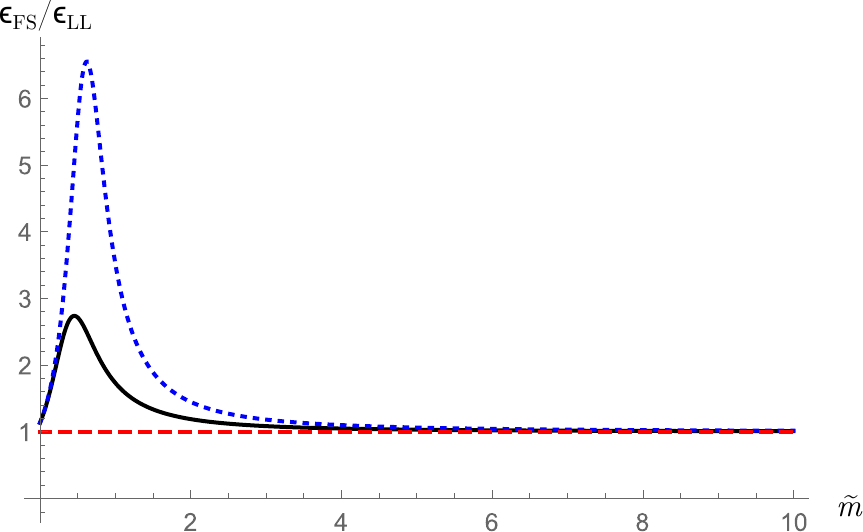}
\caption{Comparison of energies of the Fermi surface and Landau state as a function of $\tilde m^2 = m^2/ 2 v^2 m \pi$. Black solid: $\vert \tilde m_+ \vert =\vert \tilde m_-\vert =\tilde m$, Blue dotted: $m_- = 0, \tilde m_+ =\tilde m$.}
\label{fig:EnergyComparison}
\end{figure}

We can compare the mean-field energies of Fermi surface solution and Landau level solution, Fig.~\ref{fig:EnergyComparison}, and conclude that, for small and intermediate $\tilde m$, the Fermi surface state is reasonably distant in energy from the Landau level state, and one may invoke the present mean-field arguments to conclude that, upon doping, the system tends to spontaneously create emergent Landau levels. At largest $\tilde m$ the energies for Fermi surface and Landau level solutions become degenerate and making a definitive conclusion based on mean-field theory audacious.

\section{Effective quantum field theory}
\label{sec:QFT}

In this section of the supplement we summarize the main steps to derive the effective gauge theory from the microscopic theory. Using the gauge theory including fermionic matter we derive the vanishing Boltzmann weights and quantum numbers of monopole operators and finally employ the particle-vortex duality to derive the effective superfluid action.

\subsection{Determining stiffness and velocity of superflow}

The gauge propagator has no bare dynamics and to leading order it is entirely determined by RPA resummation. 

\subsubsection{Calculation of boson and fermion polarization operator}

The low-momentum, low energy propagator of bosonic spinons takes the following form upon expansion near the $\Gamma$ point~\cite{BollmannKoenig2024b} 

\begin{equation}
    D(i \omega, \v q)  = \frac{U}{\omega^2 + v_b^2 \v q^2 + m_b^2 },
\end{equation}
where $v_b = a\sqrt{UJ/4} = a\sqrt{U w_1 \mathcal G/4}$ and $ m_b^2  = U (U - U_c)$ and $U_c = 12 J$, where $J$ is the bosonic hopping. 

We can use an effective theory of coherent state bosons to describe the vicinity of the transition
\begin{equation}
    S[\Phi] = \int d\tau d^2 x \ \vert(\vec \nabla_b +i\vec a_b) \Phi \vert^2 +   m_b^2 \vert \Phi \vert^2 ], \label{eq:BosonAction}
\end{equation}
where $\vec \nabla_b = (\partial_x v_b, \partial_y v_b, \partial_\tau)$ and we have also absorbed $v_b$ into the vector field $\vec a_b = (v_b a_x, v_b a_y, a_\tau)$. The field $\Phi$ has dimensions 1/(length$^2 \times$ energy)$^{1/2}$.

Next, we consider fermions, which can be described by 

\begin{equation}
    S[\Psi] = \int d\tau d^2 x \ \sum_{\kappa, \alpha}\bar \Psi[( \vec \nabla_f - i (\vec a_f + \alpha \vec A_f))\cdot \vec \sigma + \alpha m_{\kappa}] \Psi,
\end{equation}
where we have absorbed $\sigma_z$ into $\bar \Psi$. Note that, here, $\vec \nabla_f = (\partial_x v_f, \partial_y v_f, \partial_\tau)$ and $v_f \sim \sqrt{3} t_1/2, m_{\kappa = \pm} = \sqrt{27} t_2 \pm t_0$. Note that the masses are unequal in the two valleys and moreover there are a total four species of fermions in our problem. Also note that different spin species couple with opposite sign to $A$. Again, we here tacitly absorbed $v_f$ into $a_f, A_f)$.

Integrating out the bosonic degrees of freedom, we find the effective action $S_\mathrm{eff} = 1/2 \int_q a_\mu(q) a_\nu(-q) \Pi^{\mu \nu}(q)$ with the polarization bubble ( we momentarily set $v_b = 1$.)
\begin{align}
    \Pi^{\mu \nu}(p) &= \frac{\delta^{\mu \nu} p^2 - p^\mu p^\nu}{16 \pi m_b}  \notag \\ &\times \left(- \frac{4 m_b^2}{p^2} + \left(\frac{2 m _b}{p} + \left(\frac{2m_b}{p}\right)^3\right) \arctan \frac{p}{2m_b} \right) \nonumber\\
    &\approx \frac{p^2 \delta^{\mu \nu}- p^\mu p^\nu}{24 \pi m_b}.
\end{align}

To illustrate the fermionic calculation we now momentarily set $v_f = 1$, and use $\vec a_\alpha = \vec a +\alpha \vec A$. The integration of fermions leads to an effective action
\begin{align}
    S_{\rm eff} & = - \sum_{\alpha, \kappa }\mathrm{tr} \ln[\hat G^{-1}_{\alpha, \kappa} - i \vec a_{\alpha} \cdot \vec \sigma] \\
    &\simeq \text{const.} - \sum_{\alpha, \kappa }\mathrm{tr}[\hat G_{\alpha, \kappa} \vec a_{\alpha} \cdot \vec \sigma\hat G_{\alpha, \kappa} \vec a_{\alpha} \cdot \vec \sigma]/2.
\end{align}
Here, Green's function and $a_{\alpha}$ insertions are to be read as operators and $\Tr$ includes integration in space-time.  We find for a given fermion species (index suppressed)
\begin{align}
    \Pi_{\mu \nu}(q) &\equiv - \int_p \mathrm{tr} \left[\sigma^\mu G(p-q) \sigma^\nu G(p) \right] \notag \\
    & = -\frac{q^\mu q^\nu - q^2 \delta^{\mu \nu}}{8 \pi |m|} \notag \\  
    &\times \left( \frac{4 m^2}{q^2} + \frac{2|m|}{q} \left(1  - \frac{4 m^2}{q^2} \right) \right) \arctan \left( \frac{q}{2|m|} \right) \notag \\
    &+ \frac{m}{2\pi |q|} q^\alpha \epsilon^{\mu \nu \alpha} \arctan \left( \frac{|q|}{2m} \right) \\
    &\approx \frac{q^2 \delta^{\mu \nu} - q^\mu q^\nu}{6 \pi |m|} + \epsilon^{\mu \nu \alpha} q^\alpha \frac{m}{|m|} \frac{1}{4 \pi} .\label{eq:PiF}
\end{align}

\subsubsection{Ioffe-Larkin-like addition of contributions}
The non-topological, symmetric contribution (indicated by round brackets around indices) to the polarization bubble presented in Eq.~\eqref{eq:Polops} leads to a Maxwell term

\begin{equation}
    S_{\rm mw} = \int d\tau d^2 x \ \sum_{\substack{\lambda =\\ b, f_+, f_-}} \sum_{\mu \nu}\frac{[\tilde f_{\mu \nu}^{(\lambda)}]^2}{4 g_\lambda^2 v_\lambda^2} \equiv \int d\tau d^2 x \ \sum_{\mu \nu} \frac{\tilde f_{\mu \nu}^2}{4g^2 c^2},
    \label{eq:SmwApp}
\end{equation}

where we introduced $g_b^2 = 12 \pi m_b, g_{f_\pm}^2 = 3 \pi \vert m_\pm \vert/2$ and $v_{f_\pm} =v_f$. We here used non-standard field strength tensors obtained from the $\partial_{f/b}$ derivatives and $a_{f/b}$  rescaled fields
\begin{align}
\tilde f_{\mu \nu}^{(\lambda)} &= \partial^\mu_\lambda a^\nu_\lambda - \partial^\nu_\lambda a^\mu_\lambda = \left (\begin{array}{c c c}
    0 & v_\lambda^2 b & -v_\lambda e_y \\
    -v_\lambda^2 b & 0 & v_\lambda e_x \\
    v_\lambda e_y & - v_\lambda e_x &0 \\
\end{array} \right)_{\mu \nu},\\
\tilde f_{\mu \nu} &= \partial^\mu a^\nu - \partial^\nu a^\mu \equiv \left (\begin{array}{c c c}
    0 & c^2 b & -c e_y \\
    -c^2 b & 0 & c e_x \\
    c e_y & - c e_x &0 \\
\end{array} \right)_{\mu \nu}.
\end{align}

By comparison we identify
\begin{align}
    \frac{1}{g^2} & = \sum_\lambda \frac{1}{g_\lambda^2}, \\
    \frac{c^2}{g^2} & = \sum_\lambda \frac{v_\lambda^2}{g_\lambda^2}
\end{align}
which is equivalent to
\begin{subequations}
\begin{align}
    c & = \sqrt{ \frac{\sum_\lambda {v_\lambda^2}/{g_\lambda^2}}{\sum_\lambda {1}/{g_\lambda^2}}} \notag \\
     &= g \sqrt{\sum_\lambda v_\lambda^2/g_\lambda^2}.
\end{align}
\end{subequations}
Thus the effective Maxwell term takes the form
\begin{equation}
    S_{\rm mw} = \int d\tau d^2 x \ \frac{\v e^2 + c^2 b^2}{2g^2} = \int d\tau d^2 x \ \frac{\v e^2/c + c b^2}{2\tilde g^2}.
\end{equation}

As a last step, we wrote 
\begin{align}
    \frac{1}{\tilde g^2}  = \frac{c}{g^2}.
\end{align}

We remark that $g^2$ is an energy and determines the stiffness, while $\tilde g^2$ has dimensions of $1/\text{length}$ and is the actual gauge coupling. 

This concludes the derivation of Eq.~\eqref{eq:StiffnessSpeedOfLight} of the main text.

\subsection{Stiffness for finite doping and finite magnetic field}

\subsubsection{Disorder allowed field penetration.}

Here we consider the superconducting state, but in a magnetic field. This is not possible in the clean case~\ref{sec:BField}, but disorder changes the situation.

When the chemical potential vanishes, the density in the $\alpha$ sector is
\begin{equation}
    n_{\alpha} = \frac{\alpha b +B }{2\pi}.
\end{equation}
We occupy $\nu_-$ (empty $\nu_+$) Landau level states of the positive (negative) zeroth Landau level of the $\alpha = -$ ($\alpha = +$) sector, with $0 < \nu_\pm \ll 1$. Note that, in view of disorder these partial filling goes to the Lifshitz tails of Landau levels which are Anderson localized, and does not alter the topological state under consideration. We interpret the extra localized charge associated to non-trivial filling fractions $\nu_+,\nu_-$ as pinned vortices. 

We thus find
\begin{align}
    n_{\alpha = +1} & = \frac{b+B}{\pi} \left (\frac{1}{2} - \nu_+\right ),\\
    n_{\alpha = -1} & = \frac{b-B}{\pi} \left (-\frac{1}{2} + \nu_+\right ).
\end{align}
The additional energetic cost is 
\begin{equation}
\Delta E = 2\bar \nu m, 
\end{equation}
so we search a solution with minimal $\bar \nu = (\nu_- + \nu_+)/2$, with both $\nu_-, \nu_+$ positive.

We will use the notation $\Delta \nu = \nu_+ - \nu_-$ to express the condition of vanishing emergent charge and externally imposed energy as

\begin{subequations}
\begin{align}
     0 &= \frac{b+B}{\pi} \left (\frac{1}{2} - \nu_+\right ) +\frac{b-B}{\pi} \left (-\frac{1}{2} + \nu_- \right ) \notag \\
     & = [- \Delta \nu b + (1- 2\bar \nu)B]/\pi, \label{eq:GaugeCond}\\
     n &= \frac{b+B}{\pi} \left (\frac{1}{2} - \nu_+\right ) +\frac{b-B}{\pi} \left (\frac{1}{2} - \nu_-  \right ) \notag \\
     & = [b(1- 2\bar \nu) - B \Delta \nu]/\pi.
\end{align}
\end{subequations}

This leads to
\begin{align}
    \Delta \nu & = \frac{B}{b} (1- 2 \bar \nu) \\
   \Rightarrow 2 \bar \nu & = 1 - \frac{b\pi n}{b^2-B^2} .
\end{align}
Putting this back into the condition for $\Delta \nu$ leads to
\begin{equation}
    \Delta \nu  = \frac{B \pi n}{b^2 - B^2}. \label{eq:DeltanuCond}
\end{equation}
Depending on the field sign we either occupy only $\nu_+$ or only $\nu_-$ to minimize $\bar \nu$, i.e.

\begin{align}
        \nu_+ = \frac{B \pi n}{b^2 - B^2}, &\nu_- = 0, \quad \text{if } B>0, \\
                \nu_- = \frac{\vert B  \pi n \vert}{b^2 - B^2}, &\nu_+ = 0, \quad \text{if } B<0. 
\end{align}

Thus, $b$ in the presence of $B \neq 0$ is given by the condition
\begin{equation}
    b = \pi n + \vert B \vert + \mathcal O(B^2).
\end{equation}
We remind the reader that we have assumed small occupation $0<\nu_\pm \ll 1$ of the Lifshitz tails of the Landau levels, hence we can only treat the limit $\vert B \vert  \ll  n  $. 

 Note that the entire calculation was performed under the assumption $b \pm B >0$, corresponding to $n>0$. The case of negative doping leads to virtually the same result, except that all fields and the role of $\alpha = + 1$, $\alpha = -1$ are reversed.

\subsubsection{Estimate of corrections of stiffness}
Following the results of the previous section and section Sec.~\ref{sec:FiniteDoping} we estimate the density-density response in the presence of a field and finite doping, i.e. in the presence of $B$ and $b$. 
We concentrate on $\varphi = \pi/2$. {Here, we do not repeat the entire calculation of $\Pi_{\mu \nu}(q)$ but instead concentrate on the simplest case of current-current response}

\begin{align}
    \Pi(i \omega; \v x, \v x') & = \int \frac{d \epsilon}{2\pi} \tr[G(i\epsilon_+, \v x, \v x')  \sigma_xG(i\epsilon_-, \v x', \v x)\sigma_x], \label{eq:PiX}
\end{align}
where here we use $\epsilon_{\pm} = \epsilon \pm \omega/2$ and the standard imaginary time Green's function in Lehmann representation
\begin{equation}
    G(i\epsilon, \v x, \v x') = \sum_{\lambda}\frac{\Psi_\lambda(\v x)\Psi^\dagger_\lambda(\v x'   )}{i \epsilon - \epsilon_\lambda}.
\end{equation}
The energy eigenvalues $\epsilon_{\lambda}$ to exact eigenfunctions $\Psi_\lambda(\v x)$ of the potentially disordered Hamiltonian are generally unknown. Returning briefly to the clean case without magnetic field, we highlight that the Fourier transform of Eq.~\eqref{eq:PiX} yields the expression $\Pi_{11}(q)$ in Eq.~\eqref{eq:PiF}. 

Instead, here we consider the limit in which we have emergent magnetic field $b$, whose energy scale $v_f \sqrt{b}$ is large as compared to physical magnetic field $B$ 
and than disorder (which is measured in the spread of the Landau level broadening). As we here only determine the corrections to the gauge coupling (i.e. we assume approximate Lorentz-invariant structure of the polarization operator and just determine the coefficient) we concentrate on 
\begin{align}
    \Pi(i \omega) &=    \Pi_{11}(i \omega, \v q = 0) \notag \\
    & = \frac{1}{\mathcal A} \int d^2 x d^2 x'  \Pi(i \omega; \v x, \v x')\notag  \\
    & = \frac{1}{\mathcal A} \int d^2 x d^2 x' \sum_{\lambda, \lambda'} \int d^2x d^2x' \notag \\
    &\frac{[\Psi^\dagger_\lambda(\v x) \sigma_x \Psi_{\lambda'}(\v x)][\Psi^\dagger_{\lambda'}(\v x') \sigma_x \Psi_{\lambda}(\v x')]}{[i \epsilon_+ - \epsilon_\lambda][i \epsilon_- - \epsilon_\lambda]}\notag  \\
    & = i \frac{1}{\mathcal A} \int d^2 x d^2 x' \sum_{\epsilon_\lambda >0, \epsilon_{\lambda '} <0}  \notag \\
    & \times \left [\frac{1}{\omega - i (\epsilon_{\lambda} - \epsilon_{\lambda'})}  - \frac{1}{\omega + i (\epsilon_{\lambda} - \epsilon_{\lambda'})}  \right ] \notag \\
    &\times \Big [\Psi^\dagger_\lambda(\v x) \sigma_x \Psi_{\lambda'}(\v x)][\Psi^\dagger_{\lambda'}(\v x') \sigma_x \Psi_{\lambda}(\v x')\Big ]
\end{align}

Here, $\mathcal A$ is the area of the system and the energy is measured with respect to the Fermi energy, which sits in the Lifshitz tails, cf. Fig.~\ref{fig:LLsMaintext}. For example, in the setup as in said figure, the smallest energy difference in the $\alpha = -1$ sector vanishes (contributions from within the same Landau level). At the same time, these intra-Landau level contributions are negligible in the thermodynamic limit, because the matrix elements are suppressed in view of the exponentially localized wave functions $\Psi_{\lambda}(\v x)$.

In the limit of small $\nu_{\pm} \ll 1$, where only the most strongly localized tail-states are occupied, we thus expect $\Pi(i \omega)$ to be well approximated by simply taking the Fermi energy at zero energy. Moreover, if disorder is weak, one may take into account the level broadening in a self-consistent manner leading to semi-circular density of states of Landau levels~\cite{EfetovBook}. The effect of this level broadening, which is much smaller than the inter-Landau-level distance, will be strongly subleading in the calculation of $\Pi(i \omega)$. We thus evaluate $\Pi(i\omega)$ using clean Landau level eigenstates of the clean Hamiltonian instead of $\Psi_\lambda(\v x)$. For notational simplicity we set the speed of fermions $v_f = 1$ in the remainder of this section, including the following formula

\begin{align}
    \Pi(i \omega) &= i\mathcal B \sum_{\epsilon_\eta >0, \epsilon_{\eta'} <0} \vert \langle {\eta \vert \sigma_x \vert \eta'}  \rangle\vert^2 \\
     & \times \left [\frac{1}{\omega - i (\epsilon_{\eta} - \epsilon_{\eta'})}  - \frac{1}{\omega + i (\epsilon_{\eta} - \epsilon_{\eta'})}  \right ] \\
     & \simeq   \mathcal B\sum_{\epsilon_\eta >0, \epsilon_{\eta'} <0}  \omega^2 \frac{\vert \langle {\eta \vert \sigma_x \vert \eta'}  \rangle\vert^2 }{(\epsilon_\eta - \epsilon_{\eta'})^3}.
\end{align}
The prefactor $\mathcal B$ stems from the degeneracy $\mathcal B \mathcal  A$ of the Landau levels. At the ``$\simeq$" sign we dropped  a divergent constant (as we also  did in Eq.~\eqref{eq:PiF}), which can be technically achieved using Pauli-Villars regularization.

For simplicity we here calculate the contribution of a single fermion flavor, e.g. Fig.~\ref{fig:LLsMaintext} at $\alpha = -1$ (the zeroth Landau level is above zero) and $\mathcal B > 0$. In this case
\begin{equation}
    \epsilon_\eta = \vert m \vert \delta_{\eta=0} + \text{sign}(\eta) \sqrt{m^2 + 2\mathcal B \vert \eta \vert} \delta_{\eta \neq 0} .
\end{equation}
The corresponding wave functions are 
\begin{equation}
    \ket{\eta}  = \begin{cases}
        \frac{1}{2(\epsilon_\eta^2 - \vert m \vert \epsilon_\eta)}\left ( \begin{array}{c}
             \sqrt{B} \sqrt{\vert \eta \vert }\ket{ \vert \eta \vert}\rangle  \\
             (\vert m \vert - E_\eta) \ket{ \vert \eta \vert - 1}\rangle 
        \end{array}\right ), \quad n \neq 0, \\
        \left ( \begin{array}{c}
             \ket{0}\rangle \\
            0
        \end{array}\right ), \quad \eta = 0.
    \end{cases}
\end{equation}
Here, $\ket{\eta}\rangle $, $\eta \geq 0$ are regular Landau level eigenfunctions. The evaluation of the matrix element leads to a constraint between $\eta'$ and $\eta$ and ultimately to

\begin{align}
    \Pi(i \omega) & = \mathcal B \omega^2 \sum_{\eta \geq 0} \frac{m^2 + \mathcal \epsilon_\eta \epsilon_{\eta + 1}}{2 \epsilon_\eta \epsilon_{\eta + 1}} \frac{1}{(\epsilon_\eta + \epsilon_{\eta +1})^3}.
\end{align}

We evaluate this expression in two limits. First, consider the limit of $m^2 \gg \mathcal B$ (and we use $\omega_c = \mathcal B/m$) 
\begin{align}
    \Pi(i\omega) & \simeq \mathcal B\sum_{\eta \geq 0} \frac{\omega^2}{(\epsilon_\eta + \epsilon_{\eta +1})^3} 
\notag \\
    & =\omega^2  \frac{\mathcal B}{8\omega_c^3}   \left [ -\frac{1}{2} \psi^{(2)}\left ( \frac{\vert m \vert}{ \omega_c}  + \frac{1}{2}\right ) \right ] \notag \\
    & \simeq\omega^2  \frac{\mathcal B}{8\omega_c^3} \frac{\omega_c^2}{2 \vert m \vert^2} \frac{1}{1 + \frac{\omega_c^2}{4 \vert m \vert^2}} 
\end{align}

Parametrically, this result is consistent with the calculation around Eq.~\eqref{eq:PiF}. Next, consider $\mathcal B \gg m^2$, so that 

\begin{align}
    \Pi(i \omega) &\simeq \frac{\omega^2}{\sqrt{\mathcal B}} \sum_{\eta \geq 0} \frac{1}{16 (\sqrt{\eta} + \sqrt{\eta + 1
    })^3} \notag \\
    & \approx \frac{1.2}{16} \frac{\omega^2}{\sqrt{\mathcal B}}.
\end{align}

Thus, at up to coefficients of order one which beyond the accuracy of this calculation, the stiffness in the presence of doping and magnetic field induced $\mathcal B$ may be  replaced by $m \rightarrow \sqrt{m^2 + \mathcal C \vert \mathcal B \vert}$ which is an interpolating function, with $\mathcal C \sim 1$ and $\mathcal B = b \pm B$. We sum over both $\alpha$ channels and again expand in small $\vert B /n \vert$, and thus estimate

\begin{align}
    K(B,b) &\sim \sqrt{m^2 + \mathcal C   \vert b \vert }  \notag \\
    &= \sqrt{m^2 + \mathcal C  (\pi \vert n \vert + \vert B \vert)}  + \mathcal O(B^2).
\end{align}

This implies that superconductor linearly improves upon inclusion of doping or magnetic field. This concludes the derivation of the field and doping dependent behavior quoted in Eq.~\eqref{eq:StiffnessCorrections} of the main text.

\subsection{Effective field theory and duality}
The parity-odd term in \eqref{eq:Polops} can be seen to correspond to a Chern-Simons term for the emergent gauge field that is induced by integrating out the fermions in a topologically non-trivial band. In the problem at hand, the fermions do not only couple to the emergent gauge field, but also carry (opposite) electromagnetic charges, i.e.~the two bands carry (gauge, EM) charges $(+1,+1)$ and $(+1,-1)$. Upon coupling to the emergent electromagnetic gauge field $A^\mu$ and integrating out the fermions, with thus get a lowest-order response theory of the form $S_\mathrm{eff} = S_\mathrm{\rm mw}[a] + S_\mathrm{\rm MW}[A] + S_\mathrm{CS}[a;A]$ with
\begin{multline} \label{eq:SCSApp}
    S_\mathrm{CS} = \int d^3 x \left(-i \frac{1}{4 \pi}
    \right)\epsilon^{\mu \nu \rho}\big[C_1 (A_\mu + a_\mu) \partial_\nu (A_\rho + a_\rho) \\ + C_2 (-A_\mu + a_\mu) \partial_\nu (-A_\rho + a_\rho)  \big]
\end{multline}
Using that $C_1 = - C_2  = C = 1$, one immediately finds $S_\mathrm{CS} = (- i C/ \pi) \int d^3 x \epsilon^{\mu \nu \rho} A_\mu \partial_\nu a_\rho + (\text{boundary terms})$.

\subsubsection{Duality mapping}

To make this supplement self-contained, we now summarize the main steps of the duality transformation from QED$_3$ to action of the phase field $\phi$, Eq.~\eqref{eq:PhaseTheory}. For a pedagogic review, see also Refs.~\cite{Polyakov1987} and~\cite{Tong2018}.

We consider the emergent U(1) gauge theory composed of $S_{\rm mw}$ and $S_{\rm CS}$, Eqs.~\eqref{eq:SmwApp}, \eqref{eq:SCSApp}. In the absence of monopoles (if the gauge field is non-compact), there is a conserved current $\partial_\mu j^\mu_\mathrm{top} = 0$
\begin{equation}
	j^\mu_\mathrm{top} = \frac{1}{2\pi} \epsilon^{\mu \nu \rho} \partial_\nu a_\rho.
\end{equation}
We have normalized the current so that $j^0_\mathrm{top} = \frac{1}{2\pi} (\partial_x a_y - \partial_y a_x)$ implies that a $2 \pi$ flux of the emergent gauge field carries topological charge $1$.
It is useful to write
\begin{equation}
	j^\mu_\mathrm{top} = \frac{1}{2\pi} \epsilon^{\mu \nu \rho} \partial_\nu a_\rho = \frac{1}{4\pi} \epsilon^{\mu \nu \rho } f_{\nu \rho},
\end{equation}
which implies that the CS term can be written as $\mathcal{S}_\mathrm{CS} = - {2 i  C} A_\mu j^\mathrm{\mu}_\mathrm{top} = -\frac{ i  C}{2 \pi} \epsilon^{\mu \nu \rho} A_\mu f_{\nu \rho}$.
Note that the topological charge conservation is equivalent to the Bianchi identity.
In contrast, if we some finite number of monopoles, $j^\mu_\mathrm{top}$ is no longer conserved. Instead, one has
\begin{equation}
	\partial_\mu j^\mu_\mathrm{top} = m(x)
\end{equation}
where $m(x) = \sum_i n_i \delta(x-x_i)$ is some \emph{fixed} function that corresponds to a monopole of charge $n_i$ at position $x_i$.

For such a fixed configuration, we can avoid having to integrate over non-trivial gauge backgrounds instead formulate a path integral for $f_{\mu \nu}$ with an appropriate constraint,
\begin{multline}
	Z[m,A] = \int \mathcal{D}[f] \, 
	\delta(\partial_\mu j^\mu_\mathrm{top} - m) e^{-\int d^3 x \, \mathcal{S}_\mathrm{CS}[f;A] + \mathcal{S}_\mathrm{mw}[a]} .
\end{multline}
We now introduce a field $\phi$ to impose the delta function, normalization of $j_\mathrm{top}$ implies that we should identify $\phi \simeq \phi + 2\pi$. 

Because above action is quadratic in $f$, we can integrate out the field strength of the dynamical gauge field exactly.
To do so, we partially integrate the second-to last term once, and write the action as 
\begin{equation}
	\mathcal{L} = \frac{ i }{4\pi} \underbrace{\left(\partial_\mu \phi - \frac{C}{2} A_\mu \right)}_{\equiv D_\mu \phi} \ \epsilon^{\mu \nu \rho} f_{\nu \rho} + \frac{1}{4 g^2} f_{\mu \nu } f^{\mu \nu} +  i  \phi m(x). \label{eq:phiaLagrangian}
\end{equation}

Integration of the quadratic emergent electromagnetic field leads to
\begin{widetext}
\begin{equation} \label{eq:stueckelberg}
	Z[m,A] = \int \mathcal{D}[\phi] \exp\left[ - \int d\tau d^2 x  \, \frac{g^2}{8 \pi^2} \left( \partial_\mu \phi - 2 C A_\mu \right) \left( \partial^\mu \phi - 2 C A^\mu \right) +  i  m(x) \phi(x) \right].
\end{equation}
\end{widetext}
In the case of $m \equiv 0$ and setting $A\equiv 0$, we see that $\phi$ is a gapless mode -- this is the ``dual photon'' of the dynamical U(1) gauge theory in 2+1 dim. which can be understood as a Goldstone mode of the spontaneously broken U(1)$_\mathrm{top}$ symmetry (acting as U(1)$_\mathrm{top}:\phi \to \phi + c$) for some $c$.

\subsection{Fermionic calculation: Zero modes trapped at monopoles}

For notational convenience, we set the speed of fermions to unity in this section.

\subsubsection{Zero modes at a monopole}

We consider a monopole-antimonopole pair with string along the (imaginary) time direction and placed at the origin of the system. For illustration, we consider the following monopole event:  We consider a hole of width $L/2\pi$  (i.e.~circumference $L$) in the spatial plane and denote the coordinate along the perimeter of the hole as $x$. We will ramp a flux through this hole. The low-energy action for this configuration contains helical 1D fermions running along the perimeter which have an action

\begin{align}
S = \int_{\tau,x} \bar \psi \underbrace{[D_\tau -i  D_x \tau_z]}_{\mathcal K} \psi,
\end{align}
where $\tau_z$ is the space of right/left movers, which coincides with the space of colors (note that there is no further degeneracy). Further, $D_\mu = \partial_\mu + i a_\mu + i A_\mu  \tau_z$, with $\mu = (x,\tau)$.

A monopole configuration in the charge gauge field is, for example, $a_\mu =\delta_{\mu, x} \frac{2\pi}{L} f(\tau)$, where $f(\tau)$ is a kink which monotonously interpolates between $0$ and $1$ for monopole and between $1$ and $0$ for an antimonopole. The monopole anti-monopole pair is give by a kink/antikink configuration at distance $\Delta \tau$.

We need to study the non-Hermitian kernel $\mathcal K$, which can be conveniently written as (for simplicity, we briefly drop $A_\mu$)
\begin{align}
\mathcal K = i \sigma_y \underbrace{[(- i D_x, -i D_\tau)\cdot (\sigma_x, \sigma_y)]}_{\equiv H}.
\end{align}
The Hermitian part $H$ is nothing but a Hamiltonian for Dirac fermions in $2$ spatial dimensions subjected to a magnetic field $B = \nabla \wedge a = - 2\pi \dot f(\tau)/L$. In the simplest case of a linear ramp the magnetic field is  homogeneous leading to Landau like levels (but in time-circumference space). Even for a generic $f(\tau)$ we find a total flux $2\pi [f(\infty)- f(-\infty)] \in 2\pi \mathbb Z$, hence the Landau level degeneracy is $1$ per monopole charge. By Atiyah-Singer, there is at least one zero mode per flux through the system.

We next consider a single monopole anti-monopole pair and calculate the ``Hamiltonian'' $H$ associated to this field configuration. We find
that at lowest energies
\begin{equation}
H = \left (\begin{matrix}
0 & \lambda \\ \lambda &0
\end{matrix} \right),
\end{equation}
where $\lambda \propto e^{- [( \tau_{\rm monople} - \tau_{\rm antimonopole})]/X_0}$ is the hybridization of the two dynamical fermionic zero modes located at the monopole/antimonopole position. The reference scale $X_0$ is set by UV physics.

Taking the fermionic partition sum, we thus find a fugacity of an arbitrary monopole-antimonopole pair (we go back to arbitrary orientation of the Dirac string in 2+1 D space time)
\begin{equation}
y_{\rm monopole/antimonopole} \sim e^{- \vert \Delta X \vert/X_0}.
\end{equation} 
The monopoles are thus linearly confined and can not proliferate. 

\subsubsection{Calculation of the monopole-antimonopole correlator}
Next we calculate the monopole-antimonopole correlator, recovering the long range correlations discussed in~\cite{RanWen2009} in a different context and slightly different field theory. 

First remember that the partition sum is given by the sum over all monopole sectors

\begin{widetext}
\begin{align}
    Z &= \sum_{N, \lbrace n_j \rbrace} \frac{y_0^N}{X_0^{3N} N!} \int \prod_{j = 1
}^N d^3X_j \int \mathcal Da \mathcal D\psi \delta \left ( \epsilon_{\mu \nu \rho} \partial_\mu \partial_\nu a_\rho - \sum_{j =1}^N 2\pi n_j \delta(X-X_j) \right) e^{-S[\psi,a_\mu] - S_{\rm mw}}.
\end{align}
Here, $y_0$ is the monopole fugacity without the influence (i.e.~suppression) by the fermions and $n_i$ its charge. 

As we just saw, the zero modes impede any monopoles to be further away from each other than a typical distance $X_0$. Therefore, the phase space for monopole contribution to the partition sum (i.e.~free energy) is extremely small. 

Next we want to calculate the expectation value for a pair of monopole/antimonopole at distance $\Delta X= X_+-X_-$ apart. This is

\begin{align}
     C(\Delta X) &= \frac{1}{Z} \sum_{N, \lbrace n_j \rbrace} \frac{1}{X_0^{3N} N!} \int \prod_{j = 1
}^N d^3X_j \int \mathcal Da \mathcal D\psi \notag \\ &\delta \left ( \epsilon_{\mu \nu \rho} \partial_\mu \partial_\nu a_\rho - \sum_{j =1}^N  2\pi n_j \delta(X-X_j) - 2\pi \delta (X- X_+) +2\pi \delta (X- X_-) \right) e^{-S[\psi,a_\mu] - S_{\rm mw}}.
\end{align}
\end{widetext}
We expand this expression in orders of $y_0$.
We estimate
\begin{align}
    C(\Delta X) & \sim e^{-\vert \Delta X \vert/X_0} + \frac{y_0^2}{X_0^{6} 2} \int d^3R_+d^3R_- e^{-\frac{\vert X_+ - R_- \vert}{X_0}} e^{-\frac{\vert X_+ - R_- \vert}{X_0}} \notag \\
    & = e^{-\vert \Delta X \vert/X_0} + \# y_0^2\frac{X_0^6}{
    X_0^6}.
\end{align}

From this basic argument we see that there is a finite expectation value of the monopole operator, i.e.~true long range order.

\subsection{Superfluid vortices}

We next search for vortex solutions in the superfluid phase field $\phi$. Eq.~\eqref{eq:phiaLagrangian} of the previous section summarizes that effectively an electric field configuration which points radially outwards from the origin $\v e \propto \hat r$ induces a vortex in $\phi$. This is the essence of particle-vortex duality. 

In the gauge theory language a vortex (chosen to reside at the origin) thus corresponds to an emergent gauge charge at the origin. It could either be a fermionic or bosonic density. 

\subsubsection{Bosonic emergent charge accumulation}

The smallest boson charge can be $1 \tilde g$ and we focus on the latter.

The equations of motion of the $\Phi$ field is
\begin{align}
    0 & = [(-i \vec \nabla_b - \vec a)^2 + m_b^2 ]\Phi. 
\end{align}
As we are looking for a solution with finite $\Phi$, akin to the Higgs condensate, we expect that there is no $\vec a$ in the vortex core and we solve the equation without $\vec a$. 

We furthermore look for solutions with the property
\begin{equation}
    j_0 = i \Phi^* \dot{\Phi} - i \dot{\Phi}^* \Phi = \tilde \delta(x,y) ,
\end{equation}
where $\tilde \delta$ is a  function sharply peaked at the origin whose integral is normalized to unity. 

In the limit when temperature $T$ is small as compared to $m_b$ the solutions can be written in terms of the two modified Bessel functions: $ e^{-i \tau T} I_0(r m_b/v_b),e^{-i \tau T} K_0(r m_b/v_b)$. The former grows towards infinity but the function and its derivative are integrable near the origin. The latter decays towards infinity, but its derivative at the origin diverges as $1/r$. We thus combine the two into a continuous (yet not continuously differentiable), integrable function
\begin{align}
    \Phi_{\rm charge}(r) &= \Phi_0 e^{-i \tau T}  \sqrt{\frac{m_b^2 }{T v_b^2}}  \notag\\
    &[K_0(r m_b/v_b) \theta(r-r_*)+ I_0(r m_b/v_b) \theta(r_*-r)],
\end{align}
where $r =  \sqrt{x^2 + y^2}$ and $r_*$ is defined by $K_0(r m_b/v_b) = I_0(r m_b/v_b)$. It sets the scale of the puddle of emergent charge and, in dual language, the size of the vortex.

The dimensionless prefactor $\Phi_0$ is fixed by the condition $\int d^2x j_0 = 1$, where
\begin{align}
    j_0 &= \frac{ 2 \Phi_0 m_b^2}{v_b^2} \notag\\
    & \times [K_0(r m_b/v_b) \theta(r-r_*)+ I_0(r m_b/v_b) \theta(r_*-r)]^2.
\end{align}

Above $U_c$ we can drop the nonlinearity in the $\Phi$ action, so we can readily integrate Gaussian fluctuations on top of $\Phi_{\rm charge}$ along with the fermions. This restores the effective gauge coupling quoted in the main text. We then obtain the following Gau{\ss} law for the electric field outside the charge accumulation
\begin{equation}
    \vec \nabla \cdot \v e = g^2 j_0 \simeq g^2 \delta(x,y)
\end{equation}
to find, outside the charge puddle, the electric field induced by the bosonic point charge.

\subsubsection{Fermionic emergent charge accumulation} 

The smallest fermion induced emergent charge $j_0 = \sum_\alpha \bar \psi_\alpha \psi_\alpha = \tilde \delta(x,y)$ automatically implies the accumulation of a physical charge, $\sum_\alpha \alpha \bar \psi_\alpha \psi_\alpha = \pm \tilde \delta(x,y)$. Here, the typical length scale in the fermionic field is set by $v_f/m_-$ and plays the role of coherence length, here.

This in turn implies that the vortices at the topological transition out of the superfluid carry unit physical charge. Assuming these vortices proliferate at the transition suggests the a (semi-)metallic character of the state at the transition. 

\begin{figure}
    \centering
    \includegraphics[width=\columnwidth]{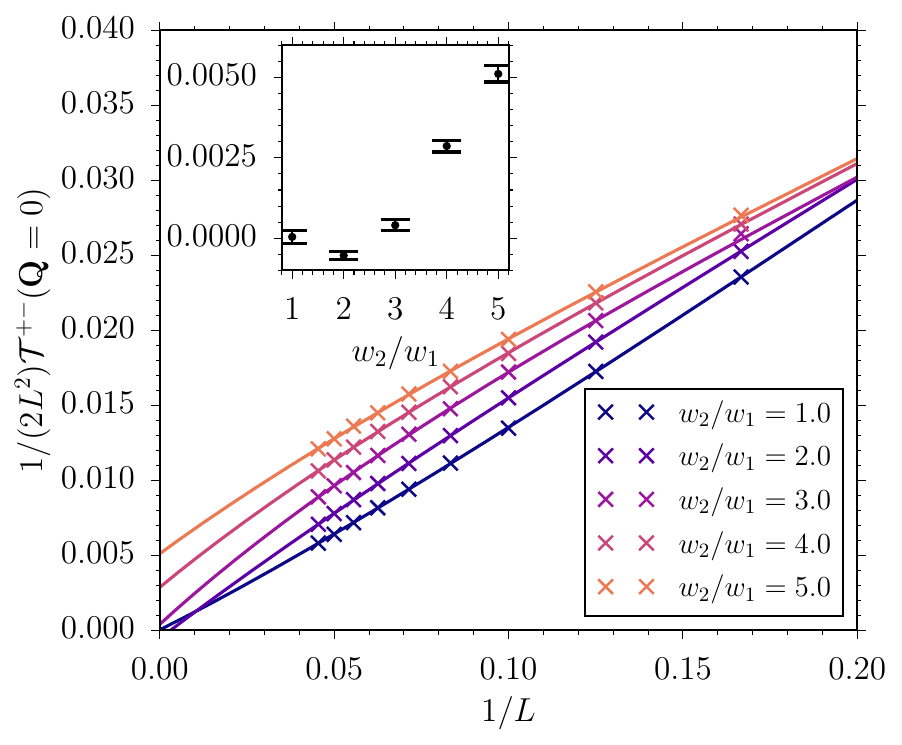}
    \caption{Scaling of the static structure factor for in-plane pseudospin correlations at zero momentum (i.e. $s$-wave pair density correlations) obtained from a Monte-Carlo evaluation of the Gutzwiller-projected ground-state wavefunction for different choices of $w_2/w_1= 5$ with $w_0$. The continuous line is a fit to a scaling function. The inset displays the $a_0$ (the intercept), corresponding to the square of the ordered moment since $\mathcal{T}^{+-}(0) \sim (t_0) (2L^2) + \dots$}
    \label{fig:gutzwiller-scaling}
\end{figure}

\section{``Anti''-Gutzwiller projection} \label{sec:gutzwiller-appendix}

We consider a given free-fermion wavefunction of the fermionic chargeon bands and project it into the subspace of empty and doubly occupied sites, which we refer to as ``anti''-Gutzwiller projection. While it is not feasible to explicitly compute the projected wavefunction beyond small system sizes, we may use Monte Carlo sampling to evaluate observables with respect to the projected wavefunction. Here, we aim to study whether the pair-correlation function exhibits off-diagonal long-range order.
In the language of pseudospin $\vec T_{\v x}$ -- see also Eq.~\eqref{Eq:Hamiltonian} --acting on the empty/doubly-occupied subspace, this corresponds to Bragg peaks for in-plane ordering at wavevector $\mathbf{Q}$, i.e. $\mathcal{T}^{+-}(\mathbf{Q}) \propto T^2 N$~\cite{KanekoGros16,WuShen19} in $\mathcal{T}^{+-}(\mathbf{Q}) = \frac{1}{N} \sum_{\v x,\v x'} \langle T_{\v x}^+ T_{\v x'}^- \rangle e^{-i \mathbf{Q} \cdot (\v x -\v x')}$.
Here, we focus on the case $\mathbf{Q} =0$, where we observe the strongest correlations.
We compute $\mathcal{T}$ on clusters of $L \times L$ unit cells ($N=2L^2$ sites) with $L=6, \dots, 22$, and then fit the data to a scaling function  $(2L^2)^{-1} \mathcal{T}^{+-}(0) \sim t_0 + t_1 L^{-1} + t_2 L^{-2}$ \cite{NeubergerZiman89}.
Here, $t_0$ corresponds to the square of the ordered moment.
We find that $t_0 > 0$, corresponding to spontaneous symmetry breaking, for $w_2/w_1 \gtrsim 3$, as shown in the inset of Fig.~\ref{fig:gutzwiller-scaling}.
For smaller values of $w_2/w_1$, we find $t_0 \approx 0$ and $t_0 < 0$, but we suggest that these results should be interpreted with caution: here, finite-size corrections might be more sizeable and including data points from larger system sizes could result in a small but finite $t_0 >0$ (note, for example, that for small $w_2/w_1$, the fit function is concave, rather than convex).

We therefore suggest that the projected wavefunction exhibits spontaneous $U(1)$ symmetry breaking for sufficiently large $w_2/w_1$ in the thermodynamic limit, but our analysis is inconclusive pertaining to its persistence in the limit of small $w_2/w_1$.

\end{document}